\documentclass[amsmath,nofootinbib,amssymb,prd]{revtex4}
\pdfoutput=1

\usepackage[utf8]{inputenc}
\usepackage{float}
\usepackage{epsfig}
\usepackage{graphicx}
\usepackage{color}
\usepackage{tabularx}
\usepackage{color}
\usepackage{verbatim}
\usepackage[inline, final]{showlabels}
\usepackage{tensor}
\usepackage{enumerate}
\usepackage{subfigure}

\usepackage{color}
\usepackage[bookmarks,colorlinks,linkcolor=webblue,citecolor=webgreen]{hyperref}	
\definecolor{webgreen}{rgb}{0, 0.5, 0} 
\definecolor{webblue}{rgb}{0, 0, 0.5} 
\definecolor{webred}{rgb}{0.5, 0, 0} 

\newcommand{\av}[1]{\left\langle #1 \right\rangle}
\newcommand{\Od}[1]{\mathcal{O}\!\left( #1 \right)}
\newcommand{\ee}{\text{e}} 
\newcommand{\ii}{\text{i}}
\newcommand{\di}{\text{d}}
\newcommand{\n}{\phantom{a}}
\newcommand{\pd}[2]{\frac{\partial #1}{\partial #2}}

\renewcommand{\v}[1]{\boldsymbol{#1}}

\renewcommand{\d}[2]{\frac{\text{d} #1}{\text{d} #2}}

\begin{document}
\preprint{}
\title{Non-comoving Cosmology}
\author{J.\,A.\,R.\,Cembranos\footnote{cembra@ucm.es}, 
        A.\,L.\,Maroto\footnote{maroto@ucm.es},
        and H.\,Villarrubia-Rojo\footnote{hectorvi@ucm.es}}
\affiliation{Departamento de  F\'{\i}sica Te\'orica and Instituto de Física de Partículas y del Cosmos IPARCOS, Universidad Complutense de Madrid, E-28040 Madrid, Spain}
\date{\today}

\begin{abstract}
    One of the fundamental assumptions of the standard $\Lambda$CDM cosmology is that,  
    on large scales,  all the matter-energy components of the Universe share a common rest frame. 
    This seems natural for the visible sector, that has been in thermal contact and tightly coupled in 
    the primeval Universe. The dark sector, on the other hand, does not have any non-gravitational 
    interaction known to date and therefore, there is no \emph{a priori} reason to impose that it is comoving 
    with ordinary matter. In this work we explore the consequences of relaxing this assumption and study 
    the cosmology of non-comoving fluids. We show that it is possible to construct a homogeneous and 
    isotropic cosmology with a collection of fluids moving with non-relativistic velocities. Our model 
    extends $\Lambda$CDM with the addition of a single free parameter $\beta_0$, the initial velocity of 
    the visible sector with respect to the frame that observes a homogeneous and isotropic universe. This 
    modification gives rise to a rich phenomenology, while being consistent with current observations for 
    $\beta_0<1.6\times 10^{-3}$ (95\% CL). This work establishes the general framework to describe a 
    non-comoving cosmology and extracts its first observational consequences for large-scale structure. 
    Among the observable effects, we find sizeable modifications in the density-velocity and 
    density-lensing potential cross-correlation spectra. These corrections give rise to deviations from
    statistical isotropy with a dipolar structure. The relative motion between the different 
    fluids also couples the vector and scalar modes, the latter acting as sources for  metric vector 
    modes and vorticity for all the species. 
\end{abstract}

\maketitle
\tableofcontents

\section{Introduction}
    The isotropy and homogeneity of the Universe on large scales are two foundational assumptions
    of the standard cosmological model, the so-called $\Lambda$CDM model. These two assumptions are usually grouped under the
    name of Cosmological Principle. All the observational evidence, ranging from the extremely isotropic
    cosmic microwave background (CMB) \cite{Ade:2015hxq, Akrami:2018vks} to the galaxy number counts and the measured expansion
    from SNIa \cite{Kolatt:2000yg, Antoniou:2010gw, Jimenez:2014jma}, supports the conclusion that 
    the Universe is very nearly isotropic on large scales.
    However, the notions of homogeneity and isotropy are inextricably linked with the election of a 
    privileged frame. For any observer moving with respect to this frame, the Universe would appear 
    anisotropic and inhomogeneous. This is precisely our situation on Earth. \\
    
    Starting with the early CMB measurements \cite{Kogut:1993ag, Lineweaver:1996xa}, a significant dipole 
    modulation, much larger than any other anisotropy, was found. This was readily interpreted as a 
    kinematical effect: a Doppler shifting effect arising from the relative motion of the Earth with 
    respect to the CMB rest frame, i.e. a frame in which the CMB looks isotropic. Recent analysis by 
    the \emph{Planck} Collaboration \cite{Aghanim:2013suk, Ade:2015hxq}
    explored other kinematical effects, like the violation of statistical isotropy induced by the observer
    motion, and reported an independent measurement of our relative velocity with respect to the CMB
    frame. This measured velocity can, given the uncertainties, fully account for the observed dipole, 
    supporting its kinematical origin. Even if it is mostly kinematical, it may still contain an 
    intrinsic contribution. Some authors have proposed searches for the intrinsic dipole, e.g. using
    spectral distortions \cite{Yasini:2016dnd}.\\
    
    A different kind of dipole should appear in the distribution of galaxies, induced by our motion with
    respect to the matter frame, i.e. a frame in which the matter distribution looks isotropic. The 
    origin of the large scale structure (LSS) dipole lies in a combination of Doppler shifting and
    aberration effects in the galaxy number counts \cite{ellis1984expected, Gibelyou:2012ri}. Unfortunately, current 
    observations can only loosely constrain its amplitude and direction, yielding a value compatible 
    with the CMB dipole \cite{Condon:1998iy, Itoh:2009vc}. Future surveys like Euclid 
    \cite{Amendola:2016saw} and SKA \cite{Maartens:2015mra} will measure it with unprecedented accuracy.\\
    
    To complete the picture, we only need to know the relative velocity between the matter and CMB frames.
    Concerning this point, $\Lambda$CDM contains the underlying assumption, that usually goes by 
    unnoticed, that both frames coincide. $\Lambda$CDM assumes matter and CMB to be \emph{comoving}.
    As we will see in this work, it is possible to relax this condition. The homogeneous and isotropic
    Robertson-Walker (RW) metric can be sourced, at the background level, using non-comoving fluids. 
    Thus, we will show that it is possible to construct a viable cosmological model for non-comoving 
    fluids, with interesting phenomenological consequences and without any flagrant isotropy violation.\\
    
    Early theoretical work concerning non-comoving fluids was mostly developed under the framework of
    tilted universes. The term was coined in the groundbreaking work by King and Ellis \cite{King:1972td}.
    The authors considered a class of homogeneous models sourced by a single moving fluid, i.e. models
    in which the fluid 4-velocity is tilted with respect to the homogeneous hypersurfaces. These tilted
    models produce homogeneous but anisotropic universes. In a different context, Coley and Tupper
    \cite{coley1986two} analyzed two-fluid cosmological models with general imperfect and non-comoving
    fluids. In order to source homogeneous and isotropic RW metrics, only very special configurations
    with radial velocities were considered. Later on, Turner \cite{Turner:1991dn} proposed a theoretical 
    mechanism to produce a mismatch between matter and CMB velocities. In Turner's tilted universes,
    the presence of a near-horizon-sized perturbation, remnant of inflation, could introduce a spatial
    gradient, driving the velocity of matter.\\
    
    The analysis of non-comoving fluids has been extended to dark energy. Given our fundamental ignorance
    about the behaviour of the dark sector, it is conceivable that it has not ever been coupled to 
    ordinary matter and that it does not share the same rest frame. Following this idea, a model of moving 
    dark energy was proposed in \cite{Maroto:2005kc}. In this case, for a dynamical dark energy fluid, 
    even if the matter and CMB frames coincide initially, they differ at late times. Different models of 
    moving homogeneous dark energy were analyzed in \cite{BeltranJimenez:2007rsj}, as well as its possible 
    impact in observables like the CMB quadrupole. The construction of a fully anisotropic model in which 
    the full dark sector, i.e dark energy and dark matter, is non-comoving with the CMB and ordinary matter 
    was carried out in \cite{Harko:2013wsa}. The authors analyzed a Bianchi I universe in which dark 
    matter and dark energy had different relative velocities with respect to the frame of ordinary 
    matter and then derived some observables, like a modified luminosity-distance relation and CMB 
    quadrupole.\\
    
    From the observational point of view, a signal of the relative motion of the matter and CMB frames 
    will be the detection of a large-scale bulk flow. In recent years, several works have claimed 
    measurements of matter flows well in excess the $\Lambda$CDM predictions 
	on different scales and at different statistical confidence levels \cite{Kashlinsky:2008ut, Ade:2013opi, Scrimgeour:2015khj}. 
	Although there seems to be a broad agreement on the direction of the flow, the amplitude is
	still  subject to  controversy \cite{Ade:2013opi}. Such flows will be an indication of the existence 
	of a cosmological preferred spatial direction. On the other hand, detected anomalies in the low 
	multipoles of the CMB  temperature power spectrum \cite{Ade:2015hxq}, such as the low-multipole 
	allignment and the dipolar or hemispherical anomalies, also suggests the presence of a preferred 
	cosmological direction \cite{Schwarz:2015cma}. This fact has triggered the search for mechanisms 
	which could break isotropy while keeping the 	predictions of the standard cosmology.\\    
    
    This work builds upon these previous studies, but we will present the first complete analysis for the
    evolution of a set of non-comoving fluids, from the early to the late Universe, both at the
    background and perturbation level. As we will see later, it is reasonable to assume that any pair
    of tightly coupled fluids share the same velocity. Hence, we can expect that photons, baryons
    and neutrinos, being in thermal contact in the early Universe, shared a common rest frame, i.e. a 
    frame in which the plasma looked isotropic. However, there is no \emph{a priori} reason to assume 
    the same about the dark sector. The dark sector, regardless of its composition, may very well possess 
    its own rest frame, with a given global velocity with respect to the visible sector. The only 
    reasonable assumption is that there is \emph{one} frame that observes a homogeneous and isotropic 
    universe, i.e. a RW background. In this paper we take seriously this possibility and prove, at the 
    background level, that
    \begin{itemize}
        \item If the dark and visible sectors are initially moving with non-relativistic 
            velocities, to first order in these velocities, it is possible to define a \emph{cosmic center 
            of mass} frame that observes a RW background. Our model only contains one additional 
            free parameter $\beta_0$, the velocity of the visible sector with respect to the cosmic center 
            of mass frame deep inside the radiation era. As we will see later, 
            $\beta_0 < 1.6\times 10^{-3}$ is a conservative limit to be in agreement with all observations.
        \item The subsequent evolution gives rise to relative velocities between all the different 
            components of the visible sector, e.g. between baryons and photons, defining each of them its
            own rest frame.
    \end{itemize}
    While at the perturbation level,
    \begin{itemize}
        \item There appear couplings, of order $\beta_0$, between scalar and vector modes.
        \item There is a production of vorticity for all the species and a net production of vorticity
            and metric vector modes.
        \item The transfer function of every cosmological quantity acquires a dipolar contribution
            of order $\beta_0$. 
        \item There is a violation of statistical isotropy of order $\beta_0$. Among the different
            observables where such an effect could be measured, the easiest to compute is the
            cross-correlation spectrum between different scalar perturbations, which acquire a dipolar
            contribution of order $\beta_0$.
    \end{itemize}
    
    In this work we will limit ourselves to LSS observables, letting for a forthcoming work the analysis
    of CMB signatures \cite{CMB_paper}.
    The structure of the paper goes as follows. Section \ref{sec:MovingFluids} assesses the problem of
    constructing a homogeneous and isotropic universe using non-comoving fluids. First, only the 
    energy-momentum tensor for a perfect fluid with bulk velocity is considered in \ref{sec:perfect_fluid},
    where the conditions for isotropy and homogeneity are discussed. Then, the formalism is extended
    to imperfect fluids in \ref{sec:kin_approach}, using the kinetic approach, where both background and
    perturbations are studied in \ref{sec:kin_background} and \ref{sec:kin_perturbations}. Section 
    \ref{sec:BoltzmannEquation} analyzes the dynamics of this model from the point of view of the
    Boltzmann equations. The free-streaming term is derived in \ref{sec:boltzmann_lhs}, while the
    collision term for the photon-baryon plasma is computed in \ref{sec:boltzmann_rhs}. The
    main evolution equations are presented in \ref{sec:boltzmann_final}. The usual scalar-vector-tensor 
    decomposition is performed in section \ref{sec:AngularAnalysis}, where we describe our approximation
    scheme. Section \ref{sec:EinsteinEquations} contains the Einstein equations, which are not modified in our case.
    In section \ref{sec:ReducedSystem}, we present a reduced version of the original system, under our
    approximation scheme. Sections \ref{sec:bulk_velocities}, \ref{sec:scalar_modes} and
    \ref{sec:vector_modes} are devoted to bulk velocities, scalar and vector modes, respectively.
    Finally, section \ref{sec:ResultsSpectra} contains the numerical solution of the aforementioned
    reduced system and discusses some observables.
    Section \ref{sec:Conclusions} gathers the main conclusions and presents some prospects for future
    work.

\section{Moving fluids}\label{sec:MovingFluids}
    \subsection{Perfect fluid with bulk velocity}\label{sec:perfect_fluid}
        \subsubsection{Physical setting}
            Let us consider a perfect fluid with energy-momentum tensor
            \begin{equation}\label{eq:Tmunu_fluid_def}
                T\indices{^\mu_\nu} = (\rho + P)u^\mu u_\nu + P\delta\indices{^\mu_\nu}\ ,
            \end{equation}
            in a flat Robertson-Walker (RW) metric
            \begin{equation}\label{eq:RW_metric}
                \di s^2 = a^2(\tau)\left(-\di\tau^2 + \delta_{ij}\di x^i\di x^j\right)\ .
            \end{equation}
            Now we will consider the situation where the fluid  possesses a bulk velocity with respect
            to the frame in which the metric takes the form \eqref{eq:RW_metric}. Parameterizing the
            four-velocity as
            \begin{equation}
                u_\mu = a\gamma(-1,\,v_i)\ ,
            \end{equation}
            from the normalization condition, $u_\mu u^\mu =-1$, we have
            \begin{equation}
                \gamma = \frac{1}{\sqrt{1-v^iv_i}}\ ,
            \end{equation}
            where the spatial indices in $v_i$ are lowered and raised using $\delta_{ij}$. With this
            parameterization the components of the energy-momentum tensor are
            \begin{subequations}\label{eq:Tmunu_components}
            \begin{align}
                T\indices{^0_0} &= -\rho - (\rho + P)\gamma^2v^2\ ,\\
                T\indices{^0_i} &= (\rho + P)\gamma^2 v_i\ ,\\
                T\indices{^i_j} &= P\delta\indices{^i_j} + (\rho + P)\gamma^2v^iv_j\ .
            \end{align}
            \end{subequations}
            Since the non-diagonal components are not zero, this moving fluid cannot act as a source for
            the geometry \eqref{eq:RW_metric}. Let us show how to construct a valid source for the homogeneous
            and isotropic metric \eqref{eq:RW_metric} using a collection of fluids.\\
            
            \paragraph{Isotropy.} For non-relativistic fluids, the only non-diagonal component, to first
                order in $v$, is
                \begin{equation}
                    T\indices{^0_i} = (\rho + P)v_i\ .
                \end{equation}
                If instead of a single fluid we have several fluids in relative motion, they can act
                as a source for \eqref{eq:RW_metric} if they satisfy
                \begin{equation}\label{eq:CM_condition}
                    T\indices{^0_i} = \sum_s {T_s}\indices{^0_i} = \sum_s(\rho_s + P_s)v_{s\,i} = 0\ .
                \end{equation}
                The physical content of this condition is that of a kind of \emph{center of mass} frame
                condition. An isotropic source can be constructed, to first order in $v$, out of two
                non-relativistic fluids if the net flux of momentum of one fluid is counterbalanced by
                that of the other fluid. We will see later that this constraint is conserved in time, 
                so it can be implemented with an appropiate choice of the initial conditions. In 
                \cite{BeltranJimenez:2007rsj} it is discussed how to transform to this frame, starting
                from an arbitrary configuration of the fluids.\\
            
            \paragraph{Homogeneity.} Homogeneity is easily implemented when the fluids are at rest, but
                we need to be cautious in our context. Consider two observers locally related by a boost
                \begin{itemize}
                    \item $(\tau, \v{x})$, $\mathcal{O}$ frame in which \eqref{eq:CM_condition} is
                        satisfied and the metric takes the form \eqref{eq:RW_metric}.
                    \item $(\tilde{\tau}, \v{\tilde{x}})$, $\tilde{\mathcal{O}}$ frame moving with respect
                        to $\mathcal{O}$ with velocity $\v{\beta}$.
                \end{itemize}
                In the $\tilde{\mathcal{O}}$ frame, the transformed coordinates are obtained applying a local
                Lorentz transformation
                \begin{equation}\label{eq:LocalLorentz_transformation}
                    \di\tilde{x}^\mu = \Lambda\indices{^\mu_\nu}(\beta)\di x^\nu\ ,
                \end{equation}
                and the metric looks inhomogeneous
                \begin{equation}
                    \di s^2 = a^2\big(\tau(\tilde{\tau}, \tilde{\v{x}})\big)\left(-\di\tilde{\tau}^2
                        +\delta_{ij}\di \tilde{x}^i\di \tilde{x}^j\right)\ .
                \end{equation}
                The same applies to other time-dependent quantities like $\rho$ and $P$. To provide
                a consistent source, we will require that the energy-momentum tensor of each fluid is
                homogeneous in the $\mathcal{O}$ frame, i.e. the frame that observes an isotropic and
                homogeneous metric, \emph{not} in the comoving frame with the fluid.\\
                                            
            Notice that on the tangent space at a given space-time point, we can always define an
            orthonormal basis $\v{e}_a$ given by
            \begin{equation}
                \v{e}_a = e\indices{^\mu_a}\pd{}{x^\mu}\ ,\qquad e\indices{^\mu_a}=a^{-1}(\tau)\delta\indices{^\mu_a}\ .
            \end{equation}            
            and the corresponding orthonormal basis in the $\tilde{\mathcal{O}}$ frame reads
            \begin{equation}\label{eq:vierbein_background}
                \v{\tilde{e}}_a = \tilde{e}\indices{^\mu_a}\pd{}{\tilde{x}^\mu}\ ,
                    \qquad \tilde{e}\indices{^\mu_a}=a^{-1}(\tau(\tilde{x}))\delta\indices{^\mu_a}\ .
            \end{equation}
            Thus, as expected, the two basis are just related by a Lorentz transformation on the tangent
            space
            \begin{equation}
                \v{\tilde{e}}_a = \left(\Lambda^{-1}(\beta)\right)\indices{^b_a}\v{e}_b\ .
            \end{equation}
                
            To sum up, we can source a flat RW metric \eqref{eq:RW_metric} with a collection of 
            non-relativistic moving fluids as long as
            \begin{itemize}
                \item We are in the center of mass frame, where
                    \begin{equation}
                        \sum_s (\rho_s + P_s)v_s^i = 0\ .
                    \end{equation}
                \item The energy-momentum tensor is homogeneous in that frame
                    \begin{equation}
                        \partial_iT\indices{^\mu_\nu} = 0\ .
                    \end{equation}
            \end{itemize}
            
        \subsubsection{Evolution}
            Finally, in this subsection we will analyze the evolution of a perfect fluid with bulk 
            velocity. Assuming homogeneity
            \begin{equation}
                \partial_i T\indices{^\mu_\nu} = 0\ ,
            \end{equation}
            the conservation equation
            \begin{equation}
                \nabla_\mu T\indices{^\mu_\nu} = 0 \ ,
            \end{equation}
            for a flat RW metric in conformal time \eqref{eq:RW_metric} yields the equations of motion
            \begin{align}
                \partial_0\left(a^{3}T\indices{^0_0}\right) &= \mathcal{H}a^3T\indices{^i_i}\ ,
                    \label{eq:Tmunu_cons_energy}\\
                \partial_0\left(a^4T\indices{^0_i}\right) &= 0\ , \label{eq:Tmunu_cons_momentum}
            \end{align}
            where $\mathcal{H}\equiv \dot{a}/a$ is the conformal Hubble parameter and 
            $\dot{\n}\equiv \partial_0\equiv \partial/\partial\tau$. As it was anticipated, 
            \eqref{eq:Tmunu_cons_momentum} proves that the center of mass constraint 
            \eqref{eq:CM_condition} is conserved in time. Writing explicitly the components 
            \eqref{eq:Tmunu_components}, the equations of motion can be rewritten in terms of $v_i$ and
            $\rho$. The equation for the velocity can be expressed
            as
            \begin{equation}
                \partial_0v_i - \partial_0\log\left((1+w)a^4\gamma^2\rho\right)v_i = 0\ ,
            \end{equation}
            where $w\equiv P/\rho$ is the equation of state of the fluid.
            In the absence of interactions, the velocity does not change its direction, so we only need
            to follow the evolution of its magnitude.
            Combining \eqref{eq:Tmunu_cons_energy} and \eqref{eq:Tmunu_cons_momentum}, it is possible
            to obtain
            \begin{align}
                \dot{\rho} &= \frac{(v^2-3)(1+w)}{1-wv^2}\mathcal{H}\rho + \frac{\dot{w}}{1-wv^2}v^2\rho\ ,
                    \label{eq:Tmunu_rho_eq}\\
                \dot{v} &= \frac{(1-v^2)(3w-1)}{1-wv^2}\mathcal{H}v 
                    + \frac{\dot{w}}{1+w}\frac{1-v^2}{1-wv^2} v\ ,
                    \label{eq:Tmunu_v_eq}
            \end{align}
            assuming $w\neq -1$.
            It is worth particularizing these results to two kind of fluids.
            \begin{itemize}
                \item \emph{Radiation}, $w = 1/3$.
                    \begin{align}
                        \dot{\rho} &= -4\mathcal{H}\rho\ ,\\
                        \dot{v} &= 0\ .
                    \end{align}
                    The fluid moves with constant velocity and with the usual scaling $\rho\propto a^{-4}$.
                \item \emph{Matter}, $w = 0$.
                    \begin{align}
                        \dot{\rho} &= (v^2-3)\mathcal{H}\rho \ ,\\
                        \dot{v} &= -(1-v^2)\mathcal{H}v\ .
                    \end{align}
                    The equations can be solved analitically in this case
                    \begin{align}
                        \rho &= \frac{\rho_0}{a^2\sqrt{v_0^2 + a^2(1-v_0^2)}}\ ,\\
                        v &= \frac{v_0}{\sqrt{v_0^2 + a^2(1-v_0^2)}}\ , \label{eq:RCDM_v}\\
                        \gamma^2\rho &= \frac{\rho_0}{a^4(1-v_0^2)}\sqrt{v_0^2+a^2(1-v_0^2)} \ , \label{eq:RCDM_rho}
                    \end{align}
                    where $v_0$ and $\rho_0$ are the velocity and density today. If the fluid starts with 
                    ultrarelativistic initial conditions, it behaves as radiation $\gamma^2\rho\propto a^{-4}$
                    until the velocity drops down and it enters the non-relativistic regime. In the
                    non-relativistic regime, to first order in $v$, the velocity slows down with the
                    expansion $v\propto a^{-1}$ and the density scales as usual 
                    $\gamma^2\rho\simeq\rho\propto a^{-3}$.
            \end{itemize}
            Analytic expressions for a generic equation of state $w(a)$ can be obtained in the regime
            of small velocities
            \begin{align}
                \rho &= \rho_0 \exp\left(-3\int\frac{\di a}{a}\left(1+w\right)\right) + \Od{v^2}\ ,\\
                v &= \frac{v_0(1+w_0)}{a^4(1+w)}\exp\left(3\int\frac{\di a}{a}\left(1+w\right)\right) + \Od{v^2}\ ,
            \end{align}
            where $w_0$ is the value of the equation of state today. For the particular case $w=\text{const.}$, 
            we have \cite{Maroto:2005kc, BeltranJimenez:2007rsj}
            \begin{align}
                \rho &= \rho_0\, a^{-3(1+w)} + \Od{v^2}\ ,\\
                v &= v_0\, a^{-(1-3w)} + \Od{v^2}\ .
            \end{align}
        
    \subsection{Kinetic approach}\label{sec:kin_approach}
        Non-relativistic particles and the photon-baryon plasma in the early Universe can be well
        approximated as perfect fluids, and analyzed using only their energy-momentum tensor. However, 
        to describe the free-streaming of neutrinos or the details of decoupling, the fluid approximation
        is not enough and we must use a more general kinetic approach. In this section we follow closely 
        the presentation of \cite{Ma:1995ey}.\\
        
        The phase-space of the system is described by
        \begin{itemize}
            \item Three positions $x^i$.
            \item Three conjugate momenta $P_i$. These conjugate momenta are defined as the spatial
                components of the four-momentum
                \begin{equation}\label{eq:mass_shell}
                    P^\mu \equiv m\frac{\di x^\mu}{\di \lambda}\ , \qquad P_\mu P_\nu g^{\mu\nu} = -m^2\ ,
                \end{equation}
                where $\di\lambda \equiv \sqrt{-\di s^2}$ is the proper time and the spatial index $i$ has
                been lowered with the full metric $g_{\mu \nu}$.
        \end{itemize}
        
        The number of particles per unit of phase-space volume is 
        \begin{equation}
            \di N = g_* f\left(\tau, x^i, P_j\right) \frac{\di^3 x\, \di^3 P}{(2\pi)^3}\ ,
        \end{equation}
        where $g_*$ is the number of internal degrees of freedom, e.g. the number of helicity states, and 
        $f$ is the phase-space distribution function. In the kinetic approach, the 
        energy-momentum tensor can be defined as \cite{Ma:1995ey}
        \begin{equation}\label{eq:Tmunu_def}
			T_{\mu \nu} =\frac{g_*}{(2\pi)^3} \int \di^3 P\, (-g)^{-1/2} \frac{P_\mu P_\nu}{P^0}f(\tau, 
			    x^i, P_j)\ ,
		\end{equation}
		where $g$ is the determinant of the metric $g_{\mu\nu}$. In a cosmological setting, the next step
		would be to particularize these definitions to a RW metric \eqref{eq:RW_metric} and then to 
		study perturbations over the metric and the distribution function. We will split the discussion
		in two parts. Since our main modification with respect to standard comoving cosmology concerns 
		the definition of the unperturbed distribution function, we will focus on the background in the 
		next section \ref{sec:kin_background}. Section \ref{sec:kin_perturbations} contains the complete 
		treatment of perturbations.
        
	    \subsubsection{Background}\label{sec:kin_background}
	        Starting again with the metric 
	        \begin{equation}\label{eq:RW_metric2}
	            \di s^2 = a^2(\tau)\left(-\di\tau^2 + \delta_{ij}\di x^i\di x^j\right)\ ,
	        \end{equation}
	        and denoting the momentum as
	        \begin{subequations}
	        \begin{align}
	            \epsilon &\equiv -P_0\ ,\\
	            q_i &\equiv P_i\ ,
	        \end{align}
	        \end{subequations}
	        such that $q^i\equiv q_i$, we obtain from the mass-shell condition \eqref{eq:mass_shell}
	        \begin{equation}
	            \epsilon^2 = m^2a^2 + q^2\ .
	        \end{equation}
	        We introduce again two related frames.
	        \begin{itemize}
	            \item $\mathcal{O}$, frame in which the metric takes the form \eqref{eq:RW_metric2}.
	            \item $\tilde{\mathcal{O}}$, frame moving with respect to $\mathcal{O}$ with velocity $\v{\beta}$.
	        \end{itemize}
	        The local Lorentz transformation \eqref{eq:LocalLorentz_transformation} that connects both 
	        frames yields
	        \begin{subequations}\label{eq:Lorentz_transf_def}
	        \begin{align}
	            \tilde{\epsilon} &\equiv \Lambda_\beta \epsilon = \gamma(\epsilon - \v{q}\cdot\v{\beta})\ ,\\
	            \tilde{q}^i &\equiv \Lambda_\beta q^i = \mathcal{P}^i_{j}q^j - \gamma\epsilon\beta^i\ ,\\
	            \v{\beta}\cdot\tilde{\v{q}} &= \gamma (\v{q}\cdot\v{\beta}-\epsilon\beta)\ .
	        \end{align}
	        \end{subequations}
	        We have defined
	        \begin{subequations}\label{eq:Lorentz_defs}
	        \begin{align}
	            \mathcal{P}^i_j &\equiv \delta^i_j + (\gamma - 1)\hat{\beta}^i\hat{\beta}_j\ ,\\
	            \gamma &\equiv (1-\beta^2)^{-1/2}\ ,
	        \end{align}
	        \end{subequations}
	        where $\hat{\beta}^i$ is a unit vector along $\v{\beta}$ and every spatial index has been 
	        lowered or raised with $\delta_{ij}$. Next we consider a homogeneous distribution function 
	        in the $\mathcal{O}$ frame
	        \begin{equation}
	             f(\tau, \v{x}, \v{q}) = f_0(\tau, \v{q})\ .
	        \end{equation}
	        With this distribution function, we can define the usual fluid quantities
	        \begin{alignat}{2}\label{eq:def_fluid_quantities_bg}
	            \rho &\equiv a^{-4} g_*\int\frac{\di^3 q}{(2\pi)^3}\,\epsilon\, f_0\ ,&\qquad\qquad
	                P &\equiv a^{-4} g_*\int\frac{\di^3 q}{(2\pi)^3}\, \frac{q^2}{3\epsilon}\, f_0\ ,\nonumber\\
	            Q^i &\equiv a^{-4} g_*\int\frac{\di^3 q}{(2\pi)^3}\, q^i\,f_0\ ,&\qquad\qquad
	                n &\equiv a^{-3} g_*\int\frac{\di^3 q}{(2\pi)^3}\, f_0\ ,\nonumber\\
	            \Pi^{ij} &\equiv a^{-4} g_*\int\frac{\di^3 q}{(2\pi)^3}\left(\frac{q^iq^j}{\epsilon}
	                    -\frac{q^2}{3\epsilon}\delta^{ij}\right) f_0\ ,&\qquad\qquad
	                V^i &\equiv a^{-3} g_*\int\frac{\di^3 q}{(2\pi)^3}\, \frac{q^i}{\epsilon}\, f_0\ ,
	        \end{alignat}
	        that represent the usual energy, momentum, shear tensor, pressure, number and velocity
	        densities of the fluid. To relate this set of quantities with those computed in the 
	        boosted $\tilde{\mathcal{O}}$ frame, we can either use their tensorial character under
	        local Lorentz transformations or the fact that $f$ transforms as a scalar
	        \begin{equation}
	            \tilde{f}_0(\tilde{\tau}, \v{\tilde{x}}, \v{\tilde{q}}\big) = f_0\big(\tau(\tilde{\tau}, \tilde{\v{x}}), \v{q}(\v{\tilde{q}})\big)\ ,
	        \end{equation}
	        where, from now on, we will denote $\tilde{f}_0(\tilde{\tau}, \v{\tilde{x}}, \v{\tilde{q}}\big)$ just as $\tilde{f}_0(\tau,\v{\tilde{q}})$.
	        With this property and the Lorentz-invariant volume element we can write, for instance,
	        \begin{equation}
	            \tilde{Q}^i = a^{-4}g_*\int\frac{\di^3\tilde{q}}{(2\pi)^3}\,\tilde{q}^i\,\tilde{f}_0(\tau, \tilde{\v{q}})
	                = a^{-4}g_*\int\frac{\di^3q}{(2\pi)^3}\frac{(\Lambda_\beta\epsilon)(\Lambda_\beta q^i)}{\epsilon}
	                    f_0(\tau, \v{q})\ .
	        \end{equation}
	        Using this procedure, one can obtain
	        \begin{subequations}\label{eq:boost_fluid_quantities}
	        \begin{align}
	            \tilde{\rho} &= \rho + \gamma^2\left(\beta_k\beta_l\Pi^{kl}-2\beta_kQ^k\right)
	                    + \gamma^2\beta^2(\rho + P)\ ,\\
	            \tilde{Q}^i &= \gamma\mathcal{P}^i_j\left(Q^j-\beta_k\Pi^{kj}\right)
	                    -\gamma^2\beta^i\left(\rho+P-Q^j\beta_j\right)\ ,\label{eq:boost_momentum}\\
	            \tilde{\Pi}^{ij}&= \left(\mathcal{P}^i_k\mathcal{P}^j_l-\frac{1}{3}\gamma^2\beta_k\beta_l
	                        \delta^{ij}\right)\Pi^{kl}
	                    - \gamma\left(\beta^i\mathcal{P}^j_k+\beta^j\mathcal{P}^i_k
	                        -\frac{2}{3}\gamma\delta^{ij}\beta_k\right)Q^k
	                    + \gamma^2\left(\beta^i\beta^j-\frac{1}{3}\beta^2\delta^{ij}\right) (\rho + P)\ ,\\
	            \tilde{P} &= P + \frac{1}{3}\gamma^2\left(\beta_k\beta_l\Pi^{kl} - 2\beta_kQ^k\right)
	                    + \frac{1}{3}\gamma^2\beta^2(\rho+P)\ ,\\
	            \tilde{n} &= \gamma (n-V^j\beta_j)\ ,\\
	            \tilde{V}^i &= \mathcal{P}^i_jV^j- \gamma \beta^i n \label{eq:boost_flux}\ .
	        \end{align}
	        \end{subequations}
	        It is important to stress that the preceeding relations hold as well if the quantities are
	        defined with the full distribution function $f$, instead of using just the background part
	        $f_0$, and we will make use of them when we study perturbations.\\
	        
	        There remains the question of how to describe the moving fluids of section \ref{sec:perfect_fluid}
	        in terms of a distribution function. We will describe the different constituents of the universe
	        with an unperturbed distribution function that satisfies
	        \begin{equation}\label{eq:f0_main_condition}
	            f_0(\tau, \v{q}) = \tilde{f}_0(\tau, \tilde{q})\ .
	        \end{equation}
	        That is, the distribution function is homogeneous in the $\mathcal{O}$ frame, i.e. the frame
	        that observes a homogeneous and isotropic universe, and isotropic in the $\tilde{\mathcal{O}}$
	        frame, i.e. the frame comoving with the fluid. This parallels the discussion in section
	        \ref{sec:perfect_fluid} and allows us to describe a fluid moving with velocity $\v{\beta}$. 
	        The condition \eqref{eq:f0_main_condition} is the main physical assumption in our work.
	        For instance, applying it to a black-body spectrum for massless particles, using 
	        \eqref{eq:Lorentz_transf_def} we obtain the usual boosted distribution function
	        \begin{equation}
	            \tilde{f}_0(\tilde{q}) = \frac{1}{\ee^{\,\tilde{q}/\tilde{T}}-1} 
	                = \frac{1}{\ee^{\,q/T(\v{q})}-1} = f_0(\v{q})\ ,\qquad 
	                T(\v{q})\equiv \gamma \left(1-\frac{\v{q}\cdot\v{\beta}}{q}\right)\tilde{T}\ .
	        \end{equation}
	        
	        If the distribution function satisfies \eqref{eq:f0_main_condition} we have
	        \begin{equation}
	            \tilde{\Pi}_{ij}=0\ ,\qquad \tilde{V}_i = \tilde{Q}_i = 0\ ,
	        \end{equation}
	        so we are indeed in the comoving frame with the (perfect) fluid. In this case, the 
	        relation between both sets of fluid variables is
	        \begin{subequations}
	        \begin{align}
	            \rho &= \tilde{\rho} + \gamma^2\beta^2(\tilde{\rho} + \tilde{P})\ ,\\
	            Q^i &= \gamma^2\beta^i\left(\tilde{\rho}+\tilde{P}\right)\ ,\\
	            \Pi^{ij}&= \gamma^2\left(\beta^i\beta^j-\frac{1}{3}\delta^{ij}\beta^2\right) 
	                (\tilde{\rho} + \tilde{P})\ ,\\
	            P &= \tilde{P} + \frac{1}{3}\gamma^2\beta^2(\tilde{\rho}+\tilde{P})\ ,\\
	            n &= \gamma \tilde{n}\ ,\\
	            V^i &= \gamma \beta^i \tilde{n}\ .
	        \end{align}
	        \end{subequations}
	        Using the definition of the energy-momentum tensor \eqref{eq:Tmunu_def}, we can write its
	        components as
	        \begin{subequations}
	        \begin{align}
	            T\indices{^0_0} &= -\rho = -\tilde{\rho} - \gamma^2(\tilde{\rho}+\tilde{P})\beta^2\ ,
	                \label{eq:background_T_00}\\
	            T\indices{^0_i} &= Q_i = \gamma^2(\tilde{\rho}+\tilde{P})\beta_i\ ,
	                \label{eq:background_T_0i}\\
	            T\indices{^i_j} &= P\delta^i_j + \Pi^i_j = \tilde{P}\delta^i_j +\gamma^2(\tilde{\rho}
	                +\tilde{P})\beta^i\beta_j\ .
	                \label{eq:background_T_ij}
	        \end{align}
	        \end{subequations}
	        These expressions agree with the ones obtained for a perfect fluid \eqref{eq:Tmunu_components}. 
	        Both approaches are equivalent at this level. To first order in $\beta$ we have
	        \begin{subequations}
	        \begin{align}
	            T\indices{^0_0} &= -\tilde{\rho}\ ,\\
	            T\indices{^0_i} &= (\tilde{\rho}+\tilde{P})\beta_i\ ,\\
	            T\indices{^i_j} &=  \tilde{P} \delta^i_j\ .
	        \end{align}
	        \end{subequations}
	        The full energy-momentum for a collection of fluids is homogeneous, as it was imposed in 
	        \eqref{eq:f0_main_condition}, and it is also isotropic, to first order in $\beta$, if the 
	        velocities of the fluids satisfy the constraint
	        \begin{equation}\label{eq:CM_condition2}
	            \sum_s (\tilde{\rho}_s + \tilde{P}_s)\beta_s^i = 0\ ,
	        \end{equation}
	        which is the same condition obtained in \eqref{eq:CM_condition}. It is clear from 
	        \eqref{eq:background_T_0i} that a similar constraint can always be imposed, to all orders
	        in $\beta$, to achieve $T\indices{^0_i}=0$ but this is not enough to source a RW geometry.
	        Already to second order in $\beta$, \eqref{eq:background_T_ij} contains a quadrupolar
	        anisotropy that cannot be compensated by the other fluids. In this case, we should go one
	        step further and consider a Bianchi universe. However, for the values of $\beta$ that we will
	        consider, this quadrupole lies well below the observed value \cite{BeltranJimenez:2007rsj}. 
	        Therefore, in this work we will restrict ourselves to first order and a RW background.
        
	    \subsubsection{Perturbations}\label{sec:kin_perturbations}
	        Our starting point now is a perturbed flat RW metric
	        \begin{equation}\label{eq:RW_perturbed}
	            \di s^2 = a^2(\tau)\Big(-(1-A)\di\tau^2 + 2B_i\di\tau\,\di x^i 
	                + (\delta_{ij}+H_{ij})\di x^i\di x^j\Big)\ .
	        \end{equation}
	        Reparameterizing the momentum as
	        \begin{subequations}\label{eq:P_parameterization}
	        \begin{align}
	            P_0&\equiv -aE + \delta P_0\ ,\\
	            P_i&\equiv a\left(\delta^j_i +\frac{1}{2}H^j_i\right)p_j\ ,
	        \end{align}
	        \end{subequations}
	        from the mass-shell condition \eqref{eq:mass_shell} we obtain
	        \begin{subequations}
	        \begin{align}
	            E^2 &= m^2 + p^2\ ,\\
	            \delta P_0 &= \frac{1}{2}A\,aE  + ap_iB^i\ .
	        \end{align}
	        \end{subequations}
	        These expressions can be regarded just as redefinitions but they have a very simple 
	        physical interpretation in terms of the vierbein \eqref{eq:vierbein_background}. 
	        Our choice of momentum, $P_\mu = e\indices{_\mu^a}p_a$, corresponds to a choice of vierbein 
	        with components
	        \begin{subequations}\label{eq:vierbein_components}
	        \begin{align}
	            e\indices{_0^0} &= a\left(1-\frac{1}{2}A\right)\ ,\\
	            e\indices{_i^0} &= 0\ ,\\
	            e\indices{_0^i} &= aB^i\ ,\\
	            e\indices{_i^j} &= a\left(\delta^i_j+\frac{1}{2}H^i_j\right)\ .
	        \end{align}
	        \end{subequations}
	        With these definitions, $p_i$ are the momenta measured by a locally inertial observer at
	        a fixed spatial position. It is convenient to work with a closely related set of variables
	        defined as
	        \begin{equation}\label{eq:q_p_relation}
	            q_i\equiv ap_i\ ,\qquad \epsilon \equiv aE\ .
	        \end{equation}
	        As we will see, written in terms of $q_i$, the Boltzmann equation in section \ref{sec:BoltzmannEquation}
	        does not contain a zero-order term. 
	        We perturb the phase-space distribution accordingly
	        \begin{equation}
	            f(\tau, \v{x}, \v{q}) = f_0(\tau, \v{q}) + \delta f(\tau, \v{x}, \v{q})\ ,
	        \end{equation}
	        and define the corresponding perturbed fluid variables
	        \begin{alignat}{2}\label{eq:def_perturbed_fluid}
	            \delta\rho &\equiv a^{-4} g_*\int\frac{\di^3 q}{(2\pi)^3}\,\epsilon\, \delta f\ ,&\qquad\qquad
	                \delta P &\equiv a^{-4} g_*\int\frac{\di^3 q}{(2\pi)^3}\, \frac{q^2}{3\epsilon}\, \delta f\ ,\nonumber\\
	            \delta Q^i &\equiv a^{-4} g_*\int\frac{\di^3 q}{(2\pi)^3}\, q^i\,\delta f\ ,&\qquad\qquad
	                \delta n &\equiv a^{-3} g_*\int\frac{\di^3 q}{(2\pi)^3}\, \delta f\ ,\nonumber\\
	            \delta\Pi^{ij} &\equiv a^{-4} g_*\int\frac{\di^3 q}{(2\pi)^3}\left(\frac{q^iq^j}{\epsilon}
	                    -\frac{q^2}{3\epsilon}\delta^{ij}\right) \delta f\ ,&\qquad\qquad
	                \delta V^i &\equiv a^{-3} g_*\int\frac{\di^3 q}{(2\pi)^3}\, \frac{q^i}{\epsilon}\, \delta f\ .
	        \end{alignat}
	        The components of the perturbed energy-momentum tensor are
	        \begin{subequations}
	        \begin{align}
	            \delta T\indices{^0_0} &= -\delta\rho + B^iQ_i\ ,\\
	            \delta T\indices{^0_i} &= \delta Q_i +\frac{1}{2}AQ_i + \frac{1}{2}H^j_iQ_j\ ,\\
	            \delta T\indices{^i_j} &= \delta P\delta^i_j + \delta\Pi^i_j -B^iQ_j 
	                + \frac{1}{2}\left(H^k_j\Pi^i_k-H^i_k\Pi^k_j\right)\ .
	        \end{align}
	        \end{subequations}
	        Expressing the background quantities in the $\tilde{\mathcal{O}}$ frame
	        \begin{subequations}
	        \begin{align}
	            \delta T\indices{^0_0} &= -\delta\rho +\gamma^2(\tilde{\rho}+\tilde{P})B_i\beta^i\ ,\\
	            \delta T\indices{^0_i} &= \delta Q_i +\frac{1}{2}\gamma^2(\tilde{\rho}+\tilde{P})
	                \left(A\delta^j_i + H^j_i\right)\beta_j\ ,\\
	            \delta T\indices{^i_j} &= \delta P\delta^i_j + \delta\Pi^i_j 
	                -\gamma^2(\tilde{\rho}+\tilde{P})B^i\beta_j
	                +\frac{1}{2}\gamma^2(\tilde{\rho}+\tilde{P})\left(\beta^iH^k_j\beta_k 
	                - \beta_jH^i_k\beta^k\right)\ .
	        \end{align}
	        \end{subequations}
	        It is easy to show that, to first order in $\beta$, once we apply the center of mass condition
	        \eqref{eq:CM_condition2}, every metric variable cancels out and does not appear in the definition 
	        of the total energy-momentum tensor.
	        
	        \subsubsection*{Non-relativistic limit}
	            We end this section with an aside about the non-relativistic (NR) limit ($q/\epsilon \ll 1$). In the next sections
	            only massless (photons and massless neutrinos) and NR particles (baryons and cold dark matter)
	            will be taken into account. In particular, we will only keep the first NR order, 
	            neglecting the pressure and sound speed of baryons.\\
	            
	            Although this kind of approximation is standard, we must be careful when taking the NR limit
	            in a moving frame. The proper way to account for this limit is to take it in the frame comoving
	            with the fluid, i.e. $\tilde{\mathcal{O}}$. To first NR order we have
	            \begin{equation}
	                \delta\tilde{\Pi}^{ij}\simeq 0\ ,\qquad \delta \tilde{P}\simeq 0\ .
	            \end{equation}
	            However, in the $\mathcal{O}$ frame, using \eqref{eq:boost_fluid_quantities}, it can be
	            seen that we do have pressure and anisotropic stress
	            \begin{subequations}
	            \begin{align}
	                \delta P &\simeq \frac{2}{3}\gamma^2\beta_k\delta \tilde{Q}^k 
	                    + \frac{1}{3}\gamma^2\beta^2\delta\tilde{\rho}\ ,\\
	                \delta \Pi^{ij} &\simeq \gamma\left(\beta^i\mathcal{P}^j_k+\beta^j\mathcal{P}^i_k
	                    -\frac{2}{3}\gamma \delta^{ij}\beta_k\right)\delta\tilde{Q}^k
	                    +\gamma^2\left(\beta^i\beta^j-\frac{1}{3}\beta^2\delta^{ij}\right)\delta\tilde{\rho}\ .
	            \end{align}
	            \end{subequations}
	            To first NR order and to first order in $\beta$, we have the following useful results
	            \begin{alignat}{2}
	                \delta P &\simeq  \frac{2}{3}\beta_k\delta Q^k\ ,&\qquad\qquad 
	                    \delta V^i &\simeq \frac{\delta Q^i}{m}\ ,\nonumber\\
	                \delta \Pi^{ij} &\simeq  \left(\beta^i\delta^j_k+\beta^j\delta^i_k 
	                        - \frac{2}{3}\delta^{ij}\beta_k\right)\delta Q^k\ ,&\qquad\qquad 
	                    \delta \tilde{V}^i &\simeq \frac{1}{m}\left(\delta Q^i - \beta^i\delta\rho\right)\ ,\nonumber\\
	                \delta \tilde{n} &\simeq  \frac{\delta \tilde{\rho}}{m} \simeq 
	                        \frac{1}{m}\left(\delta\rho-2\beta_k\delta Q^k\right)\ ,&\qquad\qquad
	                    \delta n &\simeq \frac{1}{m}\left(\delta\rho -\beta_k\delta Q^k\right)\ .
	            \end{alignat}
	            Note that the preceeding results hold as well for the corresponding unperturbed quantities,
	            see \eqref{eq:def_fluid_quantities_bg}.
	            Finally, the full energy-momentum tensor for a non-relativistic species to first order in
	            $\beta$ is
	            \begin{subequations}
	            \begin{align}
	                T\indices{^0_0}+\delta T\indices{^0_0} &\simeq -\tilde{\rho}
	                    -\delta\rho+\tilde{\rho}\beta^iB_i\ ,\\
	                T\indices{^0_i}+\delta T\indices{^0_i} &\simeq m\,\delta V_i
	                    +\tilde{\rho}\left(\delta^j_i+\frac{1}{2}\delta^j_iA+\frac{1}{2}H^j_i\right)\beta_j\ ,\\
	                T\indices{^i_j}+\delta T\indices{^i_j} &\simeq 
	                    m \left(\beta^i\delta V_j - \beta_j\delta V^i\right) + \tilde{\rho}\beta_jB^i\ .
	            \end{align}
	            \end{subequations}
        
\section{Boltzmann equation}\label{sec:BoltzmannEquation}
    After writing the energy-momentum tensor in terms of the distribution function, we need to compute
    its time evolution. This information is encoded in the Boltzmann equation, which in the locally
    Minkowskian frame takes the form
    \begin{equation}\label{eq:BoltzmannEq_definition}
        \frac{Df}{\di t} = \mathcal{C}[f]\ .
    \end{equation}
    where $\di t\equiv e\indices{_\mu^0}\di x^\mu$ is the time measured by the locally Minkowskian 
    observer, which for the vierbein choice in \eqref{eq:vierbein_components} reads 
    $\di t= a(1-\frac{1}{2}A)\di\tau$. The left-hand side, the so-called Liouville operator, describes 
    the free streaming of particles in phase space. It is defined as
    \begin{equation}\label{eq:Liouville_definition}
        \frac{Df}{\di t} \equiv \pd{f}{t} + \d{x^i}{t}\pd{f}{x^i} + \d{q^i}{t}\pd{f}{q^i}\ .
    \end{equation}
    This operator contains the information about the space-time geometry, through its effects on the
    geodesics of the particles. The functional on the right-hand side of \eqref{eq:BoltzmannEq_definition} 
    is the so-called collision term. It takes into account how the number of particles per unit of 
    phase-space volume change due to collisions, i.e. local interactions. Hence, it does not contain 
    information about the underlying geometry \cite{Bernstein:1988bw}. The collision term takes the same 
    form as in flat space-time when written in terms of the momenta measured by a locally inertial 
    observer $p_a$ defined above. The final form of the Boltzmann equation in conformal time is
    \begin{equation}\label{eq:BoltzmannEq_definition2}
        \frac{Df}{\di \tau} = a\left(1-\frac{1}{2}A\right)\mathcal{C}[f]\ .
    \end{equation}
    The next section is devoted to the left-hand side of the Boltzmann equation, i.e. the Liouville operator,
    particularizing to massless and non-relativistic particles. This analysis exhausts all the information
    needed to follow the evolution of non-interacting particles, like CDM and neutrinos. However, to 
    describe the photon-baryon plasma, we must move on to the full Boltzmann equation. Section 
    \ref{sec:boltzmann_rhs} studies the interaction between photons and electrons.\\
    
    \subsection{Liouville operator}\label{sec:boltzmann_lhs}
        In order to compute the time derivatives appearing in \eqref{eq:Liouville_definition}, we need the geodesics in the metric \eqref{eq:RW_perturbed}. The whole computation
        of the geodesics can be found in detail in the appendix \ref{sec:geodesics}. Using the 
        definition of the four-momentum \eqref{eq:mass_shell} and the parameterization 
        \eqref{eq:P_parameterization}, the final results are
        \begin{subequations}\label{eq:final_geodesics}
        \begin{align}
	        \d{x^i}{\tau} &= \frac{q^i}{\epsilon}\Big(1-\frac{1}{2}A\Big)-B^i
	            -\frac{1}{2\epsilon}H^{i}_kq^k\ ,\label{eq:geodesic_x}\\
            \d{q^i}{\tau} &= \frac{1}{2}\epsilon \partial_iA + q^jC_{ij} 
                + \frac{q^jq^k}{\epsilon}D_{ijk}\ ,\label{eq:geodesic_q}
        \end{align}
        \end{subequations}
        where the following combinations of metric variables have been defined
        \begin{subequations}\label{eq:geo_tensor_defs}
        \begin{align}
            C_{ij} &\equiv \partial_iB_j - \frac{1}{2}\dot{H}_{ij}\ ,\\
            D_{ijk} &\equiv \frac{1}{2}(\partial_iH_{jk}-\partial_kH_{ij})\ .
        \end{align}
        \end{subequations}      
        The left-hand side of the Boltzmann equation \eqref{eq:BoltzmannEq_definition2} is evaluated
        in the $\mathcal{O}$ frame, where the geodesics are computed,
        \begin{equation}\label{eq:Liouville_def}
            \frac{Df}{\di\tau} \equiv \pd{f}{\tau} + \d{x^i}{\tau}\pd{f}{x^i} + \d{q^i}{\tau}\pd{f}{q^i}\ .
        \end{equation}
        We will assume that the distribution function takes the form
        \begin{align}\label{eq:f0_main_condition2}
            f(\tau, \v{x},\v{q}) &= f_0(\tau, \v{q}) + \delta f(\tau, \v{x}, \v{q})\nonumber\\
                                 &= \tilde{f}_0(\tau, \Lambda_\beta \epsilon) 
                                     + \delta f(\tau, \v{x}, \v{q})\ ,
        \end{align}
        where $\tilde{f}_0(\tau, \tilde{\epsilon})$ is the standard isotropic distribution in the 
        $\tilde{\mathcal{O}}$ frame. Since \eqref{eq:geodesic_q} is already first order in perturbations, 
        the left-hand side of the Boltzmann equation can be recast as
        \begin{equation}\label{eq:Liouville_def_expanded}
            \frac{Df}{\di\tau} = \pd{f_0}{\tau} + \pd{\delta f}{\tau} + \d{x^i}{\tau}\pd{\delta f}{x^i}
                +\d{q^i}{\tau}\pd{f_0}{q^i}\ .
        \end{equation}
        We will restrict our discussion to massless and non-relativistic massive particles.
        
        \subsubsection{Massless particles}
            For massless particles the Lorentz transformations \eqref{eq:Lorentz_transf_def} take a simpler form
            \begin{subequations}
            \begin{align}
                \tilde{q} &= \gamma\left(1-\hat{n}\cdot\v{\beta}\right)q\ ,\\
                \tilde{q}^i &= \left(\mathcal{P}^i_jn^j-\gamma \beta^i\right)q\ ,
            \end{align}
            \end{subequations}
            where we have splitted the momentum into direction and magnitude
            \begin{equation}
                q^i\equiv q\, n^i\ ,\qquad n_i\delta^{ij}n_j=1\ .
            \end{equation}
            Using the Boltzmann equation, it can be directly checked that the unperturbed distribution 
            function in the $\tilde{\mathcal{O}}$ frame only depends on $\tilde{q}$, if there is no 
            zero-order collision term,
            \begin{equation}
                \tilde{f}_0(\tau,\tilde{q}) = \tilde{f}_0(\tilde{q})\ .
            \end{equation}
            It is convenient to work with the reduced phase-space density, integrating out the dependence 
            on the momentum magnitude, defined as
            \begin{equation}
                \mathcal{F}(\tau, \v{x}, \hat{n}) \equiv \frac{1}{\tilde{\mathcal{N}}}
                    \int q^3\di q\;\delta f(\eta, \v{x}, \v{q})\ ,
            \end{equation}
            where the constant $\tilde{\mathcal{N}}$ is related to the comoving energy density
            \begin{equation}\label{eq:normalization_photon}
                \tilde{\mathcal{N}} \equiv \int \tilde{q}^3\di \tilde{q}\; \tilde{f}_0(\tilde{q})
                    = \frac{2\pi^2}{g_*}a^4\tilde{\rho}\ .
            \end{equation}
            Plugging the geodesic equation \eqref{eq:final_geodesics} into the right-hand
            side of \eqref{eq:Liouville_def_expanded} and integrating over the
            momentum magnitude, we obtain, to all orders in $\beta$,
            \begin{align}\label{eq:LhsMassless_full}
                \int q^3\di q\;\frac{Df}{\di\tau} = \tilde{\mathcal{N}}\Bigg[\pd{}{\eta}\frac{1}{\gamma^4
                    (1-\hat{n}\cdot\v{\beta})^4} &+ \dot{\mathcal{F}} + n^i\partial_i\mathcal{F}
                    - \frac{4}{\gamma^4(1-\hat{n}\cdot\v{\beta})^5}\Bigg(\frac{1}{2}n^i\partial_iA
                    +n^in^jC_{ij} \nonumber\\
                &-\frac{1}{2}\beta^i\partial_iA - \beta^in^jC_{ij} -\beta^in^jn^kD_{ijk}\Bigg)\Bigg]\ .
            \end{align}
            Expanding it to first order in $\beta$, we have
            \begin{align}\label{eq:LhsMassless_first_order}
                \int q^3\di q\; \frac{Df}{\di\tau} = \tilde{\mathcal{N}}\Bigg[4\hat{n}\cdot\dot{\v{\beta}}
                    &+\dot{\mathcal{F}} 
                    + n^i\partial_i\mathcal{F} -4\Bigg(\frac{1}{2}n^i\partial_iA + n^in^jC_{ij}\Bigg)
                    (1+5\hat{n}\cdot\v{\beta})\nonumber\\
                &+4\beta^i\left(\frac{1}{2}\partial_iA+ n^jC_{ij}+n^jn^kD_{ijk}\right)\Bigg]\ .
            \end{align}
            The first moments of the angular distribution can be obtained performing the appropiate
            integrals
            \begin{align}
                \int\frac{\di\Omega}{4\pi}\int q^3\di q\;\frac{Df}{\di\tau} 
                    &= \tilde{\mathcal{N}}\left[\dot{\delta}+\frac{4}{3}\partial_i\delta v^i-\frac{4}{3}C_{ij}
                    \delta^{ij}-\frac{4}{3}\beta^i\partial_iA-\frac{4}{3}D_{ijk}\delta^{ij}\beta^k\right]\ ,
                    \label{eq:density_massless}\\
                \int\frac{\di\Omega}{4\pi}\,n^i\int q^3\di q\;\frac{Df}{\di\tau}
                    &= \frac{4}{3}\tilde{\mathcal{N}}\left[\dot{\beta}^i+\delta\dot{v}^i 
                    + \frac{3}{4}\partial_j\pi^{ij}
                    +\frac{1}{4}\partial^i\delta - \frac{1}{2}\partial^iA
                    -(\delta^{ij}\beta^k+\delta^{jk}\beta^i)C_{jk}\right]\ ,
                    \label{eq:momentum_massless}
            \end{align}
            where we have defined
            \begin{subequations}\label{eq:fluidphotons_def}
            \begin{align}
                \delta &\equiv \int\frac{\di\Omega}{4\pi}\,\mathcal{F} 
                    = \frac{\delta\rho}{\tilde{\rho}}\ ,
                    \label{eq:density_def}\\
                \delta\v{v} &\equiv \frac{3}{4}\int\frac{\di\Omega}{4\pi}\,\hat{n}\,\mathcal{F}
                    = \frac{\delta\v{Q}}{\tilde{\rho}+\tilde{P}}\ ,
                    \label{eq:velocity_def}\\
                \pi^{ij} &\equiv \int\frac{\di\Omega}{4\pi}\left(n^in^j-\frac{1}{3}\delta^{ij}\right)
                    \mathcal{F}
                    = \frac{\delta\Pi^{ij}}{\tilde{\rho}}\ .
                    \label{eq:sheartensor_def}
            \end{align}
            \end{subequations}
            
        \subsubsection{Massive particles}
            The results for massive particles are much more involved. In this case, we must use the
            full expressions for the Lorentz transformations \eqref{eq:Lorentz_transf_def} and the
            geodesics \eqref{eq:final_geodesics}. However, since we will focus on non-relativistic
            particles, we can simplify the analysis restricting it to the first moments of the 
            Boltzmann equation. Summing up, in this section we compute the evolution of the number 
            density, energy and velocity perturbations.\\ 
            
            The energy density contrast, equation of state and sound speed are defined as
            \begin{equation}\label{eq:w_cs2_defs}
                \delta \equiv \frac{\delta\rho}{\tilde{\rho}}\ ,\qquad
                w \equiv \frac{\tilde{P}}{\tilde{\rho}}\ ,\qquad
                c_\text{s}^2 \equiv \frac{\delta P}{\delta \rho}\ .
            \end{equation}
            Note that in these definitions the background quantities are referred to the 
            $\tilde{\mathcal{O}}$ frame while the perturbed quantities are defined in the $\mathcal{O}$
            frame. The election of intermediate variables is a matter of choice, the only condition
            being that we write the energy-momentum tensor consistently in terms of these variables. We
            stick to this convention throughout this work. The final results for the first moments of
            the distribution are
            \begin{itemize}
                \item \emph{Number density}.
                    \begin{align}\label{eq:rel_liouville_number}
                        a^{-3}g_*\int\frac{\di^3q}{(2\pi)^3} \frac{Df}{\di\tau} = a^{-3}\pd{}{\tau}\left(\gamma
                            \,a^3\tilde{n}\right) + a^{-3}\pd{}{\tau}\left(a^{3}\delta n\right)+ \partial_i\delta V^i -\gamma\tilde{n}
                            \left[\frac{1}{2}\beta^i\partial_iA + \delta^{ij}C_{ij}+D_{ijk}\delta^{ij}
                            \beta^k\right]\ .
                    \end{align}
                \item \emph{Energy}.
                    \begin{align}\label{eq:rel_liouville_energy}
                        a^{-4}g_*\int\frac{\di^3q}{(2\pi)^3}\epsilon\frac{Df}{\di\tau} = &\pd{}{\tau}\left[
                            \gamma^2(\tilde{\rho}+\beta^2\tilde{P})\right]
                            +3\mathcal{H}(\tilde{\rho}+\tilde{P})\left(1+\frac{4}{3}\beta^2\gamma^2\right)
                            +\partial_i\delta Q^i\nonumber\\
                        &+ \tilde{\rho}\left(\dot{\delta}+3\mathcal{H}\delta (c_\text{s}^2-w)\right)
                            +\delta\left(\dot{\tilde{\rho}}+3\mathcal{H}(\tilde{\rho}+\tilde{P})\right)
                            \nonumber\\
                        &-\gamma^2(\tilde{\rho}+\tilde{P})\left[\beta^i\partial_iA + \delta^{ij}C_{ij}
                            +\beta^i\beta^jC_{ij}+ D_{ijk}\delta^{ij}\beta^k\right]\ .
                    \end{align}
                \item \emph{Momentum}.
                    \begin{align}\label{eq:rel_liouville_momentum}
                        a^{-4}g_*\int\frac{\di^3q}{(2\pi)^3}q^i \frac{Df}{\di\tau} = &a^{-4}\pd{}{\tau}\left[\gamma^2
                            \beta^i\,a^{4}(\tilde{\rho}+\tilde{P})\right] + a^{-4}\pd{}{\tau}\left(a^{4}\delta Q^i \right)
                            + \partial_j\left(\delta \Pi^{ij} + \delta^{ij} \delta P\right)\nonumber\\
                        &-\gamma^2(\tilde{\rho}+\tilde{P})\left[\frac{1}{2}\left(\delta^{il}
                            +\beta^i\beta^l\right)\partial_lA+\left(\delta^{jk}\beta^i
                            +\delta^{ij}\beta^k\right)\left(C_{jk}+D_{jkl}\beta^l\right)\right]\ .
                    \end{align}
            \end{itemize}
            These results are exact to all orders in $\beta$, for relativistic and non-relativistic
            particles alike. They can be shown to reproduce \eqref{eq:density_massless} and
            \eqref{eq:momentum_massless} for massless particles. Next we define
            \begin{equation}\label{eq:dn_dv_defs}
                \delta_n\equiv\frac{\delta n}{\tilde{n}}\ ,\qquad 
                \delta v^i\equiv \frac{\delta V^i}{\tilde{N}}\ .
            \end{equation}
            Assuming that the zero-order Boltzmann equation (without collisions) is satisfied, so we keep 
            only cosmological perturbations and terms with $\beta$, and expanding to first NR order and 
            to first order in $\beta$ we have
            \begin{itemize}
                \item \emph{Number density}.
                     \begin{equation}\label{eq:lioville_number_density}
                         a^{-3}g_*\int\frac{\di^3 q}{(2\pi)^3}\frac{Df}{\di\tau} \simeq \tilde{N}\left\{
                             \dot{\delta_n} +\partial_i\delta v^i -\frac{1}{2}\beta^i\partial_i A
                             -\delta^{ij}C_{ij}-D_{ijk}\delta^{ij}\beta^k\right\}\ .
                     \end{equation}
                \item \emph{Energy}.
                    \begin{equation}\label{eq:liouville_density_NR}
                        a^{-4}g_*\int\frac{\di^3 q}{(2\pi)^3}\epsilon\frac{Df}{\di\tau}\simeq \tilde{\rho}\left\{
                            \dot{\delta}+2\mathcal{H}\beta_k\delta v^k +\partial_i\delta v^i
                            -\beta^i\partial_iA-\delta^{ij}C_{ij}-D_{ijk}\delta^{ij}\beta^k\right\}\ .
                    \end{equation}
                \item \emph{Momentum}.
                    \begin{align}\label{eq:liouville_momentum_NR}
                        a^{-4}g_*\int\frac{\di^3 q}{(2\pi)^3}q^i\frac{Df}{\di\tau}\simeq \tilde{\rho}\,\bigg\{
                            \dot{\beta}^i +\mathcal{H}\beta^i + \delta\dot{v}^i
                            &+\left(\beta^i\delta^j_k+\beta^j\delta^i_k\right)\partial_j\delta v^k
                                +\mathcal{H}\delta v^i\nonumber\\                            
                            &-\frac{1}{2}\partial^iA-\left(\delta^{jk}\beta^i+\delta^{ij}\beta^k\right)
                                C_{jk}\bigg\}\ .
                    \end{align}
            \end{itemize}
            
    \subsection{Collision term}\label{sec:boltzmann_rhs}
        This section is devoted to the calculation of the collision term for Compton scattering between
        electrons and photons. The notation in this section is slightly different. As we mentioned 
        before, the collision term must be written in terms of the momenta $p_i$ measured by a 
        locally inertial observer at a fixed spatial position. They are related to the momenta $q_i$ we 
        have been using as defined in \eqref{eq:q_p_relation}, i.e. $q_i=p_i/a$.\\
        
        The standard physical assumptions underlying the derivation of the collision term are 
        \begin{itemize}
            \item When written in terms of the momenta $p_i$, the collision term is the same as in flat 
                space, since it takes into account local information where the curvature 
                effects are not important. In the same way, the matrix element $\mathcal{M}$ is computed 
                using quantum field theory (QFT) in flat space.
            \item The temperature of the plasma is low enough so that the electrons are non-relativistic.
                We keep only the first order correction in the non-relativistic expansion, so we keep the
                electron velocity but neglect its pressure and sound speed. We consider the NR limit of 
                Compton scattering, i.e. Thomson scattering.
            \item Electrons and protons are much more tightly coupled between them than to the photons. The
                velocities of free electrons, protons and the full baryonic velocity are
                the same throughout the evolution.
            \item The angular dependence of Thomson scattering is neglected. This angular dependence has 
                proven important for 1\% accuracy and especially for polarization, but we will not take it
                into account in this work. We use the angle-averaged matrix element instead.
            \item The number of internal degrees of freedom is \emph{not} included in the definition of $f$, i.e.
                the equilibrium distributions correspond to the usual Bose-Einstein and Fermi-Dirac
                distributions.
            \item The medium is diluted enough so that we can neglect the quantum statistical factors
                $(1\pm f)$ responsible for the Bose enhancement and Pauli blocking effects. We do not
                take into account any plasma effect from finite temperature QFT.
        \end{itemize}
        
        With these qualifications, our starting point is the following definition of the collision term
        for the process $e(p_e)+\gamma (p)\leftrightarrow e(p_e')+\gamma (p')$
        \begin{equation}\label{eq:origin_collision}
			\mathcal{C}[f(p)]=\frac{1}{4p}\int \mathcal{D}p_e\mathcal{D}p'\mathcal{D}p'_e 
			    (2\pi)^4\delta (p^\mu +p_e^\mu - p'^\mu -p_e'^\mu)
			    \Big[f(\v{p'})f_e(\v{p'}_e)-f(\v{p})f_e(\v{p}_e)\Big]\sum_\text{spins} |\mathcal{M}|^2\ ,
		\end{equation}
		where $\mathcal{D}p=\frac{\di \v{p}^3}{(2\pi)^32E}$ is the Lorentz-invariant phase-space volume
		element and the dependence of the distribution functions on space-time coordinates has been 
		omitted since it does not play any role.	In our setting, we must implement the fact that the 
		fluids are moving. The collision term is defined in the cosmic center of mass, $\mathcal{O}$ 
		frame, and in this frame both photons and electrons have their own bulk velocity. We will 
		represent it schematically as
		\begin{equation}
		    \mathcal{C}[f] = \frac{1}{4p}\int\mathcal{D}p_e\mathcal{D}p'\mathcal{D}p'_e(2\pi)^4
                \delta (p^\mu +p_e^\mu - p'^\mu -p_e'^\mu)\Big[\bar{f}(\Lambda_\beta\v{p}')\tilde{f}_e(\Lambda_{\beta_e}\v{p}_e')
                -\bar{f}(\Lambda_\beta\v{p})\tilde{f}_e(\Lambda_{\beta_e}\v{p}_e)\Big]\sum_\text{spins}
                |\mathcal{M}|^2\ ,
		\end{equation}
		where $\bar{f}$ and $\tilde{f}_e$ are the distribution functions of photons and electrons in their
		frame, moving with bulk velocities $\v{\beta}$ and $\v{\beta}_e$, respectively, with respect
		to the $\mathcal{O}$ frame.\\
		
		Previously, the $\tilde{\mathcal{O}}$ frame was defined as the frame comoving with the fluid. In
		this case, we are facing two moving fluids. We take $\tilde{\mathcal{O}}$ to be the frame
		moving with velocity $\v{\beta}_e$ with respect to $\mathcal{O}$, i.e. the frame comoving with
		the \emph{electrons}. Performing the integration in this frame, we have
		\begin{equation}
		\mathcal{C}[f] = \frac{1}{4p}\int\mathcal{D}\tilde{p}_e\mathcal{D}\tilde{p}'\mathcal{D}\tilde{p}_e'(2\pi)^4
                \delta (\tilde{p}^\mu +\tilde{p}_e^\mu - \tilde{p}'^\mu -\tilde{p}_e'^\mu)\left[
                \bar{f}(\Lambda_\beta\Lambda_{\beta_e}^{-1}\v{\tilde{p}}')\tilde{f}_e(\v{\tilde{p}}_e')
                -\bar{f}(\Lambda_\beta\Lambda_{\beta_e}^{-1}\v{\tilde{p}})\tilde{f}_e(\v{\tilde{p}}_e)\right]
                \sum_\text{spins}|\mathcal{M}|^2\ .
		\end{equation}
        The previous two equations may seem devoid of any additional content with respect to
        \eqref{eq:origin_collision}. As they stand, without defining $f$ and $f_e$, they correspond
        just to a renaming of functions and reshuffling of variables. The physical content lies in
        \eqref{eq:f0_main_condition2}, i.e. in the structure and relation of the background distribution
        function in $\mathcal{O}$ and $\tilde{\mathcal{O}}$. The $\tilde{\mathcal{O}}$ frame is 
        comoving with the electrons and observes a standard isotropic equilibrium distribution. It is
        in this frame that we can perform the usual NR expansion \cite{Dodelson:2003ft} to get
        \begin{equation}\label{eq:collision_starting_point}
            \mathcal{C}[f]=\frac{\sigma_T}{4\pi p}\int\tilde{p}'\di\tilde{p}'\di\tilde{\Omega}'
                \left[\tilde{n}_e^\text{full}\delta (\tilde{p}-\tilde{p}')
                +\tilde{n}_e\v{\tilde{u}}^\text{full}_e\cdot
                (\v{\tilde{p}}-\v{\tilde{p}}')\pd{\delta (\tilde{p}-\tilde{p}')}{\tilde{p}'}\right]
                \left(\bar{f}(\Lambda_\beta\Lambda_{\beta_e}^{-1}\v{\tilde{p}}')
                -\bar{f}(\Lambda_\beta\Lambda_{\beta_e}^{-1}\v{\tilde{p}})\right)\ ,
        \end{equation}
        where $\sigma_T$ is the Thomson cross section and we have defined
        \begin{equation}\label{eq:ne_def}
            \tilde{n}_e^\text{full}\equiv 2\int\frac{\di^3\tilde{p}_e}{(2\pi)^3}\tilde{f}_e(\v{\tilde{p}}_e)\ ,
            \qquad
            \tilde{n}_e\v{\tilde{u}}_e^\text{full} \equiv 2\int\frac{\di^3\tilde{p}_e}{(2\pi)^3}
                \frac{\v{\tilde{p}}_e}{\tilde{E}_{p_e}}\tilde{f}_e(\v{\tilde{p}}_e)\ ,
        \end{equation}
        where $\bar{f}$, $\tilde{f}_e$ correspond to the full distribution functions. Just as we did for the left-hand
        side in section \ref{sec:boltzmann_lhs}, we will split the distribution functions into a 
        background and a perturbation part and integrate out the magnitude of the photon momentum. The
        whole process, to all orders in $\beta$, is detailed in the appendix \ref{sec:full_collision}.
        Here we present only the final results. To first order in $\beta$, we get
        \begin{align}\label{eq:coll_photon}
            \frac{1}{\tilde{\mathcal{N}}}\int q^3\di q\;\mathcal{C}[f] = \tilde{n}_e\sigma_T\,
                \Bigg[&-(1-\hat{n}\cdot \v{\beta}_e)\mathcal{F}_\gamma 
                + (1+3\hat{n}\cdot\v{\beta}_e)\delta_\gamma
                -\frac{8}{3}\v{\beta}_e\cdot\delta\v{v}_\gamma -4\hat{n}\cdot\Delta\v{\beta} 
                - \frac{8}{3}\delta\v{v}_e\cdot\Delta\v{\beta}\nonumber\\
            &+4\hat{n}\cdot\delta\v{v}_e\,\hat{n}\cdot\Delta\v{\beta}+4\delta\v{v}_e\cdot(\hat{n}
                -\v{\beta}_e + 4\hat{n}\, \hat{n}\cdot\v{\beta}_e) 
                -4\hat{n}\cdot\v{\beta}\,\frac{\delta n_e}{\tilde{n}_e}\Bigg]\ ,
        \end{align}
        where $\tilde{n}_e$ is the number density 
        of free electrons in the $\tilde{\mathcal{O}}$ frame, $\delta n_e$ and $\delta\v{v}_e$ are defined in 
        \eqref{eq:dn_dv_defs} and the difference of velocities is
        \begin{equation}
            \Delta\v{\beta} \equiv \v{\beta}-\v{\beta}_e\ .
        \end{equation}
        Since, to first order in $\beta$, the background quantities like $\rho$ or $n$ concide in 
        the $\mathcal{O}$ and $\tilde{\mathcal{O}}$ frames, we will drop the distinction.
        The first two moments of the photon collision term are
        \begin{align}
            \frac{1}{\tilde{\mathcal{N}}}\int\frac{\di\Omega}{4\pi}\int q^3\di q\;\mathcal{C}[f(\v{p})]
                &= -\frac{4}{3}n_e\sigma_T\Big[\v{\beta}_e\cdot(\delta\v{v}_\gamma 
                    -\delta\v{v}_e) + \delta\v{v}_e\cdot\Delta\v{\beta}\Big]\ ,
                    \label{eq:coll_photon_density}\\
            \frac{1}{\tilde{\mathcal{N}}}\int\frac{\di\Omega}{4\pi}\,n^i\int q^3\di q\;\mathcal{C}[f(\v{p})]
                &= -\frac{4}{3}n_e\sigma_T\left[\delta v_{\gamma}^i-\delta v^i_e + \Delta\beta^i
                    +\beta^i\frac{\delta n_e}{n_e}-\beta^i_e\delta_\gamma - \frac{3}{4}\beta_{e\,j}
                    \pi_\gamma^{ij}\right]\ .
                    \label{eq:coll_photon_velocity}
        \end{align}
        
        \subsubsection*{Conserved quantities}
            We have just computed the collision term \emph{for photons}. However, the whole
            plasma is described by the coupled system
            \begin{subequations}
            \begin{align}
                \frac{Df}{\di t} &= \mathcal{C}[f,f_e]\ ,\\
                \frac{Df_e}{\di t} &= \mathcal{C}_e[f,f_e]\ .
            \end{align}
            \end{subequations}
            We ought to compute the collision term for electrons $\mathcal{C}_e$ as well. Not surprisingly,
            both terms are not independent. In fact, we can make use of some conservation laws derived
            from the full Boltzmann equation to save us most of the work. Following \cite{Dodelson:2003ft},
            we write both collision terms with the compact notation
            \begin{subequations}
            \begin{align}
		    		c_{e\gamma} &=\frac{1}{2}\frac{1}{2E_{p_e}}\frac{1}{2p}\frac{1}{2E_{p'_e}}\frac{1}{2p'}
		    		    (2\pi)^4\delta (p^\mu +p_e^\mu - p'^\mu -p_e'^\mu)
        			    \Big[f(\v{p'})f_e(\v{p'}_e)-f(\v{p})f_e(\v{p}_e)\Big]\sum_\text{spins} |\mathcal{M}|^2\ ,\\
		    		\mathcal{C}_e[f_e(p_e)] & \equiv \int\frac{\di^3p}{(2\pi)^3}\frac{\di^3p'}{(2\pi)^3}\frac{\di^3p'_e}{(2\pi)^3}\ c_{e\gamma} \equiv \langle 
		    						c_{e\gamma}\rangle_{pp'p'_e}\ ,\\
		    		\mathcal{C}[f(p)] &\equiv \langle c_{e\gamma}\rangle_{p'p_e'p_e}\ .
		    	\end{align}   
		    	\end{subequations}
		    Integrating over all the momenta, it is easy to see that we have
		    	\begin{align}
                \langle c_{e\gamma}\rangle_{p_epp_e'p'} &= 0\ ,\\
                \langle (p+E_{p_e}) c_{e\gamma}\rangle_{p_epp_e'p'} &= 0\ ,\\
                \langle (\v{p}+\v{p}_e) c_{e\gamma}\rangle_{p_epp_e'p'} &= 0\ ,
		    	\end{align}
		    	corresponding to the conservation of the number of particles, energy and momentum. Using
		    	these results, the following equalities hold
		    	\begin{align}
		    	    \int\frac{\di^3 p_e}{(2\pi)^3}\,E_{p_e}\,\mathcal{C}_e[f_e(p_e)] 
		    	        &= -\int\frac{\di^3 p}{(2\pi)^3}\,p\,\mathcal{C}[f(p)]\ ,
		    	        \label{eq:conserved_energy}\\
		    	    \int\frac{\di^3 p_e}{(2\pi)^3}\,\v{p}_e\,\mathcal{C}_e[f_e(p_e)] 
		    	        &= -\int\frac{\di^3 p}{(2\pi)^3}\,\v{p}\,\mathcal{C}[f(p)]\ .
		    	        \label{eq:conserved_momentum}
		    	\end{align}
		    	This means that we can compute the first two moments of the Boltzmann equation for electrons,
		    	the only ones that we will need since they are non-relativistic, from the first two moments
		    	of the photons, already computed in \eqref{eq:coll_photon_density} and 
		    	\eqref{eq:coll_photon_velocity}.
        
    \subsection{Boltzmann equation for different components}\label{sec:boltzmann_final}
        \subsubsection{Photons}
            The reduced Boltzmann equation for photons is obtained combining the Liouville operator 
            \eqref{eq:LhsMassless_first_order} and the collision term \eqref{eq:coll_photon}. To zero
            order in cosmological perturbations, it describes the evolution of the bulk velocity $\beta$
            \begin{equation}\label{eq:beta_g}
		        \dot{\beta}^i = -an_e\sigma_T\Delta\beta^i\ .
		    \end{equation}
            To first order in cosmological perturbations and $\beta$, we get the evolution of
            the reduced phase-space density
		    \begin{align}\label{eq:BoEq_photons}
		        \dot{\mathcal{F}}_\gamma &+ n^i\partial_i\mathcal{F}_\gamma 
		            -4\left(\frac{1}{2}n^i\partial_iA + n^in^jC_{ij}\right)
		            (1+5\hat{n}\cdot\v{\beta})
		            +4\beta^i\left(\frac{1}{2}\partial_iA+ n^jC_{ij}+n^jn^kD_{ijk}\right)
		            \nonumber\\
		        &= an_e\sigma_T\Bigg[-(1-\hat{n}\cdot\v{\beta}_e)\mathcal{F}_\gamma 
		            + (1+3\hat{n}\cdot\v{\beta}_e)\delta_\gamma-\frac{8}{3}\v{\beta}_e\cdot\delta\v{v}_\gamma
		            -\frac{8}{3}\delta\v{v}_b\cdot\Delta\v{\beta}\nonumber\\
		        &\quad\qquad\qquad+4\hat{n}\cdot\delta\v{v}_b\,\hat{n}
		            \cdot\Delta\v{\beta}+4\delta\v{v}_b\cdot(\hat{n}-\v{\beta}_e + 4\hat{n}\, \hat{n}\cdot
		            \v{\beta}_e)-4\hat{n}\cdot\v{\beta}\,\frac{\delta n_e}{n_e}+2\hat{n}\cdot\Delta\v{\beta}A\Bigg]\ .
		    \end{align}
		    Since protons and electrons form a single tightly coupled baryonic fluid,
		    we have substituted $\delta \v{v}_e$ with $\delta \v{v}_b$, the velocity of baryons.
		    The evolution of the fluid variables can be obtained performing the appropiate angular
		    integrals. The equations for the density, combining 
		    \eqref{eq:density_massless} and \eqref{eq:coll_photon_density}, and the velocity, combining
		    \eqref{eq:momentum_massless} and \eqref{eq:coll_photon_velocity}, are
		    \begin{align}
		        \dot{\delta}_\gamma+\frac{4}{3}\partial_i\delta v^i_\gamma-\frac{4}{3}C_{ij}
		            \delta^{ij}-\frac{4}{3}\beta^i\partial_iA-\frac{4}{3}D_{ijk}\delta^{ij}\beta^k
		        &= -\frac{4}{3}an_e\sigma_T\Big[\v{\beta}_e\cdot(\delta\v{v}_\gamma 
		                        -\delta\v{v}_b) + \delta\v{v}_b\cdot\Delta\v{\beta}\Big]\ ,
		                        \label{eq:BoEq_dg}\\
		        \delta\dot{v}^i_\gamma + \frac{3}{4}\partial_j\pi^{ij}_\gamma
		            +\frac{1}{4}\partial^i\delta_\gamma - \frac{1}{2}\partial^iA
		            -(\delta^{ij}\beta^k+\delta^{jk}\beta^i)C_{jk}
		        &= -an_e\sigma_T\left[\delta v_{\gamma}^i-\delta v^i_b +\beta^i\frac{\delta n_e}{n_e}-\beta^i_e
		            \delta_\gamma - \frac{3}{4}\beta_{e\,j}\pi_\gamma^{ij}-\frac{1}{2}\Delta\beta^iA\right]\ .
		            \label{eq:BoEq_vg}
		    \end{align}
		    
        \subsubsection{Baryons}
            The evolution of the baryon density can be found using the left-hand side 
            \eqref{eq:liouville_density_NR} and energy conservation \eqref{eq:conserved_energy}.
            For the velocity, we must use the left-hand side \eqref{eq:liouville_momentum_NR} and
            momentum conservation \eqref{eq:conserved_momentum}. As mentioned before, since they are much 
            more tightly coupled between them than to photons, electrons and protons form a single 
            baryonic fluid. We use $\v{\beta}_e$ to denote the baryon velocity. To zero order in cosmological perturbations, we find the evolution of the 
            bulk velocity $\beta_e$
            \begin{equation}\label{eq:beta_e}
		        \dot{\beta}_e^i+\mathcal{H}\beta_e^i 
		            = \frac{4\rho_\gamma}{3\rho_b}an_e\sigma_T\Delta\beta^i\ .
		    \end{equation}
		    To first order in cosmological perturbations and $\beta$, the evolution of the first two
		    moments of the distribution is
            \begin{align}
		        \dot{\delta}_b&+2\mathcal{H}\v{\beta}_e\cdot\delta \v{v}_b +\partial_i\delta v^i_b-\beta_e^i\partial_iA 
		                -\delta^{ij}C_{ij}-D_{ijk}\delta^{ij}\beta_e^k\nonumber\\
		            &= \frac{4\rho_\gamma}{3\rho_b}an_e\sigma_T\Big[\v{\beta}_e\cdot(\delta\v{v}_\gamma
		                -\delta\v{v}_b) + \delta\v{v}_b\cdot\Delta\v{\beta}\Big]\ ,
		                \label{eq:BoEq_db}\\
		        \delta\dot{v}_b^i & +\mathcal{H}\delta v_b^i
		                +\left(\beta_e^i\delta^j_k+\beta_e^j\delta^i_k\right)\partial_j\delta v_b^k 
		                -\frac{1}{2}\partial^iA -\left(\delta^{jk}\beta_e^i
		                +\delta^{ij}\beta_e^k\right)C_{jk}\nonumber\\
		            &= \frac{4\rho_\gamma}{3\rho_b}an_e\sigma_T\left[\delta v_{\gamma}^i-\delta v^i_b 
		                +\beta^i\frac{\delta n_e}{n_e}-\beta^i_e\delta_\gamma - \frac{3}{4}\beta_{e\,j}
		                \pi_\gamma^{ij} -\frac{1}{2}\Delta\beta^iA\right]\ .
		                \label{eq:BoEq_vb}
		    \end{align}		    
		    
        \subsubsection{Massless neutrinos}
            Since we will neglect both the mass and coupling of neutrinos, they only free-stream
	        with the same left-hand side as photons. The equation for $\beta_\nu$ is
		    \begin{equation}\label{eq:beta_nu}
		        \dot{\beta}^i_\nu = 0\ .
		    \end{equation}
	        The equation for the evolution of the reduced phase-space density is
	        \begin{align}\label{eq:BoEq_nu}
		        \dot{\mathcal{F}}_\nu + n^i\partial_i\mathcal{F}_\nu 
		            -4\left(\frac{1}{2}n^i\partial_iA + n^in^jC_{ij}\right)(1+5\hat{n}\cdot\v{\beta}_\nu)
		            +4\beta_\nu^i\left(\frac{1}{2}\partial_iA+ n^jC_{ij}+n^jn^kD_{ijk}\right)
		            =0\ .
		    \end{align}		    
		    
        \subsubsection{Cold dark matter}
            Cold dark matter behaves as collisionless non-relativistic matter, i.e. it just follows
	        the same equations as baryons without interactions. The equation for $\beta_c$ is
	        \begin{equation}\label{eq:beta_c}
		        \dot{\beta}_c^i+\mathcal{H}\beta_c^i = 0\ .
		    \end{equation}
		    The relevant equations for the perturbations are
	        \begin{align}
		        \dot{\delta}_c+2\mathcal{H}\v{\beta}_c\cdot\delta \v{v}_c +\partial_i\delta v^i_c
		            -\beta_c^i\partial_iA -\delta^{ij}C_{ij}-D_{ijk}\delta^{ij}\beta_c^k &= 0\ ,
		            \label{eq:BoEq_dc}\\
		        \delta\dot{v}_c^i +\mathcal{H}\delta v_c^i
		            +\left(\beta_c^i\delta^j_k+\beta_c^j\delta^i_k\right)\partial_j\delta v_c^k
		            -\frac{1}{2}\partial^iA-\left(\delta^{jk}\beta_c^i
		            +\delta^{ij}\beta_c^k\right)C_{jk}&= 0\ .
		            \label{eq:BoEq_vc}
		    \end{align}		    
        
        \subsubsection{Total fluid}
            The total energy-momentum tensor, adding all the components, does not contain any explicit
            $\beta$ contribution after enforcing the cosmic center of mass condition. The conservation
            of the total energy-momentum tensor, which is a direct consequence of the Einstein equations, 
            \begin{equation}
                \nabla_\mu T\indices{^\mu_\nu} = 0\ ,
            \end{equation}
            gives us the conservation and Euler equations for the total fluid
            \begin{align}
                \dot{\delta} + 3\mathcal{H}(c^2_\text{s}-w)\delta + (1+w)\partial_i\left(\delta v^i-B^i\right)
                    + \frac{1}{2}(1+w)\dot{H}\indices{^i_i} &= 0\ ,\label{eq:BoEq_dtot}\\
                \delta\dot{v}_i+ \mathcal{H}(1-3w)\delta v_i + \frac{\dot{w}}{1+w}\delta v_i
                    + \frac{1}{1+w}\partial_i\left(c^2_\text{s}\delta\right) 
                    + \frac{1}{1+w}\partial_j\pi\indices{^j_i}-\frac{1}{2}\partial_i A &= 0\ ,
            \end{align}
            where the different variables are defined in the same way as for the individual components,
            but using the total energy-momentum tensor, e.g.
            \begin{align}
                \delta &\equiv \frac{1}{\rho}\sum_s\rho_s\delta_s\ , \\
                c^2_\text{s}\delta &\equiv \frac{1}{\rho}\delta P \equiv\frac{1}{\rho} \sum_s \delta P_s\ .
            \end{align}

\section{Multipole analysis}\label{sec:AngularAnalysis}
    The cosmological perturbations can be classified according to their transformation rules
    under the group of spatial rotations. This yields the usual splitting in scalar, vector and
    tensor perturbations, the only ones that contribute to the Einstein equations. Additionally, in the
    standard cosmological perturbation theory, the three types of perturbations are decoupled at the 
    linear level, a fact known as decomposition theorem \cite{Kodama:1985bj}. Since, in this case, 
    the only angular contribution comes from factors of the form $(\hat{n}\cdot\hat{k})$, i.e. the 
    angle between the line of sight and the direction of the Fourier mode, it is customary to write a 
    multipole expansion for the scalar part of the Boltzmann equation \eqref{eq:BoEq_photons} in terms 
    of Legendre polynomials \cite{Ma:1995ey}
    \begin{equation}\label{eq:Pl_expansion}
        \mathcal{F}(\tau, \v{k}, \hat{n}) = \sum_{\ell=0}^\infty (-\ii)^\ell (2\ell + 1)
            \mathcal{F}_\ell(\tau, \v{k})P_\ell(\hat{n}\cdot\hat{k})\ .
    \end{equation}
    The Boltzmann equation then unfolds into a whole hierarchy of coupled differential equations for the
    coefficients $\mathcal{F}_\ell$. The vector modes are usually neglected altogether since, even if
    initially present, they rapidly decay. The tensor modes are predicted in small quantities in many
    inflationary scenarios and their evolution must be followed when studying polarization effects.\\
    
    The key difference in our scenario is the existence of a new direction $\hat{\beta}$, introducing
    new angular dependencies in the Boltzmann equation \eqref{eq:BoEq_photons}. This means that we must
    resort to a full decomposition in terms of spherical harmonics of the form
    \begin{equation}\label{eq:Ylm_expansion}
        \mathcal{F}(\tau, \v{k}, \hat{n}) = \sqrt{4\pi}\sum_{\ell=0}^\infty\sum^\ell_{m=-\ell} 
            (-\ii)^{\ell+m} \sqrt{2\ell + 1}\, \mathcal{F}^m_\ell(\tau, \v{k})Y^m_\ell(\hat{n})\ ,
    \end{equation}
    where the coefficients have been defined to match the previous ones for the scalar $m=0$ modes.
    In particular, in our case, the decomposition theorem no longer holds. As it can be checked, in 
    addition to the usual coupling between $\ell-1$ and $\ell+1$ modes, the term 
    $(\hat{n}\cdot\hat{\beta})$ introduces new couplings between the modes $m-1$ and $m+1$. It is 
    possible to write down the new hierarchy of coupled differential equations for the modified 
    Boltzmann equation \eqref{eq:BoEq_photons} and it is important for a correct computation of
    CMB anisotropies \cite{CMB_paper}. However, in this work we are mainly concerned with LSS 
    observables and we do not need to trace the evolution of ultrarelativistic species with high
    accuracy.     
    Therefore, we will be working under an approximation scheme that allows us to consider only a subset of 
    these equations. Consequently, we present in this section a self-contained simplified derivation 
    of the system of equations that we will solve in later sections, bypassing the full 
    multipole decomposition.
    
    \subsection{Scalar-vector-tensor decomposition}
        Any spatial vector, in particular the velocity, can be decomposed into a divergence and a 
        divergenceless part
        \begin{equation}
            \delta v_i = \partial_i v^\text{S} + \chi_i\ ,\qquad \partial_i\chi^i = 0\ ,
        \end{equation}
        where $v^\text{S}$ is the scalar part of the velocity and $\v{\chi}$ is the vector part, the
        vorticity. In Fourier space, it can be written as
        \begin{equation}
            \delta v_i = -\frac{\ii\hat{k}^i}{k}\theta + \chi_i\ ,\qquad \theta\equiv -k^2v^\text{S}\ .
        \end{equation}
        A spatial traceless tensor can be decomposed in a similar way
        \begin{equation}
            \pi_{ij} = \left(\partial_i\partial_j - \frac{1}{3}\delta_{ij}\partial^k\partial_k\right)\pi^\text{S}
                + 2\partial_{(i}\pi^\text{V}_{j)} + \pi^\text{T}_{ij}\ ,
        \end{equation}
        where again the vector part $\v{\pi}^\text{V}$ is divergenceless and $\pi^\text{T}_{ij}$ is the
        tensor part, satisfying
        \begin{equation}
            \partial^i\pi_{ij}^\text{T}=0\ ,\qquad \delta^{ij}\pi_{ij}^\text{T}=0\ .
        \end{equation}
        Alternatively, we can write it in Fourier space as
        \begin{equation}
            \pi_{ij} = -2\left(\hat{k}_i\hat{k}_j-\frac{1}{3}\delta_{ij}\right)\sigma
                +\ii k\left(\hat{k}_i\pi^\text{V}_j+\hat{k}_j\pi^\text{V}_i\right) + \pi^\text{T}_{ij}\ ,
            \qquad\sigma\equiv \frac{k^2}{2}\pi^\text{S}\ ,
        \end{equation}
        according to the notation of \cite{Ma:1995ey} for the scalar part of the shear tensor, $\sigma$. 
        Adapting the notation of \cite{Mukhanov:1990me} for a generic gauge, the metric perturbations 
        can be decomposed as
        \begin{align}\label{eq:RW_decomposed}
            \di s^2 = a^2(\tau)\Bigg\{-&(1+2\psi)\di\tau^2 + 2(\partial_i B-S_i)\di x^i\di\tau \nonumber\\
             &+\Big(\delta_{ij}-2\phi\delta_{ij}+2\partial_i\partial_jE + (\partial_iF_j+\partial_jF_i)
             +h_{ij}\Big)\di x^i\di x^j\Bigg\}\ ,
        \end{align}
        where $\v{S}$ and $\v{F}$ are vector perturbations, i.e. divergenceless vectors, and
        $h_{ij}$ is a tensor perturbation, i.e. a divergenceless and traceless tensor. Our previously
        defined variables for a general metric perturbation \eqref{eq:RW_perturbed} now take the form
        \begin{subequations}\label{eq:metric_main_defs}
        \begin{align}
	        A &= -2\psi\ ,\\
	        B_i &= \partial_i B -S_i\ ,\\
	        H_{ij} &= -2\phi\delta_{ij} + 2\partial_i\partial_jE+\partial_iF_j +\partial_jF_i + h_{ij}\ .
        \end{align}
        \end{subequations}
        From now on it will be convenient to work in Fourier space and to choose a basis adapted
        to the previous decomposition. The components of the line-of-sight vector $\hat{n}$ are
        \begin{subequations}
        \begin{align}
            \hat{n} &= \sin\theta\cos\phi\,\hat{x} + \sin\theta\sin\phi\,\hat{y}+\cos\theta\,\hat{z}\ ,\\
                    &= \frac{1}{\sqrt{2}}\ee^{\ii\phi}\sin\theta\,\hat{e}_+ 
                        +\frac{1}{\sqrt{2}}\ee^{-\ii\phi}\sin\theta\,\hat{e}_-
                        +\cos\theta\,\hat{k}\ ,\\
            &= \sqrt{\frac{4\pi}{3}}\left(-Y^{+1}_1\hat{e}_++Y^{-1}_1\hat{e}_-+Y^0_1\hat{k}\right)\ ,
        \end{align}
        \end{subequations}
        where we have chosen the so-called helicity basis \cite{Durrer:2008eom}
        \begin{subequations}
        \begin{align}
            \hat{k} &\equiv \hat{z}\ ,\\
            \hat{e}_+ &\equiv \frac{1}{\sqrt{2}}(\hat{x}-\ii\hat{y})\ ,\\
            \hat{e}_- &\equiv \frac{1}{\sqrt{2}}(\hat{x}+\ii\hat{y})\ ,
        \end{align}
        \end{subequations}
        and our convention for the spherical harmonics matches those of \cite{Durrer:2008eom} or
        \cite{arfken1999mathematical}.
    
    \subsection{Lower moments evolution}
        The first moments of the Boltzmann equation have already been obtained in the previous sections,
        with $\ell=0$ corresponding to the density \eqref{eq:BoEq_dg} and $\ell=1$ to the
        velocity \eqref{eq:BoEq_vg}. The next moment $\ell=2$ can be obtained via direct integration 
        of the Boltzmann equation \eqref{eq:BoEq_photons} and corresponds to the shear tensor
        \eqref{eq:sheartensor_def}. Performing the appropiate integral, we have
        \begin{align}
	        \dot{\pi}_{ij} &+ \partial_k\int\frac{\di\Omega}{4\pi}n^kn_in_j\mathcal{F}_\gamma
	                -\frac{4}{9}\delta_{ij}\partial_k\delta v^k_\gamma 
	                -\frac{4}{3}\left(\beta_{(i}\partial_{j)}-\frac{1}{3}\delta_{ij}\beta^k\partial_k\right)A 
	                -\frac{8}{15}\left(C_{(ij)}-\frac{1}{3}\delta_{ij}\delta^{kl}C_{kl}\right)\nonumber\\
	            &\qquad+\frac{8}{15}\beta^m\left(D_{m(ij)}-\frac{1}{3}\delta_{ij}D_{mkl}\delta^{kl}\right)\nonumber\\
	        &= -an_e\sigma_T\Bigg[\pi_{ij}-\beta^k_e\int\frac{\di\Omega}{4\pi}n_kn_in_j\mathcal{F}_\gamma 
	                +\frac{4}{9}\v{\beta}_e\cdot\delta\v{v}_\gamma\delta_{ij} 
	                -\frac{32}{15}\left(\delta v_{b\,(i}\beta_{e\, j)}-\frac{1}{3}\delta_{ij}\delta\v{v}_b\cdot
	                \v{\beta}_e\right)\nonumber\\
	            &\qquad\qquad\qquad-\frac{8}{15}\left(\delta v_{b\,(i}\Delta\beta_{j)}
	                -\frac{1}{3}\delta_{ij}\delta\v{v}_b\cdot\Delta\v{\beta}\right)\Bigg]\ .
	    \end{align}
	    To obtain the scalar, vector and tensor parts of this expression, we compute the components as
	    \begin{equation}
	        \pi_{33}\equiv \hat{k}^i\hat{k}^j\pi_{ij}\ ,\qquad 
	            \pi_{3+}\equiv \hat{k}^i\hat{e}_+^j\pi_{ij}\ ,\qquad 
	            \pi_{++}\equiv\hat{e}_+^i\hat{e}_+^j\pi_{ij}\ .
	    \end{equation}
	    The projection of a vector $\v{V}$ in the helicity basis is computed in a similar way
	    \begin{equation}
	        V_3\equiv \hat{k}\cdot\v{V},\qquad
	        V_+\equiv \hat{e}_+\cdot\v{V},\qquad V_-\equiv \hat{e}_-\cdot\v{V}\ .
	    \end{equation}
	    Projecting the equations of motion we obtain
	    \begin{align}
	        \dot{\pi}_{33} &+ \ii k\int\frac{\di\Omega}{4\pi}(\hat{n}\cdot\hat{k})^3\mathcal{F}_\gamma 
	            - \frac{4}{9}\theta_\gamma -\frac{8\ii}{9}\v{\beta}\cdot\v{k}\,A 
	            - \frac{8}{15}\left(C_{33}-\frac{1}{3}C\indices{^k_k}\right)
	            + \frac{8}{15}\left(\beta^mD_{m33}-\frac{1}{3}\beta^mD_{mkl}\delta^{kl}\right)\nonumber\\
	        &= -an_e\sigma_T\Bigg[\pi_{33} -\int\frac{\di\Omega}{4\pi}
	            (\v{\beta}_e\cdot\hat{n})(\hat{k}\cdot\hat{n})^2\mathcal{F}_\gamma +\frac{4}{9}\v{\beta}_e\cdot\delta
	            \v{v}_\gamma - \frac{32}{15}\left(\delta v_b^3\beta_e^3-\frac{1}{3}\delta\v{v}_b\cdot
	            \v{\beta}_e\right)\nonumber\\
	        &\qquad\qquad\qquad-\frac{8}{15}\left(\delta v_b^3\Delta\beta_3 -\frac{1}{3}\delta\v{v}_b\cdot
	            \Delta\v{\beta}\right)\Bigg]\ ,\\
	        \dot{\pi}_{3+}&+\ii k\int\frac{\di\Omega}{4\pi}(\hat{n}\cdot\hat{k})^2(\hat{n}\cdot\hat{e}_+)
	            \mathcal{F}_\gamma - \frac{2}{3}\ii\beta_+kA-\frac{4}{15}\left(C_{3+}+C_{+3}\right)
	            +\frac{4}{15}\beta^j\left(D_{j3+}+D_{j+3}\right)\nonumber\\
	        & = -an_e\sigma_T\left[\pi_{3+}-\int\frac{\di\Omega}{4\pi}(\v{\beta}_e\cdot\hat{n})
	            (\hat{n}\cdot\hat{k})(\hat{n}\cdot\hat{e}_+)\mathcal{F}_\gamma 
	            -\frac{16}{15}\left(\delta v_b^3\beta_e^++\delta v_b^+\beta_e^3\right)
	            -\frac{8}{15}\left(\delta v_b^3\Delta\beta^++\delta v_b^+\Delta\beta^3\right)\right]\ ,\\
            \dot{\pi}_{++}&+\ii k\int\frac{\di\Omega}{4\pi}(\hat{n}\cdot\hat{k})(\hat{n}\cdot\hat{e}_+)^2
                \mathcal{F}_\gamma
	            -\frac{8}{15}C_{++} + \frac{8}{15}\beta^mD_{m++}\nonumber\\
	        &= -an_e\sigma_T\Bigg[\pi_{++} 
	            -\int\frac{\di\Omega}{4\pi}(\v{\beta}_e\cdot\hat{n})(\hat{n}\cdot\hat{e}_+)^2\mathcal{F}_\gamma
	            -\frac{32}{15}\delta v_b^+\beta_e^+ - \frac{8}{15}\delta v_b^+\Delta\beta_+\Bigg]\ ,
	    \end{align}
	    where it is easy to check that the projections correspond to the scalar part and one of the
	    two vector and tensor helicities
	    \begin{equation}
	        \pi_{33}=-\frac{4}{3}\sigma\ ,\qquad \pi_{3+} = \ii k\pi_{+}^\text{V}\ , \qquad \pi_{++}=\pi_{++}^\text{T}\ .
	    \end{equation}	    
	    There remains to perform a couple of angular integrals, writing down the appropiate coefficients
	    of the expansion \eqref{eq:Ylm_expansion} and to substitute the metric variables defined in the
	    previous section \eqref{eq:RW_decomposed}. After these simplifications, and rearranging terms, 
	    the first moments of the Boltzmann equation for photons are\\
	    
	    \paragraph*{Scalar.} $(m=0,\ \ell = 0,1,2)$
	        \begin{align}
	            \dot{\delta}_\gamma&+\frac{4}{3}\theta_\gamma +\frac{4k^2}{3}(B-\dot{E})-4\dot{\phi}
	                +\frac{8}{3}\ii\,(\v{\beta}\cdot\v{k})(\psi-\phi)+\frac{2k^2}{3}\v{\beta}\cdot\v{F}\nonumber\\
		        &= -\frac{4}{3}an_e\sigma_T\left[\v{\beta}_e\cdot(\v{\chi}_\gamma-\v{\chi}_b)
		            +\v{\chi}_b\cdot\Delta\v{\beta}-\frac{\ii}{k}\hat{k}\cdot\Big(
		            \v{\beta}_e\,(\theta_\gamma-\theta_b)
		            +\theta_b\,\Delta\v{\beta}\Big)\right]\ ,
        		        \label{eq:scalar_modes_dg}\\
	            \dot{\theta}_\gamma &-\frac{k^2}{4}(\delta_\gamma - 4\sigma_\gamma) -k^2\psi 
	                - 4\ii (\v{\beta}\cdot\v{k})\dot{\phi} + 2\ii k^2(\v{\beta}\cdot\v{k})(B-\dot{E})
	                -k^2\v{\beta}\cdot\left(\v{S}+\frac{1}{2}\v{F}\right)\nonumber\\
	            &= -an_e\sigma_T\Bigg[\theta_\gamma - \theta_b - \ii \v{k}\cdot\left(\v{\beta}_e(\delta_\gamma
	                -\sigma_\gamma) - \v{\beta}\frac{\delta n_e}{n_e}\right) 
                    - \frac{3\ii k}{4}\v{\beta}\cdot\v{\pi}^\text{V}_\gamma + \ii\psi\Delta\v{\beta}\cdot\v{k}\Bigg]\ ,
                    \label{eq:scalar_modes_thg}\\
                \dot{\sigma}_\gamma &- \frac{4}{15}\theta_\gamma + \frac{3k}{10}\mathcal{F}^0_3
	                -\frac{4\ii}{15}(\v{\beta}\cdot\v{k})\left(\phi+5\psi\right)
	                -\frac{4}{15}k^2(B-\dot{E})-\frac{2}{15}k^2\v{\beta}\cdot\v{F}\nonumber\\
	            &= -an_e\sigma_T\Bigg[\sigma_\gamma-\frac{4\ii}{15k}\hat{k}\cdot\left(\v{\beta}_e\theta_\gamma 
	                + 4\v{\beta}_e\theta_b
	                +\Delta\v{\beta}\,\theta_b\right) + \frac{3\ii}{10}\left((\v{\beta}_e\cdot\hat{k})\mathcal{F}_3^0
	                -\ii\sqrt{\frac{2}{3}}\Big((\v{\beta}_e\cdot\hat{e}_+)\mathcal{F}_3^{-1}
	                +(\v{\beta}_e\cdot\hat{e}_-)\mathcal{F}_3^{+1}\Big)\right)\nonumber\\
	            &\qquad\qquad\qquad -\frac{2}{15}\left(\v{\beta}\cdot\v{\chi}_\gamma + 4\v{\beta}_e\cdot\v{\chi}_b
	                +\Delta\v{\beta}\cdot\v{\chi}_b\right)\Bigg]\ .
	                \label{eq:scalar_modes_sgg}
	        \end{align}
        
        \paragraph*{Vector.} $(m=+1,\ \ell = 1,2)$
            \begin{align}
                \dot{\chi}^+_\gamma &-\frac{3k^2}{4}\pi_+^\text{V} - 4\beta_+\dot{\phi} + \beta_+k^2(B-\dot{E})
	                + \frac{\ii}{2}\v{\beta}\cdot\v{k}\,\dot{F}_+ + \frac{1}{2}\beta_-\dot{h}_{++}\nonumber\\
	            &= -an_e\sigma_T\Bigg[\chi^+_\gamma-\chi^+_b + \beta_+\frac{\delta n_e}{n_e}
	                -\beta_e^+\delta_\gamma - \frac{1}{2}\beta_e^+\sigma_\gamma
	                -\frac{3\ii}{4}(\v{\beta}_e\cdot\hat{k})\pi_+^\text{V} - \frac{3}{4}\beta_e^-\pi_{++}
	                +\psi\Delta\beta_+\Bigg]\ ,
	                \label{eq:vector_modes_vg}\\
	            \dot{\pi}^{\text{V}}_+ & +\frac{4}{15}\left(\chi_\gamma^++\sqrt{\frac{3}{2}}\mathcal{F}^{+1}_3\right)
	                +\frac{4}{15}\beta_+(\phi+5\psi) + \frac{4}{15}\left(S_++\dot{F}_+\right)
	                -\frac{2}{15}\beta_-h_{++} + \frac{2\ii}{15}\beta_3kF_+\nonumber\\
	            &= -an_e\sigma_T\Bigg[\pi_+^\text{V} + \frac{4}{15k^2}\beta_e^+\left(\theta_\gamma+k\mathcal{F}^0_3\right)
	                +\frac{1}{k}\sqrt{\frac{2}{15}}\beta_e^-\mathcal{F}^{+2}_3 + \frac{4\ii}{15k}\beta_e^3
	                \left(\chi_\gamma^++\sqrt{\frac{3}{2}}\mathcal{F}^{+1}_3\right)\nonumber\\
                &\qquad\qquad\qquad+\frac{4}{15k^2}\left(\Delta\beta_+\theta_b+4\beta_e^+\theta_b\right)
                    +\frac{4\ii}{15k}\left(\Delta\beta_3\chi_b^+ + 4\beta_e^3\chi_b^+\right)\Bigg]
                    \label{eq:vector_modes_piVg}
            \end{align}
            
        \paragraph*{Tensor.} $(m=+2,\ \ell = 2)$
            \begin{align}
                \dot{\pi}_{++}&+k\sqrt{\frac{2}{15}}\mathcal{F}^{+2}_3 +\frac{4}{15}\dot{h}_{++} 
	                + \frac{4}{15}\ii(\v{\beta}\cdot\v{k})h_{++}
	            =-an_e\sigma_T\Bigg[\pi_{++} - \frac{8}{15}\left(\beta_e^+\chi_\gamma^+ 
	                +4\beta_e^+\chi_b^+ + \Delta\beta_+\chi_b^+\right)\Bigg]\ .
	                \label{eq:tensor_modes_piT}
            \end{align}
        The corresponding results for the other helicity can be obtained substituting $-\leftrightarrow +$
        in every sub and superscript.\\
        
        Several comments are in order now. In the first place, note that all the couplings between
        scalar, vector and tensor modes are introduced by terms proportional to $\beta$. In the 
        standard case, each mode evolves independently. In the second place, note the appearance
        of terms with $\ell = 3$. What we present here are but the lowest moments of a whole 
        hierarchy of coupled differential equations. This system obeys a recurrence relation but
        it must be truncated at a finite, prefearably large, value of $\ell$. The traditional 
        line-of-sight approach \cite{Seljak:1996is} was developed to allow a truncation at 
        lower $\ell$, and it is used by every modern Boltzmann solver \cite{Lewis:2002ah, Blas:2011rf}. It will not be used here, 
        even though it can be adapted to our case \cite{CMB_paper}, since we are mainly interested in 
        the matter power spectra and not in the CMB. Instead, we will study a simplified version of the 
        system, under the following approximations.        
        \begin{itemize}
            \item We will work to first order in $\beta$ and to first order in cosmological perturbations, 
                keeping cross-products. We have been implicitly working under this assumption, since the 
                RW background is only correct to first order in $\beta$, but we will consistently carry 
                it through. 
                
            \item We will assume that there are no initial vector or tensor modes to zero order in $\beta$. 
                The assumption is justified for vector modes, since most popular models of inflation do 
                not produce them at all. On the other hand, since we have not detected tensor modes so 
                far, and we have stringent limits on their amplitude, we assume that their amplitude is 
                small enough so we can neglect them.
                
                Under this assumption, the hierarchy is simplified. Since they are zero initially
                the only production occurs through their new couplings, i.e. it is proportional to 
                $\beta$. The vector modes are then $\Od{\beta}$ and the tensor modes are $\Od{\beta^2}$.
                In general, we can neglect the backreaction of higher $m$ modes into lower $m$ modes.
                As we will see in the next section, the Einstein equations are not modified, so we
                can apply the same reasoning to the metric variables, i.e. $\v{S}$ and $\v{F}$ are 
                $\Od{\beta}$ and $h_{ij}$ is $\Od{\beta^2}$.
            
            \item The last approximation is the so-called fluid approximation. We will truncate the
                hierarchy at $\ell = 2$ for scalar and vector modes. This is a classic
                working assumption in approximate computations of the CMB, that may introduce up to
                10\% errors in computations of the CMB spectrum \cite{Hu:1995fqa}. Another source of
                error in our case is the naive truncation scheme we are using, i.e. setting to zero
                all higher moments. It is a well known fact that this truncation scheme, neglecting
                the damping produced by the transfer of power to higher moments, produces some spurious
                growth at small scales in the ultrarelativistic species. Since, in this work, we are 
                not interested in following with great precision the evolution of photons or neutrinos,
                and we have checked that they have a negligible impact in our final results, 
                we will nonetheless stick to this crude truncation scheme in our numerical solutions.
                
                To increase the accuracy of the results for ultrarelativistic species one would need to 
                evolve the full hierarchy \cite{CMB_paper}, or at least introduce a better truncation scheme 
                \cite{Ma:1995ey} or an effective viscosity in the equations of motion for 
                ultrarelativistic species \cite{Hu:1998kj}.
            
        \end{itemize}
        
\section{Einstein equations}\label{sec:EinsteinEquations}
    In this section we present the last piece of information needed to solve the system: the evolution
    of the metric perturbations. It is worth remembering that, to first order in $\beta$, the background
    quantities we are interested in, e.g. $\rho$ and $P$, are equal in the $\mathcal{O}$ and 
    $\tilde{\mathcal{O}}$ frames. Hence, as in previous sections, we will drop the distinction. 
    The full energy-momentum tensor for each component is
    \begin{subequations}
    \begin{align}
        T\indices{^0_0} + \delta T\indices{^0_0} &= -\rho-\delta\rho-(\rho+P)B_i\beta^i\ ,\\
        T\indices{^0_i} + \delta T\indices{^0_i} &= \delta Q_i+(\rho+P)\left(\delta^j_i
            +\frac{1}{2}A\delta^j_i +\frac{1}{2}H^j_i\right)\beta_j\ ,\\
        T\indices{^i_j} + \delta T\indices{^i_j} &= P\delta^i_j + \delta P\,\delta^i_j + \delta\Pi^i_j 
            +(\rho + P)\beta_jB^i\ ,
    \end{align}
    \end{subequations}
    where $\beta$ is different for each component. We must write now the Einstein equations with this
    source for the metric \eqref{eq:RW_decomposed}. For the background evolution we obtain the
    standard Friedmann equations plus a condition for the cosmic center of mass frame
    \begin{align}
        \mathcal{H}^2 &= \frac{8\pi Ga^2}{3}\sum_s\rho_s\ , \label{eq:Einstein_friedmann}\\
        0 &= \sum\v{\beta}_s(\rho_s + P_s)\ ,\label{eq:Einstein_CMframe}\\
        \dot{\mathcal{H}}+\frac{1}{2}\mathcal{H}^2 &= -4\pi Ga^2\sum_s P_s\ .\label{eq:Einstein_friedmann_accel}
    \end{align}
    Applying the condition \eqref{eq:Einstein_CMframe}, the explicit $\beta$ contributions to the
    full energy-momentum tensor vanish
    \begin{subequations}
    \begin{align}
        \delta T\indices{^0_0} &= \sum_s\delta T\indices{_s^0_0} = -\sum_s\delta\rho_s\ ,\\
        \delta T\indices{^0_i} &= \sum_s\delta T\indices{_s^0_i} = \sum_s\delta Q_{s\;i}\ ,\\
        \delta T\indices{^i_j} &= \sum_s\delta T\indices{_s^i_j} = \sum_s\left(\delta P_s\delta^i_j 
            +\delta\Pi\indices{_s^i_j}\right)\ .
    \end{align}
    \end{subequations}
    In our case the non-relativistic species have pressure and anisotropic stress of order 
    $\beta$. Splitting the sources into relativistic and non-relativistic components we have
    \begin{align}
        \delta\rho &= \sum_\text{NR}\delta\rho + \sum_\text{R}\delta\rho\ ,\\
        \delta P &= \frac{2}{3}\sum_\text{NR}\rho\,\beta_k\delta v^k + \frac{1}{3}\sum_\text{R}\delta\rho\ ,\\
        \delta Q^i &= \sum_\text{NR}\rho\,\delta v^i + \frac{4}{3}\sum_\text{R}\rho\,\delta v^i\ ,\\
        \delta\Pi^{ij} &= \sum_\text{NR}\rho\left(\beta^i\delta v^j + \beta^j\delta v^i
            -\frac{2}{3}\delta^{ij}\beta_k\delta v^k\right)+\sum_\text{R}\delta\Pi^{ij}\ .
    \end{align}
    Finally, the Einstein equations read
    \begin{itemize}
        \item $(0,0)$
            \begin{align}
                2k^2\phi+6\mathcal{H}(\dot{\phi}+\mathcal{H}\psi) 
                    -2k^2\mathcal{H}\Big(B-\dot{E}\Big) =-8\pi G a^2\delta\rho\ .
            \end{align}
        \item $(0,i)$
            \begin{align}
                \ii k(\dot{\phi}+\mathcal{H}\psi) &= -4\pi Ga^2\delta Q_3\ ,\\
                k^2(S_++\dot{F}_+) &= 16\pi Ga^2\delta Q_+\ .
            \end{align}
        \item $(i,j)$
            \begin{align}
                k^2(\phi-\psi)-k^2(\partial_\tau +2\mathcal{H})\Big(B-\dot{E}\Big)
                    &=-12 \pi Ga^2\delta \Pi_{33}\ , \\
                (\partial_\tau+2\mathcal{H})(\dot{\phi}+\mathcal{H}\psi)
                    +\psi (\dot{\mathcal{H}}-\mathcal{H}^2)
                    &=4\pi Ga^2(\delta P+\delta\Pi_{3 3})\ ,\\
                \ii k(\partial_\tau +\mathcal{H})(S_++\dot{F}_+) 
                    &= 16\pi Ga^2\delta\Pi_{+3}\ ,\\
                \frac{1}{2}\Big(\partial^2_\tau +2\mathcal{H}\partial_\tau + k^2\Big)h_{++}
                    &= 8\pi Ga^2\delta\Pi_{++}\ .
            \end{align}
    \end{itemize}
    Again, the results for the $-$ helicity can be obtained substituting $-\leftrightarrow +$
    in every sub and superscript.
    With our notation, the Newtonian gauge can be obtained just setting $B=E=0$ and the synchronous
    gauge is defined as
    \begin{subequations}\label{eq:sync_gauge_def}
    \begin{align}
        \psi &= B = 0\ ,\\
        \phi &= \eta\ ,\\
        E &= -\frac{1}{2k^2}\left(h+6\eta\right)\ .
    \end{align}
    \end{subequations}
    The Einstein equations in the synchronous gauge for scalar perturbations can be written as
    \begin{align}
        &\dot{h} -\frac{2 k^{2} \eta}{\mathcal{H}} = 3 \mathcal{H} \delta\ ,\\
        &\dot{\eta} = \frac{3\mathcal{H}^{2}}{2 k^{2}}\left(1+w\right)\theta\ ,\\
        & \ddot{h} + 6 \ddot{\eta} + 2\mathcal{H}\left(\dot{h} +6 \dot{\eta}\right) - 2 k^{2} \eta  
            = - 12 \mathcal{H}^{2} \sigma\ ,\\
        &\ddot{h}+\mathcal{H} \dot{h} = - 3 \left(1 + 3 c_\text{s}^2\right) \mathcal{H}^{2}\delta\ ,
    \end{align}
    where we have defined
    \begin{equation}
        c_\text{s}^2\delta \equiv \frac{1}{\rho}\delta P\ ,\qquad \theta \equiv \frac{1}{\ii k(\rho + P)}\delta Q_3\ ,
            \qquad \sigma \equiv -\frac{4}{3\rho}\delta \Pi_{33}\ .
    \end{equation}
    Appendix \ref{sec:gauge_transformations} contains a discussion about how the gauge transformations
    are modified for non-comoving fluids and, in particular, how to relate the Newtonian and synchronous 
    gauges.

\section{Reduced system and final equations}\label{sec:ReducedSystem}
    This section contains the final, simplified equations that will be numerically integrated.
    In the first place, the relevant equations for the evolution of the bulk velocities are presented
    in section \ref{sec:bulk_velocities}.
    The background follows the standard $\Lambda$CDM evolution to first order in $\beta$, but there is
    a first order effect on the perturbations.\\
    
    Once the evolution of the bulk velocities is known, we need to study the modified evolution of
    the perturbations. Working to first order in $\beta$ and with the approximations made at the end 
    of section \ref{sec:AngularAnalysis}, our modifications to the scalar and vector modes decouple 
    and can be treated separately. This is the subject of the last two sections \ref{sec:scalar_modes}
    and \ref{sec:vector_modes}. As mentioned before, the tensor modes are second order in $\beta$ and
    we neglect them.
    
    \subsection{Bulk velocities}\label{sec:bulk_velocities}
        The evolution of the velocities of the different fluids is governed by
        \begin{subequations}
        \begin{align}
             \dot{\v{\beta}} &= -\frac{1}{\tau_c}\Delta\v{\beta}\ ,\label{eq:beta_g2}\\
             \dot{\v{\beta}}_\nu &= 0\ ,\label{eq:beta_nu2}\\
             \dot{\v{\beta}}_e &= -\mathcal{H}\v{\beta}_e + \frac{1}{R\tau_c}\Delta\v{\beta}\ ,
                 \label{eq:beta_e2}\\
             \dot{\v{\beta}}_c &= -\mathcal{H}\v{\beta}_c\ .
        \end{align}
        \end{subequations}
        where we have defined
        \begin{alignat}{2}\label{eq:TC_beta_defs}
            R&\equiv \frac{3\rho_b}{4\rho_\gamma}\ , 
                &\qquad \Delta\v{\beta} &\equiv \v{\beta}_\gamma-\v{\beta}_e\ ,\nonumber\\
            \tau^{-1}_c &\equiv an_e\sigma_T\ , &\qquad \mathcal{A}&\equiv \frac{R}{1+R}\ ,
        \end{alignat}
        and, as we mentioned before, the initial conditions are chosen according to the constraint
        \begin{equation}\label{eq:beta_system_constraint}
            \sum_s(\rho_s+P_s)\v{\beta}_s = 0\ ,
        \end{equation}
        so the cosmic center of mass condition is maintained in the evolution. 
        Moreover, we will assume that all the bulk velocities $(\v{\beta},\v{\beta}_e, \v{\beta}_\nu,\v{\beta}_c)$ 
        are aligned along the $\hat{\beta}$ axis in the $\mathcal{O}$ frame. As we will shortly see,
        and can be inferred from \eqref{eq:beta_g2} and \eqref{eq:beta_e2}, when two species
        are tightly coupled their velocities evolve to become equal. Once a particle species decouples,
        the magnitude of its velocity evolves independently but, in the absence of additional interactions
        or other sources of anisotropy, it does not change its direction. In our scenario, we 
        assume that the whole visible sector has been in thermal equilibrium at some time so all its
        components, even if they are decoupled like the neutrinos, have velocities pointing in 
        the direction $\hat{\beta}$. The only remaining contribution is the dark sector, with DM among its
        components. The dark sector in the $\mathcal{O}$ frame counterbalance the flux of momentum of 
        the visible sector to achieve an isotropic universe, so it must point in the $-\hat{\beta}$ direction.\\
        
        CDM and neutrinos are decoupled, but the photon-baryon system must be treated with
        some care. In the tight coupling limit, $\tau_c\ll 1$, it is easy to see that the velocities 
        converge in direction and magnitude and we can look for an approximate solution of this system.
        Expanding perturbatively in the small parameter $\tau_c$ we have
        \begin{align}
            \dot{\v{\beta}} &= -\mathcal{A}\mathcal{H}\v{\beta} + \Od{\tau_c}\ ,\\
            \Delta\v{\beta} &= \mathcal{A}\mathcal{H}\v{\beta}\tau_c + \Od{\tau_c^2}\ .
        \end{align}
        The differential equation can be solved to yield
        \begin{equation}\label{eq:beta_evol_TC}
            \v{\beta} = \frac{\v{\beta}_0}{1+R}\ ,
        \end{equation}
        where $\v{\beta}_0$ is the initial velocity of the visible sector in the $\mathcal{O}$ frame, the 
        only additional free parameter in our model. In a similar way, the neutrino and CDM equations can 
        be solved to give
        \begin{align}
            \v{\beta}_\nu &= \v{\beta}_0\ ,\\
            \v{\beta}_c &= \v{\beta}_c^{\text{today}}a^{-1}\ .\label{eq:betac_evol}
        \end{align}
        Using the scaling \eqref{eq:betac_evol} and enforcing the constraint 
        \eqref{eq:beta_system_constraint} during the tightly coupled regime, the value of 
        $\beta_c^\text{today}$ is found to be
        \begin{equation}
            \beta_c^\text{today} = -\frac{4}{3}\beta_0\frac{\Omega_\gamma + \Omega_\nu}{\Omega_\text{cdm}}\ .
        \end{equation}
        It is important to notice that according to the evolution of $\beta_c$ \eqref{eq:betac_evol},
        early enough in time, the condition $\beta_c\ll 1$ could break down. However, this is only
        the case if the DM keeps the non-relativistic distribution at early times. 
        If the DM were light enough it could behave as a radiation-like fluid well before its bulk
        velocity reaches $\beta_c=1$. In this case, $\beta_c$ would remain constant and small.
        On the other hand, if the DM is heavy, its bulk velocity can reach the relativistic 
        regime. In this case, it is worth mentioning that even if the anisotropies that would arise 
        at the background level are $\Od{\beta_0}$ and not $\Od{\beta_0^2}$, their effects can only 
        be relevant well before the matter-dominated era, with no observational consequences.\\

        In order to avoid choosing any particular framework for the dark sector, we will not follow
        the dark matter evolution using \eqref{eq:betac_evol}. Instead, we follow
        the evolution of photons \eqref{eq:beta_g2}, neutrinos \eqref{eq:beta_nu2} and baryons
        \eqref{eq:beta_e2}, and then the momentum of the dark sector, regardless of its composition, can
        be obtained imposing the center of mass condition \eqref{eq:Einstein_CMframe}. Thus, our only 
        assumptions regarding the dark sector are that it is subdominant at early times and that it 
        behaves as cold matter plus cosmological constant at late times.
            
    \subsection{Scalar modes}\label{sec:scalar_modes}        
        Here we provide the scalar mode equations for the photon-baryon system. Neutrinos and CDM obey the
        same equations as photons and baryons, respectively, without the coupling term, i.e. $\sigma_T=0$.
        In addition to the definitions \eqref{eq:TC_beta_defs} we define
        \begin{equation}
            \Delta\theta \equiv \theta_\gamma - \theta_b\ ,
        \end{equation}        
        The evolution of the photon perturbations is described by the equations (\ref{eq:scalar_modes_dg},
        \ref{eq:scalar_modes_thg}, \ref{eq:scalar_modes_sgg}). Under the approximation scheme discussed
        at the end of section \ref{sec:AngularAnalysis}, i.e. neglecting backreaction of vector modes and
        neglecting moments higher than $\ell=2$, these equations take the form
        \begin{subequations}\label{eq:ScaSys_mixed_photons}
        \begin{align}
            \dot{\delta}_\gamma+\frac{4}{3}\theta_\gamma +\frac{4k^2}{3}(B-\dot{E})-4\dot{\phi}
	                +\frac{8}{3}\ii\,(\v{\beta}\cdot\v{k})(\psi-\phi)
	        &= -\frac{4}{3\tau_c}\left[-\frac{\ii}{k}\hat{k}\cdot\Big(
	            \v{\beta}_e\Delta\theta
	            +\theta_b\,\Delta\v{\beta}\Big)\right]\ ,\\
            \dot{\theta}_\gamma -\frac{k^2}{4}(\delta_\gamma - 4\sigma_\gamma) -k^2\psi 
	                - 4\ii (\v{\beta}\cdot\v{k})\dot{\phi} + 2\ii k^2(\v{\beta}\cdot\v{k})(B-\dot{E})
            &= -\frac{1}{\tau_c}\Bigg[\Delta\theta - \ii \v{k}\cdot\left(\v{\beta}_e(\delta_\gamma
                -\sigma_\gamma) - \v{\beta}\frac{\delta n_e}{n_e}-\Delta\v{\beta}\psi\right)\Bigg]\ ,\\
            \dot{\sigma}_\gamma - \frac{4}{15}\theta_\gamma -\frac{4}{15}k^2(B-\dot{E})
	                -\frac{4\ii}{15}(\v{\beta}\cdot\v{k})\left(\phi+5\psi\right)
            &= -\frac{1}{\tau_c}\Bigg[\sigma_\gamma-\frac{4\ii}{15k}\hat{k}\cdot\left(\v{\beta}_e\theta_\gamma 
                + 4\v{\beta}_e\theta_b
                +\Delta\v{\beta}\,\theta_b\right)\Bigg]\ .
        \end{align}
        \end{subequations}
        Neutrinos are described by the same system, without collision term.
        For baryons, we apply our approximation scheme to equations \eqref{eq:BoEq_db} and 
        \eqref{eq:BoEq_vb} obtaining
        \begin{subequations}\label{eq:ScaSys_mixed_baryons}
        \begin{align}
            \dot{\delta}_b + \theta_b -3\dot{\phi} +k^2(B-\dot{E}) 
                -2\ii\mathcal{H}\frac{(\v{\beta}_e\cdot\hat{k})}{k}\theta_b 
                +2\ii(\v{\beta}_e\cdot\v{k})(\psi-\phi)
            &= \frac{1}{R\tau_c}\left[-\frac{\ii}{k}\hat{k}\cdot\Big(
	            \v{\beta}_e\Delta\theta
	            +\theta_b\,\Delta\v{\beta}\Big)\right]\ ,\\
	        \dot{\theta}_b + \mathcal{H}\theta_b - k^2\psi + 2\ii(\v{\beta}_e\cdot\v{k})\theta_b
	            -4\ii(\v{\beta}_e\cdot\v{k})\dot{\phi} + 2\ii k^2 (\v{\beta}_e\cdot\v{k})(B-\dot{E})
	        &= \frac{1}{R\tau_c}\Bigg[\Delta\theta - \ii \v{k}\cdot\left(\v{\beta}_e(\delta_\gamma
                -\sigma_\gamma) - \v{\beta}\frac{\delta n_e}{n_e}-\Delta\v{\beta}\psi\right)\Bigg]\ .
        \end{align}
        \end{subequations}
        Again, CDM equations take the same form, but without collision term.
        The evolution of the full energy-momentum tensor is described by
        \begin{subequations}\label{eq:ScaSys_mixed_full}
        \begin{align}
            \dot{\delta} + 3 \mathcal{H} {(c_\text{s}^2-w)\delta} + \left(1+w\right) \theta - (1+w)\left(3\dot{\phi}-k^2(B-\dot{E})\right)&= 0\ ,\\
            \dot{\theta} + \left(1- 3 w\right) \mathcal{H} \theta + \frac{\dot{w}}{1+w}\theta -\frac{k^{2}}{1+w}{c_\text{s}^2\delta} +\frac{4 k^{2}}{3 \left(1+w\right)}\sigma -k^2\psi&= 0\ ,
        \end{align}
        \end{subequations}
        where
        \begin{align}
            c_\text{s}^2\delta &= \frac{1}{\rho}\left(\frac{1}{3}\rho_\gamma\delta_\gamma
                +\frac{1}{3}\rho_\nu\delta_\nu - \frac{2\ii\rho_c\beta_c\theta_c}{3k}
                -\frac{2\ii\rho_b\beta_e\theta_b}{3k}\right)\ ,\\
            \sigma &= \frac{1}{\rho}\left(\rho_\gamma\sigma_\gamma + \rho_\nu\sigma_\nu 
                + \frac{\ii\rho_c\beta_c\theta_c}{k} + \frac{\ii\rho_b\beta_e\theta_b}{k}\right)\ .
        \end{align}
        
        The system of cosmological perturbations is a system of linear differential equations. A generic
        cosmological perturbation $g(\tau,\v{k})$ can be written as a product of a primordial
        perturbation, encoding the initial condition, and a transfer function, encoding the subsequent
        evolution \cite{Liddle:2000cg}
        \begin{equation}\label{eq:g_pert}
            g(\tau,\v{k}) = T_g(\tau,\v{k})\mathcal{R}_{\v{k}}\ .
        \end{equation}
        As long as we consider only the adiabatic mode, every cosmological perturbation is proportional
        to the primordial curvature perturbation $\mathcal{R}_{\v{k}}$. Furthermore, since the system 
        is linear, it can be recast into a system for the evolution of the transfer functions with the
        substitution $g\to T_g$. It is common practice to abuse slightly of the notation and to denote
        $T_g$ as $g$, the perturbation itself, and to solve the system as if it had initial conditions
        $\mathcal{R}_{\v{k}}=1$. The information about the initial conditions is recovered later, in 
        the computation of the physical spectra, convolving the transfer function with the primordial
        spectrum. We follow this practice.\\
        
        An important difference in \eqref{eq:ScaSys_mixed_photons} and \eqref{eq:ScaSys_mixed_baryons} 
        with respect to the $\beta=0$ case is the appearance of imaginary terms. Usually, even though 
        the Fourier coefficients are generally complex, the evolution equations are real. In this case,
        both real and imaginary part of $T_g$ satisfy the same equation and with a judicious choice
        of initial global phase it can be made purely real. With the appearance of complex coefficients,
        real and imaginary parts form a coupled system with different equations of motion.\\
        
        We will assume that the global phase has been chosen so that the imaginary parts are initially
        zero, or at most $\Od{\beta}$. Even if they are initially zero, the terms proportional to 
        $\beta$ couple the imaginary to the real parts, driving them to a finite value proportional to 
        $\beta$. Then we are in the same situation as with the vector modes. The imaginary parts of the 
        scalar modes are determined by the real parts, but they do not backreact on them. The real parts 
        follow the standard cosmological evolution. Therefore, every perturbation transfer function can 
        be splitted as
        \begin{equation}\label{eq:g_transfer_splitted}
            g(\tau,\v{k}) = g^R(\tau,k) + \ii(\hat{\beta}\cdot\hat{k})g^I(\tau,k)\ ,
        \end{equation}
        where now $g^R$ and $g^I$ are purely real and do not depend on the direction of $\hat{k}$. 
        Following this definition and the previous discussion, we are in the following situation.
        \begin{itemize}
            \item The real part of the perturbations $g^R$ only contains adiabatic perturbations, 
                as in standard $\Lambda$CDM, and follows the standard evolution.
            \item The real parts act as external sources in the system for the imaginary parts,
                via contributions $\Od{\beta}$. The imaginary parts of the perturbations 
                $g^I$ are $\Od{\beta}$.
        \end{itemize}
        
        We will work in the synchronous gauge \eqref{eq:sync_gauge_def}. The modified 
        gauge-transformation properties are provided in the appendix \ref{sec:gauge_transformations}.
        Written in the synchronous gauge, the final system for the photon perturbations is
        \begin{subequations}\label{eq:ScaSys_photons_final}
        \begin{align}
            \dot{\delta}^I_\gamma + \frac{4}{3}\theta_\gamma^I + \frac{2}{3}\dot{h}^I
                -\frac{8\beta k}{3}\eta^R
	        &= \frac{4}{3k\tau_c}\Big[\beta_e\Delta\theta^R+\theta_b^R\,\Delta\beta\Big]\ ,\\
            \dot{\theta}_\gamma^I -\frac{k^2}{4}(\delta_\gamma^I - 4\sigma_\gamma^I)
                +\beta k\left(\dot{h}^R+2\dot{\eta}^R\right)
            &= -\frac{1}{\tau_c}\Bigg[\Delta\theta^I - k\left(\beta_e(\delta_\gamma^R
                -\sigma_\gamma^R) - \beta\frac{\delta n_e^R}{n_e}\right)\Bigg]\ ,\\
            \dot{\sigma}^I_\gamma - \frac{4}{15}\theta_\gamma^I
                -\frac{2}{15}\left(\dot{h}^I+6\dot{\eta}^I\right) - \frac{4\beta k}{15}\eta^R
            &= -\frac{1}{\tau_c}\Bigg[\sigma_\gamma^I-\frac{4}{15k}\left(\beta_e\theta_\gamma^R 
                + 4\beta_e\theta_b^R
                +\Delta\beta\,\theta_b^R\right)\Bigg]\ .
        \end{align}
        \end{subequations}
        The equations for neutrinos are the same, setting the collision part to zero. For the baryons 
        we have
        \begin{subequations}\label{eq:ScaSys_baryons_final}
        \begin{align}
            \dot{\delta}_b^I + \theta_b^I +\frac{1}{2}\dot{h}^I 
                -\frac{2\mathcal{H}\beta_e}{k}\theta_b^R -2\beta_ek\eta^R              
            &= -\frac{1}{kR\tau_c}\Big[\beta_e\Delta\theta^R+\theta_b^R\Delta\beta\Big]\ ,\\
	        \dot{\theta}^I_b + \mathcal{H}\theta_b^I + 2\beta_e k\theta_b^R
                +\beta_ek\left(\dot{h}^R+2\dot{\eta}^R\right)
	        &= \frac{1}{R\tau_c}\Bigg[\Delta\theta^I - k\left(\beta_e(\delta_\gamma^R
                -\sigma^R_\gamma) - \beta\frac{\delta n_e^R}{n_e}\right)\Bigg]\ .
        \end{align}
        \end{subequations}        
        And finally for the full fluid
        \begin{subequations}\label{eq:ScaSys_full_final}
        \begin{align}
            \dot{\delta}^I + 3 \mathcal{H} {(c_\text{s}^2-w)\delta^I} + \left(1+w\right) \theta^I &= -\frac{1}{2}\left(1+w\right) \dot{h}^I\ ,\\
            \dot{\theta}^I + \left(1- 3 w\right) \mathcal{H} \theta^I + \frac{\dot{w}}{1+w}\theta^I -\frac{k^{2}}{1+w}{c_\text{s}^2\delta^I} +\frac{4 k^{2}}{3 \left(1+w\right)}\sigma^I &= 0\ ,
        \end{align}
        \end{subequations}        
        where, in the synchronous gauge,
        \begin{align}
            c_\text{s}^2\delta^I &= \frac{1}{\rho}\left(\frac{1}{3}\rho_\gamma\delta_\gamma^I
                +\frac{1}{3}\rho_\nu\delta_\nu^I
                -\frac{2\rho_b\beta_e\theta_b^R}{3k}\right)\ ,\\
            \sigma^I &= \frac{1}{\rho}\left(\rho_\gamma\sigma_\gamma^I + \rho_\nu\sigma_\nu^I
                + \frac{\rho_b\beta_e\theta_b^R}{k}\right)\ .
        \end{align}
        To complete the system, we compute the variables $\eta$ and $a\dot{h}$ using a combination of the 
        Einstein equations
        \begin{align}
            \dot{\eta}^I &= \frac{3\mathcal{H}^{2}}{2 k^{2}}\left(1+w\right)\theta^I\ , \label{eq:Einstein1_final}\\      
            \ddot{h}^I+\mathcal{H} \dot{h}^I &= - 3 \left(1 + 3 c_\text{s}^2\right) \mathcal{H}^{2}\delta^I\ . \label{eq:Einstein2_final}
        \end{align}
        
        Usually, one would integrate the equations for photons, baryons, neutrinos and CDM. Then, after
        adding all the components, one would compute the sources for the metric perturbations, i.e. the
        full energy-momentum tensor. As we mentioned before, in our setup we have found it more 
        convenient to follow a different route. Instead of tracking the behaviour of the dark sector, 
        thus choosing a particular DM framework, we follow the evolution of the whole fluid 
        \eqref{eq:ScaSys_full_final}. Using these equations, the only underlying assumptions are
        \begin{itemize}
            \item The dark sector is subdominant with respect to neutrinos and photons at early times,
                i.e. before the matter-domination era. 
            \item There is a transition to a CDM behaviour at late times. 
        \end{itemize}
        Under these assumptions, the only CDM contribution to the full fluid goes into the 
        equation of state $w$, since in the synchronous gauge it does not contribute to
        $\delta P$ or $\sigma$ at first order in $\beta$. The evolution of the dark sector can be
        obtained afterwards subtracting the contributions of photons, baryons and neutrinos from the
        full fluid.\\
                
        We are in position now to numerically solve the system. Our strategy
        can be summarized as follows.
        \begin{enumerate}[I)]
            \item The background and the real part of the perturbation, labelled with $R$, follow the
                standard evolution and act as external sources in our system.
                We use {\sc class} \cite{Blas:2011rf} to precompute these sources and then solve 
                numerically the system for the imaginary parts, labelled with $I$. The equations for the 
                evolution of $\beta$ are \eqref{eq:beta_g2}, \eqref{eq:beta_nu2} and \eqref{eq:beta_e2}.
            \item The system to be solved for the imaginary parts consists of 
                \eqref{eq:ScaSys_photons_final}, \eqref{eq:ScaSys_baryons_final}, 
                \eqref{eq:ScaSys_full_final}, \eqref{eq:Einstein1_final}, \eqref{eq:Einstein2_final} 
                and a neutrino contribution equal to \eqref{eq:ScaSys_photons_final} but without 
                collisions.
            \item The initial conditions are discussed in the appendix \ref{sec:initial_conditions}, 
                where we find the analytic super-Hubble behaviour of the perturbations. In addition to
                the usual adiabatic and isocurvature modes \cite{Bucher:1999re}, we find a ``sourced
                mode'' that depends on the initial value of $\eta^R$, i.e. the standard initial 
                curvature perturbation in the adiabatic mode. If we assume that 
                $\theta_\gamma(\tau=0)=\theta_\nu(\tau=0)$, since they have been thermally coupled at
                some point, the presence of these external sources introduces an isocurvature velocity
                perturbation
                \begin{equation}
                    \theta^I_\gamma(\tau=0) = \theta^I_\nu (\tau= 0) = 2\beta_0 k \eta^R(\tau=0)\ ,
                \end{equation}
                where $\beta_0$ is the initial value of the velocity of photons and neutrinos.
                All other perturbations are assumed to be initially zero. The whole analysis and the
                series expansions are described in the appendix \ref{sec:initial_conditions}.
            \item The impact of perturbed recombination \cite{Naoz:2005pd, Lewis:2007zh} is usually 
                neglected but it will be important in our case. This subject is covered in section 
                \ref{sec:PerturbRec}.
            \item We need two approximation schemes to follow the numerical evolution, the so-called
                tight coupling approximation (TCA) and radiation streaming approximation (RSA). We use
                the same criteria and switches as {\sc class} \cite{Blas:2011rf}. The appropiate
                equations for TCA are presented in section \ref{sec:TCA}. While TCA is important at
                early times, RSA is crucial at late times, when the ultrarelativistic species are
                decoupled and start oscillating fast. In this case, it is time consuming to follow 
                every oscillation but it is not necessary since at late times the impact of 
                ultrarelativistic components on the metric potentials is negligible. We describe this 
                scheme in section \ref{sec:RSA}. 
        \end{enumerate}
        
        \subsubsection{Perturbed recombination}\label{sec:PerturbRec}
            The disturbances in the photon temperature field produce perturbations in the ionization 
            fraction of the electrons. In standard $\Lambda$CDM, this inhomogeneous recombination 
            produces second order effects in the CMB, but it has proven important at late times when 
            computing other observables like the 21 cm radiation, through its effects in the gas 
            temperature \cite{Naoz:2005pd, Lewis:2007zh}.\\
            
            Boltzmann codes like {\sc camb} \cite{Lewis:2002ah} and {\sc class} \cite{Blas:2011rf}
            have implemented perturbed recombination at late times. This implementations follow the 
            formulae of {\sc recfast} \cite{Seager:1999km}, including perturbations into the 
            recombination coefficient that effectively takes into account multilevel atom computations. 
            These codes also track the evolution of the gas temperature and its perturbations, that 
            modify significantly the baryon sound speed.\\
            
            The study of the dark ages in detail is beyond the scope of this work, but, in our case, 
            perturbed recombination plays a role in the photon-baryon system to first order in $\beta$, 
            where a perturbation in the number of \emph{free} electrons $\delta n_e$ appears, e.g. see 
            \eqref{eq:BoEq_photons}. The baryon sound speed has been neglected in our calculations, it 
            is only important at very small scales, and we will neglect the perturbations in the gas 
            temperature and the recombination coefficient as well. Defining the ionization fraction 
            \cite{Kolb:1990vq, Seager:1999km} as
            \begin{equation}
                x_e\equiv \frac{n_e}{n_b}\ ,
            \end{equation}
            where $n_e$ and $n_b$ are the number densities of free electrons and baryons, respectively, 
            we have 
            \begin{equation}\label{eq:def_dne}
                \frac{\delta n_e}{n_e}\equiv \delta_b + \delta_{x_e}\ ,
            \end{equation}
            where $\delta_{x_e}\equiv \delta x_e/x_e$ is the relative perturbation in the ionization
            fraction. Since $\delta n_e$ always appears multiplied by $\beta$, to study its evolution
            it suffices to take \eqref{eq:lioville_number_density} with $\beta=0$. After a few 
            manipulations we obtain
            \begin{equation}
                \pd{}{\tau}\left(\frac{\delta n_e}{n_e}\right)-\pd{}{\tau}\left(\frac{1}{a^3n_e}\right)
                    a^3\delta n_e + \theta_b + \frac{1}{2}\dot{h} = 0\ .
            \end{equation}
            Substituting the definition \eqref{eq:def_dne} we get the final evolution equation
            \begin{equation}
                \dot{\delta}_{x_e} + \frac{\dot{x}_e}{x_e}\Big(\delta_b + \delta_{x_e}\Big) = 0\ .
            \end{equation}
            We solve this equation for each mode with initial conditions 
            $\delta_{x_e}(\tau=\tau_\text{ini})=0$ and using the full ionization history $x_e(\tau)$ 
            provided by the thermodynamics module in {\sc class}, computed using {\sc recfast}.
            
        \subsubsection{Tight coupling expansion}\label{sec:TCA}
            At early times, the time scale of Thomson scattering, $\tau_c\equiv (an_e\sigma_T)^{-1}$, is
            much shorter than the time scales of evolution of the background, $\mathcal{H}^{-1}$, or
            that of the modes, $k^{-1}$. In this regime, the photon-baryon system becomes computationally 
            hard to solve, since it involves widely different scales. However, we can find approximate
            expressions to follow the evolution, expanding perturbatively in the small parameter
            $\tau_c$. This is known as the tight coupling approximation (TCA). In this work, as usual, 
            we will restrict ourselves to the lowest order in this expansion. Higher order terms in the 
            standard scenario, and the procedure to obtain them systematically, can be found in 
            \cite{Hu:1995en} and \cite{Blas:2011rf}.\\
            
            To obtain the equations of motion to leading order in $\tau_c$, we need to expand 
            $\Delta\theta^I$ and $\sigma^I_\gamma$ to first order in $\tau_c$, just as we did with the 
            TC expansion for $\beta$. Solving simultaneously for both quantities, and plugging in the 
            TC values for the real part of the perturbations and $\beta$, we get
            \begin{align}
                \Delta\theta^I &= \beta k\left(\delta_\gamma^R - \delta_b^R - \delta_{x_e}^R\right)
                    + \tau_c\Bigg\{\frac{k^2}{4}\mathcal{A}\delta_\gamma^I 
                    + \mathcal{A}\mathcal{H}\theta_\gamma^I    \nonumber\\ 
                    &\qquad+\beta k\left[\mathcal{A}(\mathcal{A}-1)\mathcal{H}\left(\delta_\gamma^R-\delta_b^R-\delta_{x_e}^R\right)
                    -\mathcal{A}\mathcal{H}\delta_\gamma^R + \mathcal{A}\theta_\gamma^R - \frac{4}{15}\theta_\gamma^R
                    + \mathcal{A}\dot{\delta}_{x_e}^R+\frac{\mathcal{A}}{6}\dot{h}^R
                    -\frac{2}{15}\left(\dot{h}^R+6\dot{\eta}^R\right)\right]\Bigg\}+\Od{\tau_c^2}\ ,\\
                \sigma_\gamma^I &= \frac{4\beta}{3k}\theta_\gamma^R 
                    + \frac{\tau_c}{15}\left\{4\theta_\gamma^I + \left(\dot{h}^I+6\dot{\eta}^I\right)
                    + \beta\Big[k(\mathcal{A}-5)\delta_\gamma^R + 4k\eta^R
                    + \frac{8\mathcal{A}\mathcal{H}}{k}\theta_\gamma^R\Big]\right\} +\Od{\tau_c^2}\ .
            \end{align}
            The final equations of motion of the photon-baryon plasma during the tightly coupled phase
            are
            \begin{align}
                \dot{\delta}_b^I+\theta_\gamma^I+\frac{1}{2}\dot{h}^I &=
                    \beta\left[\frac{k\mathcal{A}}{4}\delta_\gamma^R 
                    + k\left(\frac{3}{4}\delta_\gamma^R-\delta_b^R-\delta_{x_e}^R\right)
                    +2k\eta^R+\frac{2\mathcal{A}\mathcal{H}}{k}\theta_\gamma^R\right]\ ,\\
                \dot{\delta}_\gamma^I+\frac{4}{3}\theta_\gamma^I+\frac{2}{3}\dot{h}^I &=
                    \beta\left[\frac{k\mathcal{A}}{3}\delta_\gamma^R +\frac{8k}{3}\eta^R
                    +\frac{8\mathcal{A}\mathcal{H}}{3k}\theta_\gamma^R\right]\ ,\\
                \dot{\theta}_\gamma^I + \mathcal{A}\mathcal{H}\theta_\gamma^I
                    + \frac{k^2}{4}(\mathcal{A}-1)\delta_\gamma^I &=
                    \beta k\left[\mathcal{A}(1-\mathcal{A})\mathcal{H}\left(\delta_\gamma^R-\delta_b^R-\delta^R_{x_e}\right)
                    -\mathcal{A}\theta_\gamma^R -\frac{4}{3}\theta_\gamma^R - \mathcal{A}\dot{\delta}^R_{x_e}
                    -2\dot{\eta}^R - \left(1+\frac{\mathcal{A}}{6}\right)\dot{h}^R\right]\ .
            \end{align}        
            The evolution of every mode starts in a radiation-dominated phase during which TCA is valid.
            We start evolving this set of equations for each mode until we switch the TCA off, according
            to the {\sc class} switches \cite{Blas:2011rf}. Once we switch it off we evolve the full
            system, joining the solutions smoothly.
      
        \subsubsection{Radiation streaming approximation}\label{sec:RSA}
	        The evolution equations for ultrarelativistic species \eqref{eq:ScaSys_photons_final}, 
	        like neutrinos, can be combined into a single second-order differential equation
	        \begin{equation}
	            \ddot{\delta}^I_\nu = -\frac{k^2}{3}(\delta_\nu^I - 4\sigma_\nu^I) 
	                + \frac{4\beta_\nu k}{3}\left(\dot{h}^R + 4\dot{\eta}^R\right) - \frac{2}{3}\ddot{h}^I\ ,
	        \end{equation}
	        and the same applies to photons after decoupling. Once the perturbation is sub-Hubble, it
	        starts oscillating very fast. During matter domination, it is possible to ignore the 
	        contribution of ultrarelativistic species to the total density but their contribution to the
	        total velocity is still important. The radiation streaming approximation (RSA) consists on 
	        following only the non-oscillatory particular solution of these equations \cite{Blas:2011rf}.
	        Since in this regime $|\ddot{\delta}_\nu|\ll k^2|\delta_\nu|$ and $|\sigma_\nu|\ll |\delta_\nu|$, 
	        our RSA solution is
	        \begin{align}
	            \delta^I_\nu &= -\frac{2}{k^2}\ddot{h}^I + \frac{4\beta_\nu}{k}\left(\dot{h}^R+4\dot{\eta}^R\right)\ ,\\
	            \theta^I_\nu &= -\frac{1}{2}\dot{h}^I + 2\beta_\nu k\eta^R\ ,\\
	            \sigma^I_\nu &= 0\ .
	        \end{align}
	        We will apply the same equations to photons after decoupling, neglecting the small impact of 
	        reionization.
            
    \subsection{Vector modes}\label{sec:vector_modes}
        In this section, we will describe the evolution of the vector modes, following the same steps
        as in the previous section.
        Since the evolution equations, and the initial conditions, for both helicities are the same, 
        we can rewrite the vorticity for the species $s$ as
        \begin{subequations}\label{eq:single_vrt_redef}
        \begin{align}
            \v{\chi}_s &= \chi_s\left((\hat{\beta}\cdot\hat{e}_+)\,\hat{e}_-
                            + (\hat{\beta}\cdot\hat{e}_-)\,\hat{e}_+\right)\\
                        &= \chi_s\left((\hat{\beta}\cdot\hat{x})\,\hat{x}
                            + (\hat{\beta}\cdot\hat{y})\,\hat{y}\right)\ .
        \end{align}
        \end{subequations}
        Then, we do not need to distinguish between helicities and we can just write one equation for 
        $\chi_s$. The same applies to the vector part of the shear tensor $\pi^\text{V}_s$. Starting
        from \eqref{eq:vector_modes_vg} and \eqref{eq:vector_modes_piVg}, under our approximation
        scheme, i.e. neglecting tensor modes and moments higher than $\ell=2$, the evolution of the
        photon vector modes is described by
        \begin{subequations}\label{eq:vector_photons}
        \begin{align}
            \dot{\chi}_\gamma +\frac{1}{2}\beta\left(\dot{h}-2\dot{\eta}\right) -\frac{3k^2}{4}\pi^\text{V}_\gamma
                &= -\frac{1}{\tau_c}\left[\Delta \chi
                +\left(\beta\frac{\delta n_e}{n_e}-\beta_e\left(\delta_\gamma 
                +\frac{1}{2}\sigma_\gamma\right)\right)\right]\ ,
                \label{eq:chi_gamma}\\
            \dot{\pi}^\text{V}_\gamma + \frac{4}{15}\chi_\gamma + \frac{4}{15}\left(S+\dot{F}\right)
                +\frac{4}{15}\beta\eta &= -\frac{1}{\tau_c}\left[\pi^\text{V}_\gamma 
                + \frac{4}{15k^2}\left(\beta_e\theta_\gamma + 4\beta_e\theta_b + \Delta\beta\theta_b\right)\right]\ ,
                \label{eq:piVg}
        \end{align}
        \end{subequations}
        where we have defined
        \begin{equation}
            \Delta\chi \equiv \chi_\gamma - \chi_b\ .
        \end{equation}
        Again, the behaviour of neutrinos can be obtained from these equations, setting to zero the 
        collision term. From \eqref{eq:BoEq_vb} and \eqref{eq:BoEq_vc}, baryons and dark matter evolve 
        according to
        \begin{align}
            \dot{\chi}_b+\mathcal{H}\chi_b +\beta_e\theta_b 
                + \frac{1}{2}\beta_e\left(\dot{h}-2\dot{\eta}\right) 
                &= \frac{1}{\tau_cR}\left[\Delta\chi + \left(\beta\frac{\delta n_e}{n_e}
                -\beta_e\left(\delta_\gamma + \frac{1}{2}\sigma_\gamma\right)\right)\right]\ ,\label{eq:chi_b}\\
            \dot{\chi}_c+\mathcal{H}\chi_c +\frac{1}{2}\beta_c\left(\dot{h}-2\dot{\eta}\right) &= 0\ \label{eq:chi_c}.
        \end{align}
        The evolution of the total vorticity, i.e. the vorticity of the full fluid, is
        \begin{equation}
            \dot{\chi} +\mathcal{H}(1-3w)\chi + \frac{\dot{w}}{1+w}\chi -\frac{k^2}{1+w}\pi^\text{V} = 0\ ,
            \label{eq:chi_total}
        \end{equation}
        where
        \begin{equation}
            \pi^\text{V} = \frac{1}{\rho}\left(\rho_\nu\pi^\text{V}_\nu+\rho_\gamma\pi^\text{V}_\gamma
                -\frac{1}{k^2}\beta_e\theta_b\right)\ .
        \end{equation}
        Finally, the relevant Einstein equation is
        \begin{equation}
            S+\dot{F} = \frac{16\pi Ga^2}{k^2}(\rho+P)\chi\ .
            \label{eq:vector_einstein}
        \end{equation}
        Our strategy for the integration of the vector modes can be summarized as follows.
        \begin{enumerate}[I)]
            \item The background and the scalar modes evolve according to standard $\Lambda$CDM and act
                as external sources for the vector modes. We use {\sc class} to precompute the sources.
            \item The system to be integrated consists of \eqref{eq:vector_photons}, \eqref{eq:chi_b},
                \eqref{eq:chi_total}, \eqref{eq:vector_einstein} and a neutrino contribution equal
                to \eqref{eq:vector_photons} but without collisions.
            \item The initial conditions are described in the appendix \ref{sec:initial_conditions}.
            \item As in the scalar case, we have to take into account perturbed recombination and
                we need to implement the TC expansion and RSA in the vector case, as discussed in the
                next sections \ref{sec:TCA_vector} and \ref{sec:RSA_vector}.
        \end{enumerate}
        
        \subsubsection{Tight coupling expansion}\label{sec:TCA_vector}
            As in the previous section, we need to find the approximate equations to follow the tightly
            coupled phase. Performing the same manipulations, and inserting the TC solutions for $\beta$ 
            and the scalar modes, we get
            \begin{align}
                \Delta\chi &= \beta\left(\delta_\gamma - \delta_b -\delta_{x_e}\right)
                    +\tau_c\Bigg\{\mathcal{A}\mathcal{H}\chi_\gamma 
                    +\beta\bigg[\mathcal{A}(\mathcal{A}-1)\mathcal{H}\left(\delta_\gamma-\delta_b-\delta_{x_e}\right)
                    -\mathcal{A}\mathcal{H}\delta_\gamma+\frac{\mathcal{A}}{3}\theta_\gamma + \frac{2}{15}\theta_\gamma
                    +\mathcal{A}\dot{\delta}_{x_e}\nonumber\\
                &\qquad\qquad\qquad\qquad\qquad\qquad\qquad\qquad\quad
                    + \frac{\mathcal{A}}{6}\dot{h} + \frac{1}{15}\left(\dot{h}+6\dot{\eta}\right)\bigg]\Bigg\}
                    +\Od{\tau_c^2}\ ,
                    \label{eq:diff_v_TC}\\
                \pi^\text{V}_\gamma &= -\frac{4\beta}{3k^2}\theta_\gamma -\frac{4}{15} \tau_c\left\{
                    \chi_\gamma + (S+\dot{F}) - \beta\left[\frac{5}{4}\delta_\gamma - \eta
                    -\frac{\mathcal{A}}{4}\delta_\gamma-\frac{2\mathcal{A}\mathcal{H}}{k^2}\theta_\gamma\right]\right\}\ .
            \end{align}
            The equation governing the evolution of the photon vorticity during TC is
            \begin{equation}\label{eq:chi_gamma_TC}
                \dot{\chi}_\gamma = -\mathcal{A}\mathcal{H}\chi_\gamma - \beta\theta_\gamma - \frac{1}{2}\beta
                    \left(\dot{h}-2\dot{\eta}\right) + \beta \mathcal{A}\left[(1-\mathcal{A})\mathcal{H}(\delta_\gamma
                    -\delta_b-\delta_{x_e})-\frac{1}{3}\theta_\gamma- \frac{1}{6}\dot{h}-\dot{\delta}_{x_e}\right]\ .
            \end{equation}
            
        \subsubsection{Radiation streaming approximation}\label{sec:RSA_vector}
            The equations \eqref{eq:chi_gamma} and \eqref{eq:piVg} for neutrinos, or photons
            after decoupling, can be combined into a second order differential equation
            \begin{equation}
                \ddot{\chi}_\nu = -\frac{k^2}{5}\chi_\nu - \frac{k^2}{5}(S+\dot{F})-\frac{k^2}{5}\beta_\nu\eta
                    -\frac{1}{2}\beta_\nu\left(\ddot{h}-2\ddot{\eta}\right)\ .
            \end{equation}
            During the period of rapid oscillations, we are in the situation in which $|\ddot{\chi}_\nu|\ll |k^2\chi_\nu|$
            and $|\dot{\chi}_\nu|\ll |k^2\pi^\text{V}_\nu|$. From the previous equation and \eqref{eq:chi_gamma},
            the approximate non-oscillating particular solution is found to be
            \begin{align}
                \chi_\nu &= -(S+\dot{F})-\beta_\nu\eta + \frac{5}{2k^2}\beta_\nu\left(\ddot{h}-2\ddot{\eta}\right)\ ,\\
                \pi_\nu^\text{V} &= \frac{2}{3k^2}\beta_\nu\left(\dot{h}-2\dot{\eta}\right)\ .
            \end{align}
        
        \subsubsection{Semi-analytic solutions}            
            The equations obtained admit semi-analytic solutions in some regimes. During the TC regime,
            from \eqref{eq:diff_v_TC} and \eqref{eq:chi_gamma_TC},  we find, to lowest order in $\tau_c$,
            \begin{align}
                (1+R)\chi_\gamma &= -\frac{1}{2}\beta_0(h-2\eta) + 
                    \beta_0 \mathcal{A}\left(\delta_\gamma -\delta_b -\delta_{x_e}\right)
                    -\frac{\beta_0}{3}\int \left(1+\frac{\mathcal{A}}{4}\right)\theta_\gamma\di\tau +\mathcal{C}_\gamma\ ,\\
                (1+R)\chi_b &= -\frac{1}{2}\beta_0(h-2\eta)
                    -\frac{\beta_0}{1+R}\left(\delta_\gamma -\delta_b -\delta_{x_e}\right)
                    -\frac{\beta_0}{3}\int \left(1+\frac{\mathcal{A}}{4}\right)\theta_\gamma\di\tau +\mathcal{C}_\gamma\ ,
            \end{align}
            where $\mathcal{C}_\gamma$ is a constant of integration, to be set with the initial
            condition, and $\beta_0$ is the initial velocity of the photon-baryon plasma.
            Another result that can be obtained, integrating \eqref{eq:chi_c}, is the evolution
            of CDM
            \begin{align}
                a\chi_c &= -\frac{1}{2}\beta_c^\text{today}\left(h-2\eta\right)+\mathcal{C}_c\ .
            \end{align}
            As discussed before, we do not specify the behaviour of the dark sector at early times. Hence,
            we do not use this equation. We solve the system instead using the total vorticity
            \eqref{eq:chi_total} and then we obtain the vorticity of the dark sector subtracting the 
            other components.

\section{Results and discussion}\label{sec:ResultsSpectra}
    \subsection{Observables}
        Once we have constructed a consistent system of equations, we must discuss which of the 
        intermediate variables correspond to physical observables. One of the main observables in
        cosmology is the distribution of temperature anisotropies in the CMB. The CMB is very nearly
        isotropic and described, at the background level, by an equilibrium Bose-Einstein distribution
        \begin{equation}\label{eq:f0_BoseEinstein}
            f_0 = \frac{1}{\ee^{\,p/T}-1}\ .
        \end{equation}
        Deviations from this background distribution are usually parameterized as temperature
        perturbations
        \begin{equation}
            f(\tau,\v{x},p,\hat{n}) = \left[\exp\left(\frac{p}{T\left[1
                +\Theta(\tau,\v{x},\hat{n})\right]}\right)-1\right]^{-1}\ .
        \end{equation}
        The temperature perturbation can then be written in terms of the distribution function as
        \begin{equation}
            (1+\Theta)^4-1 = \frac{\displaystyle\int p^3\di p\left(f-f_0\right)}{\displaystyle\int p^3\di p\, f_0}
            \equiv \Delta\ .
        \end{equation}
        With our definition for the distribution function \eqref{eq:f0_main_condition2}, the deviations
        from \eqref{eq:f0_BoseEinstein} are
        \begin{equation}
            \Delta = \frac{1}{\gamma^4(1-\v{\beta}\cdot\hat{n})^4}-1 + \mathcal{F}_\gamma\ .
        \end{equation}
        To first order in cosmological perturbations we have
        \begin{align}\label{eq:temperature perturbation}
            \Theta &= \frac{1}{\gamma(1-\v{\beta}\cdot\hat{n})}-1 
                    + \frac{1}{4}\gamma^3(1-\v{\beta}\cdot\hat{n})^3\mathcal{F}_\gamma\ ,\nonumber\\
                   &= \v{\beta}\cdot\hat{n}
                    + \frac{1}{4}\Big(1-3(\v{\beta}\cdot\hat{n}) + \Od{\beta^2}\Big)\mathcal{F}_\gamma
                    + \Od{\beta^2}\ .
        \end{align}
        This would be the temperature perturbation observed in the $\mathcal{O}$ frame. From the Sun's 
        reference system, the observed temperature perturbation $\Theta_\odot$ is
        \begin{equation}
            1+\Theta_\odot = \gamma_\odot \left(1-\hat{n}\cdot\v{\beta}_\odot\right)(1+\Theta)\ ,
        \end{equation}
        where $\v{\beta}_\odot$ is the velocity of the Solar System in the $\mathcal{O}$ frame.
        Expanding to leading order in $\beta$ and $\beta_\odot$ we get
        \begin{equation}
            \Theta_\odot = \left(\v{\beta}-\v{\beta}_\odot\right)\cdot \hat{n} 
                + \frac{1}{4}\left(1+\hat{n}\cdot(\v{\beta}-\v{\beta}_\odot) 
                - 4(\hat{n}\cdot\v{\beta})\right)\mathcal{F}_\gamma + \Od{\beta^2}\ .
        \end{equation}
        The reduced distribution function can be decomposed schematically as
        \begin{equation}
            \mathcal{F}_\gamma (\hat{n},\v{\beta}) = \mathcal{F}_\gamma^{\Lambda\text{CDM}}(\hat{n})
                +(\hat{n}\cdot\v{\beta})\mathcal{F}_\gamma^\beta(\hat{n}) + \Od{\beta^2}\ ,
        \end{equation}
        where $\mathcal{F}_\gamma^{\Lambda\text{CDM}}$ follows the standard evolution and 
        $\mathcal{F}^\beta_\gamma$ contains our modification, i.e. the imaginary part of the scalar modes
        and the vector modes. Finally, we need to take into account the aberration effects. The direction 
        $\hat{n}_\odot$ observed from the Solar System is related to the direction $\hat{n}$ in the
        $\mathcal{O}$ frame as
        \begin{equation}
            n^i_\odot = \frac{\mathcal{P}^i_{\odot\,j}n^j-\gamma_\odot\beta^i_\odot}{\gamma_\odot
                (1-\hat{n}\cdot\v{\beta}_\odot)}\ .
        \end{equation}
        Then, to first order in $\beta$, we can express the direction as
        \begin{equation}
            n^i = n^i_\odot - \left(\delta^i_j-n^i_\odot n_{\odot\,j}\right)(\beta^{\odot\, j}_\text{CMB}-\beta^j) +\Od{\beta^2}\ ,
        \end{equation}
        where we have defined the relative velocity between the Sun and the CMB rest frame
        \begin{equation}
            \v{\beta}^\odot_\text{CMB} \equiv \v{\beta}-\v{\beta}_\odot\ .
        \end{equation}
        It is customary \cite{Aghanim:2013suk, Pant:2018smd} to express the deflection instead as 
        \begin{equation}
            \v{\beta}-(\hat{n}\cdot\v{\beta})\,\hat{n} = \v{\nabla}(\hat{n}\cdot\v{\beta})\ .
        \end{equation}
        Taking everything into account, the temperature perturbation that would be measured from Earth
        in the direction $\hat{n}_\odot$ is
        \begin{align}
            \Theta_\odot(\hat{n}_\odot) = \hat{n}_\odot\cdot\v{\beta}^\odot_\text{CMB}
                &+ \frac{1}{4}\left(1+\hat{n}_\odot\cdot\v{\beta}^\odot_\text{CMB}-4(\hat{n}_\odot\cdot\v{\beta})\right)
                \mathcal{F}_\gamma^{\Lambda\text{CDM}}\left(\hat{n}_\odot 
                    -\v{\nabla}(\hat{n}_\odot\cdot\v{\beta}^\odot_\text{CMB})
                    +\v{\nabla}(\hat{n}_\odot\cdot\v{\beta})\right)\nonumber\\
            & + \frac{1}{4}(\hat{n}_\odot\cdot\v{\beta})\mathcal{F}_\gamma^\beta(\hat{n}_\odot) + \Od{\beta^2}\ .
        \end{align}
        The first term represents the usual kinematic dipole, i.e. the Doppler-shifting effect associated 
        with the relative motion of the observer with respect to the CMB. The second term contains 
        a dipolar modulation and aberration effects. Both effects produce a kinematic mixing of the 
        multipole coefficients. The third term is a purely dynamical contribution, i.e. the effect
        of a relative motion between different species during the evolution. While recovering standard
        results \cite{Aghanim:2013suk} for $\v{\beta}=0$, in our setting we observe two kinds of new effects.
        First, the directions of the dipole, the dipolar modulation and the aberration effects do not
        coincide. This effect comes from the fact that, in our scenario, the standard $\Lambda$CDM evolution
        is recovered in the $\mathcal{O}$ frame and \emph{not} in the CMB rest frame. In standard 
        cosmology both frames coincide and this difference does not arise. The second effect is an additional
        source of statistical anisotropy, coming from the modified evolution, with a dipolar pattern.\\
        
        The CMB dipole is very well measured, with the latest \emph{Planck} value being
        $\beta^\odot_\text{CMB} = (1.23357\pm 0.00036)\times 10^{-3}$ \cite{Akrami:2018vks}. It is widely 
        accepted that its origin is mostly kinematical, 
        so it gives us a very precise measurement of our relative motion with respect to the CMB. The
        \emph{Planck} Collaboration also measured our relative motion using the kinematic
        correlations induced between different multipoles and the resultant anisotropic signal 
        \cite{Aghanim:2013suk, Ade:2015hxq}.
        Even though the uncertainties are large in this case, and there seems to be some tension 
        \cite{Ade:2015hxq}, the velocity inferred using this method is compatible with the dipole, supporting its
        kinematical origin.  The relative velocity with respect to the CMB frame is usually interpreted
        as the result of peculiar motions of the Sun and the Local Group \cite{Tully:2007ue}. However, in our
        scenario, the relative velocity would arise as a combination of the local motion, with respect
        to the matter frame, and the relative motion between the matter and CMB frames. This gives rise
        to distinctive phenomenological consequences.\\        
        
        At the background level, the non-coincidence of the CMB and matter frames produces a global
        motion of large-scale structures with respect to the CMB. This effect could be potentially 
        observed as a bulk flow on the largest scales. Measurements of bulk flows, at different scales,
        have been carried out in peculiar velocity surveys  and using the kinetic
        Sunyaev-Zeldovich (kSZ) effect \cite{Kashlinsky:2008ut, Ade:2013opi}. See, e.g., Table 5 
        of \cite{Scrimgeour:2015khj} and references therein for a collection of recent measurements. 
        Although there is a long history 
        of conflicting measurements and anomalously large flows on cosmological scales, in this work
        we adopt the reported limit of \emph{Planck} \cite{Ade:2013opi} for two reasons. In the first place, it extends to 
        the largest scales, up to 2 Gpc, where a cleaner determination of our global flow is expected.
        In the second place, it sets the more conservative bound in our parameter $\beta_0$, in the 
        sense of being the more restrictive to us. From the reported \emph{Planck} value 
        $v<254\ \text{km/s}\,(95\%\ \text{CL})$, and according to the
        time evolution of $\beta$ in Figure \ref{fig:beta_evol}, we obtain
        \begin{equation}\label{eq:beta0_constraint}
            \beta_0 < 1.6\times10^{-3}\,(95\%\ \text{CL})\ .
        \end{equation}
        Note that, since we are performing a first-order computation, all our results scale trivially 
        and we will write them explicitly in units of $\beta_0$. The previous constraint is wholly 
        compatible with the local measurements of peculiar motions mentioned above. The peculiar velocity 
        of the Local Group, and other higher order structures, with respect to the CMB is inferred from 
        the movement of the Sun with respect to both of them. The constraint \eqref{eq:beta0_constraint} 
        yields a value $\beta < 0.85\times 10^{-3}$ today, of the same order as the measured velocity 
        $\beta^\odot_\text{LG}=(1.00\pm 0.05)\times 10^{-3}$ \cite{Akrami:2018vks}, so it can be 
        accomodated without fine-tuning the directions of these relative velocities. Potentially,
        it could even constitute a component of unaccounted peculiar motions of the largest 
        structures \cite{Tully:2007ue}. This constraint also justifies our first order computation.
        In section \ref{sec:MovingFluids} it was discussed how to construct a RW background to $\Od{\beta}$
        and how the $\Od{\beta^2}$ terms introduce anisotropies, i.e. a Bianchi background. Using 
        \eqref{eq:beta0_constraint} we can see that the terms $\Od{\beta^2}$ are in fact smaller than 
        than a typical cosmological perturbation. Therefore, it is completely justified to take the 
        RW metric \eqref{eq:RW_metric2} as the background geometry.\\
        
        At the perturbation level, it can be proven that, to first order in $\beta$, our modification
        does not leave an imprint in the CMB temperature spectrum, i.e. $C_\ell$'s. To lowest order, 
        the first CMB signatures appear as deviations from statistical isotropy. It is very important
        to notice that our model produces a distinctive signature in the CMB. In the standard picture,
        as mentioned before, the motion of Earth produces a violation of statistical isotropy in the
        CMB. In our case, there would be an additional, purely dynamical, source of statistical 
        anisotropy, caused by the relative motion between matter and radiation during the evolution. 
        Both effects could in principle be disentangled. We leave this analysis for future work 
        \cite{CMB_paper}, focusing instead on LSS observables.\\
        
        The local motion of the Earth also leaves an imprint in the observed galaxy distribution
        \cite{ellis1984expected, Gibelyou:2012ri, Pant:2018smd}, even though the analysis is not straightforward in this 
        case. Upcoming galaxy surveys like Euclid \cite{Amendola:2016saw} or SKA \cite{Maartens:2015mra} 
        will measure the induced dipole with high precision. A significant difference between this dipole 
        and the CMB result would be difficult to accomodate in standard $\Lambda$CDM, but could be easily
        interpreted as the result of a relative velocity between the CMB and matter frames. Even 
        if no such difference is measured, the bulk motion can still be smaller than the local one, 
        and yet lead to observational signatures, as we will immediately see.\\
                 
        Until now, we have mainly discussed the effects on the CMB temperature perturbations.
        Another class of observables comes from the clustering of matter. The distribution and redshift 
        of galaxies give us information about density perturbations and peculiar velocities. Again, in 
        the standard matter power spectrum we do not have a first order effect. On general grounds, if 
        we consider a cosmological quantity $g$ splitted as in \eqref{eq:g_transfer_splitted}, we have
        \begin{equation}
            |g(\tau,\v{k})|^2 = |g^R(\tau,\v{k})|^2+\Od{\beta^2}\ ,
        \end{equation}
        i.e. the standard result. However, in the cross-correlation between two cosmological perturbations
        $g_1$ and $g_2$ we get a first-order dipolar contribution
        \begin{equation}
            g_1g_2^* = g_1^Rg_2^{R\,*} + \ii (\hat{\beta}\cdot\hat{k})
                \left(g_1^Ig_2^{R\, *}-g_1^Rg_2^{I\,*}\right) + \Od{\beta^2}\ .
        \end{equation}
        Every cross-correlation between cosmological quantities contains a dipole modification with
        this structure. This effect could be observed in the future in the cross-correlations
        between matter density and velocity \cite{Tegmark:2003uf}, as the precision of the surveys increases. It is
        conceivable that this effect could appear in cross-correlations between baryon and CDM 
        densities as well, even though a thorough analysis using lensing information would be in
        order. Finally, 
        the generation of vorticity, purely decaying in $\Lambda$CDM, is another distinctive feature of
        our model.\\
        
        Since the most accessible observables are related to velocity perturbations, we conclude this
        section clarifying a few points concerning our previous definitions. In particular, it is 
        important to relate the intermediate variables we have used with the physical velocities. The 
        velocity that would appear in the energy-momentum tensor of a fluid \eqref{eq:Tmunu_fluid_def} 
        is the velocity of a frame in which the momentum flux, i.e. the component $T_{0i}$, is zero 
        \cite{LandauFluidMechanics}. Using the boost-transformation properties \eqref{eq:boost_momentum} 
        we can obtain an equation for the physical velocity $U^i$ of the fluid
        \begin{equation}
            \bar{\gamma}\bar{\mathcal{P}}^i_j\left(Q^j-U_k\Pi^{kj}\right)
                    -\bar{\gamma}^2U^i\left(\rho+P-Q^jU_j\right) = 0\ ,
        \end{equation}
        where
        \begin{equation}
            \bar{\gamma}\equiv (1-U^2)^{-1/2}\ ,\qquad 
            \bar{\mathcal{P}}^i_j\equiv \delta^i_j + (\bar{\gamma}-1)\frac{U^iU_j}{U^2}\ .
        \end{equation}
        Working to first order in $\beta$ and in cosmological perturbations we have
        \begin{equation}
            U^i = \beta^i + \frac{1}{\tilde{\rho}+\tilde{P}}\left(\delta Q^i - \beta^k\delta\Pi^i_k
                -\beta^i(\delta\rho + \delta P)\right)\ .
        \end{equation}
        The physical velocity has two parts, a bulk velocity $\beta_i$ plus a peculiar contribution
        $\delta u^i$. For ultrarelativistic particles, the peculiar velocity can be expressed in terms
        of our previously defined variables \eqref{eq:fluidphotons_def} as
        \begin{equation}
            \delta u^i = \delta v^i -\frac{3}{4}\beta^k\pi_k^i-\beta^i\delta\ .
        \end{equation}
        It can be splitted into a scalar and a vector part
        \begin{equation}
            \delta u^i = -\frac{\ii\hat{k}^i}{k}\vartheta + \zeta^i\ ,
        \end{equation}
        so that we have
        \begin{subequations}\label{eq:def_phys_velocities}
        \begin{align}
            \vartheta &= \theta - \ii(\v{\beta}\cdot\v{k})(\delta - \sigma)\ ,\\
            \zeta^{\pm} &= \chi^{\pm} - (\v{\beta}\cdot\hat{e}_\pm)\left(\delta+\frac{1}{2}\sigma\right)\ .
        \end{align}
        \end{subequations}
        For non-relativistic species, the results are identical setting $\sigma=0$. It is worth noting
        that this is not the only physically sensible definition of the velocity of a fluid. It can
        alternatively be defined as the velocity of the frame in which the flux of particles
        \eqref{eq:boost_flux} is zero \cite{Weinberg:1972kfs}. Both definitions agree for non-relativistic
        fluids if the number of particles is conserved.
        
    \subsection{Time evolution and transfer functions}
\begin{figure}[t]
	        \includegraphics[scale=0.5]{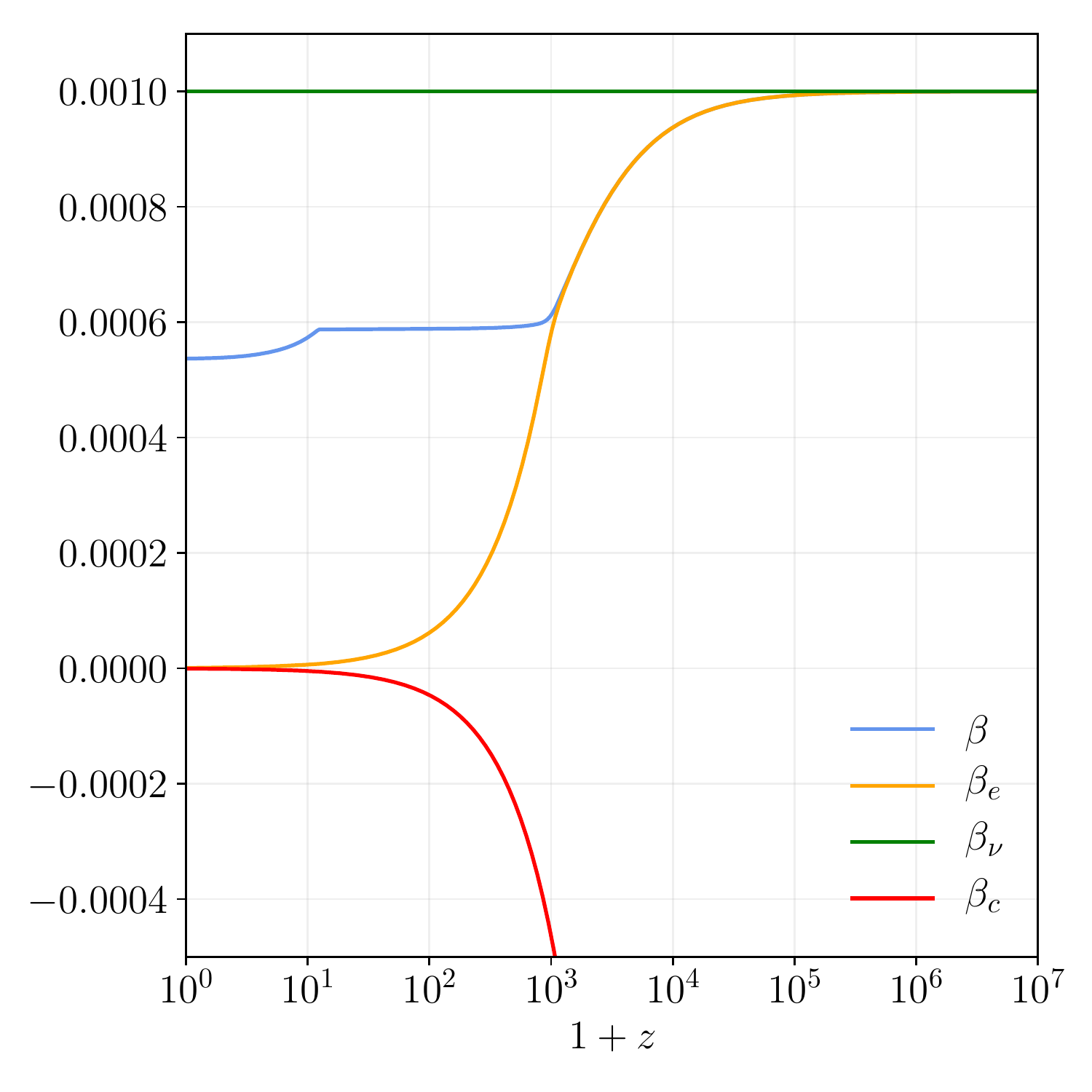}
	        \caption{Time evolution of the bulk velocities for the different components. The dark matter
	            velocity is obtained enforcing the cosmic center of mass condition 
	            \eqref{eq:Einstein_CMframe}. Massless neutrinos behave as an uncoupled ultrarelativistic 
	            species throughout the evolution and maintain a constant velocity. The photon-baryon
	            plasma behaves as a single fluid, either matter- or radiation-like, until decoupling
	            $z_\text{dec}\simeq 1090$. Around $z\simeq 11$ there is a small effect in the photon
	            velocity due to reionization.}
	        \label{fig:beta_evol}
	    \end{figure}

        The time evolution of the bulk velocities for the different components is represented in Figure
        \ref{fig:beta_evol}. All the components in the visible sector start their evolution with the 
        same velocity in the center of mass frame, and its momentum is counterbalanced by the dark sector. 
        The velocity of the neutrinos, since we are neglecting their masses, is always constant. 
        The velocity of the photon-baryon plasma is constant deep in the radiation-dominated era. Once 
        the baryonic contribution to the energy density becomes important, the plasma velocity drops 
        down as $a^{-1}$, see \eqref{eq:beta_evol_TC},
        until decoupling. After decoupling, the velocity of the baryons keeps scaling as $a^{-1}$, like
        CDM, while the photons keep a constant velocity, with a slight late-time effect from reionization.
        Today, the cosmic center of mass, i.e. the $\mathcal{O}$ frame, almost coincides with the matter 
        frame but photons and neutrinos possess a sizeable velocity.\\
        
\begin{figure}[htb]
			\subfigure{
			    \includegraphics[scale=0.5]{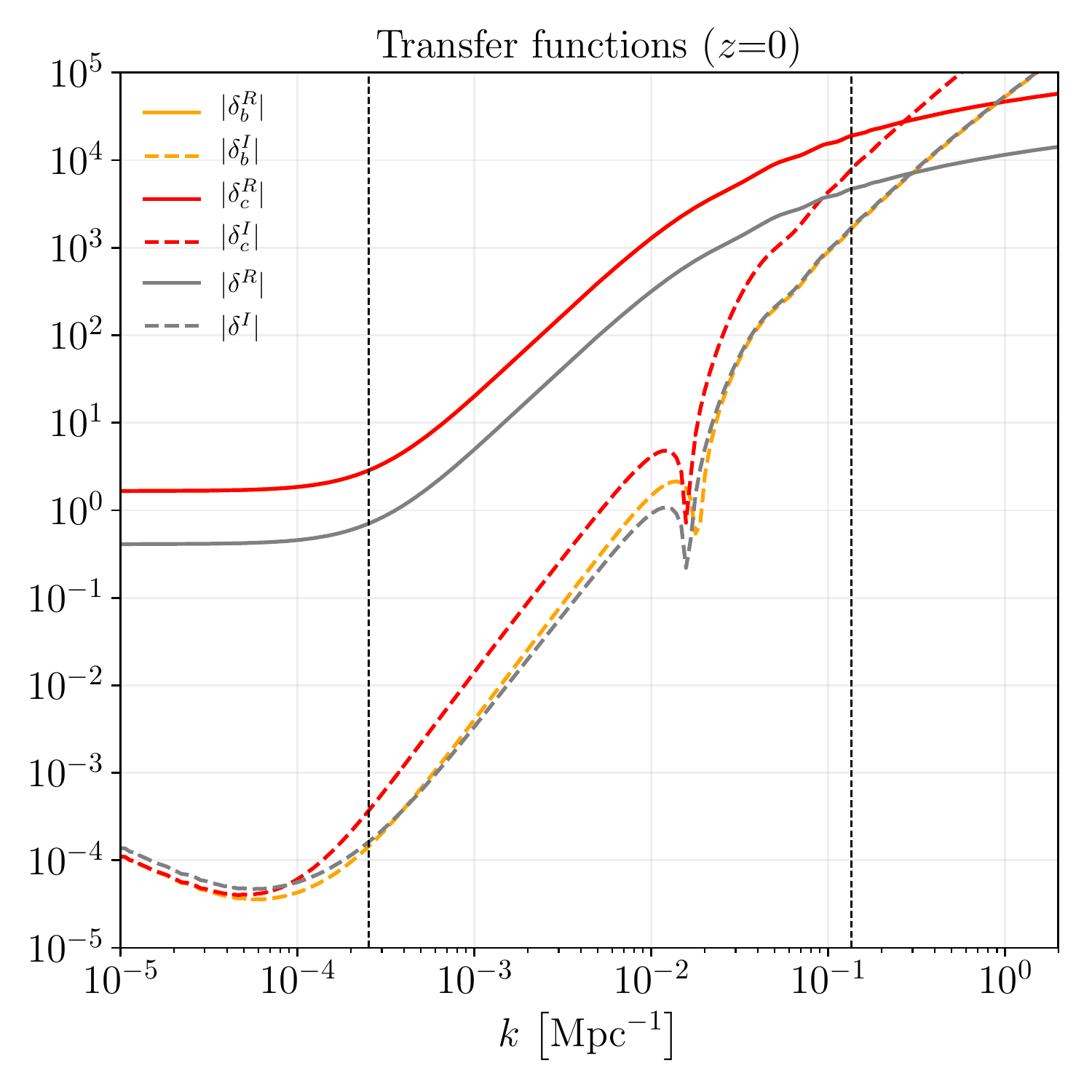}
			}
			\subfigure{
			    \includegraphics[scale=0.5]{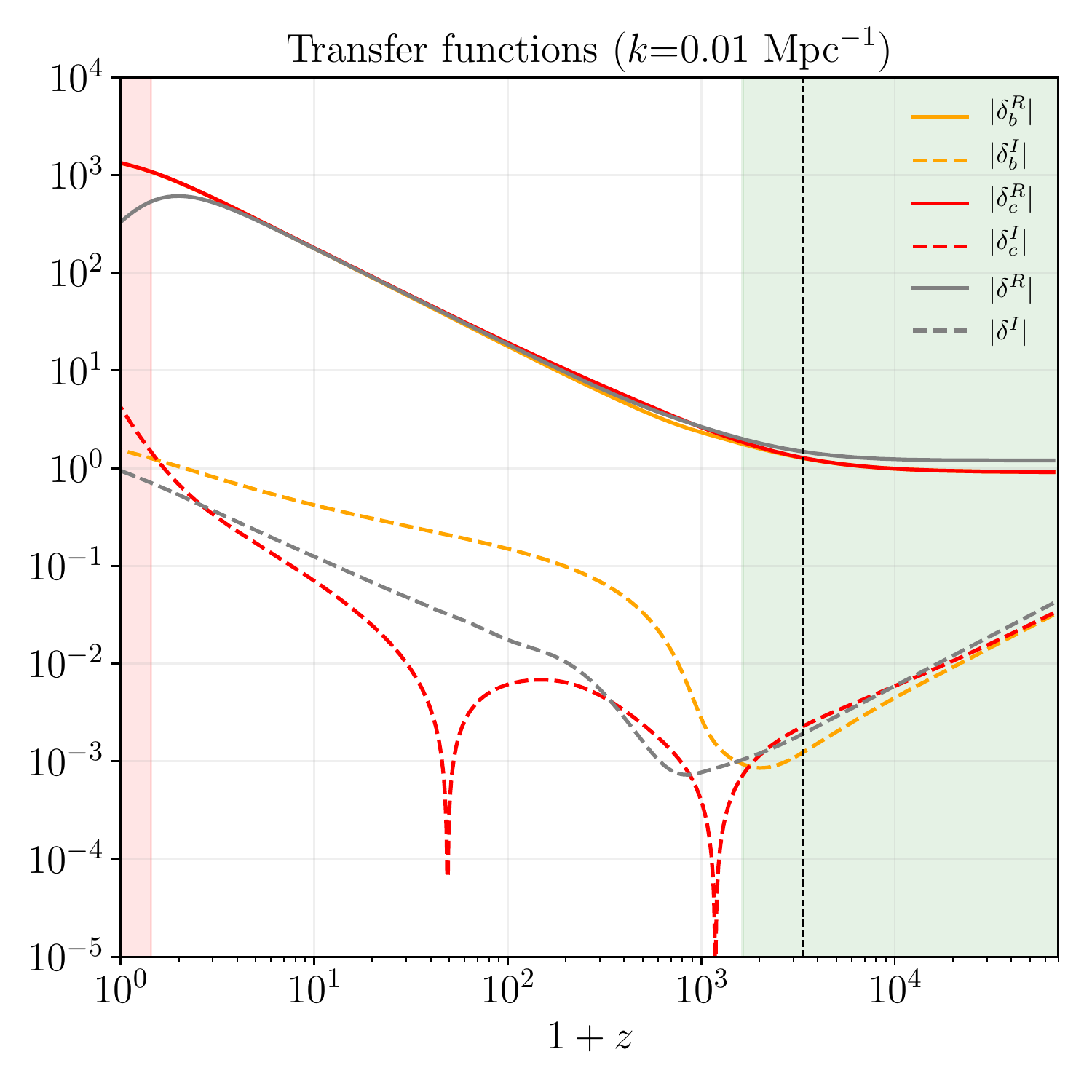}
			}
			\caption{Transfer functions, with initial curvature perturbation normalized to one, 
			    in the Newtonian gauge. Both panels represent the evolution of the density contrast, 
			    both the standard and our modification. The imaginary parts are proportional to $\beta_0$.
			    We show the results for $\beta_0=10^{-3}$.
			    (Left) The vertical line indicates the super-Hubble and non-linearity scales, respectively.
			    (Right) The vertical line marks the horizon crossing. In the red and green shaded regions
			    the RSA and TCA, respectively, are switched on.}
		    \label{fig:density_transfer}
        \end{figure}

\begin{figure}[htb]
			\subfigure{
			    \includegraphics[scale=0.5]{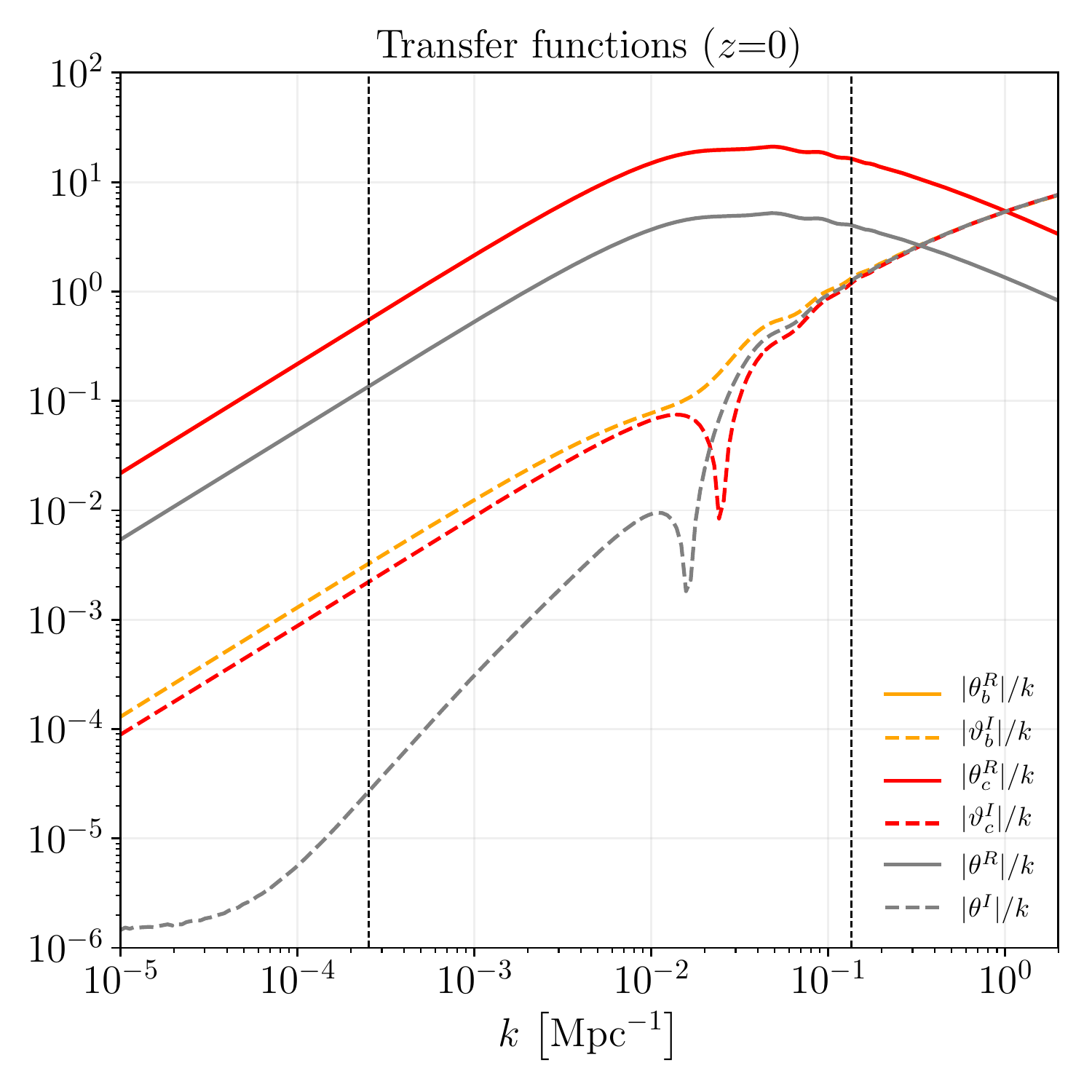}
			}
			\subfigure{
			    \includegraphics[scale=0.5]{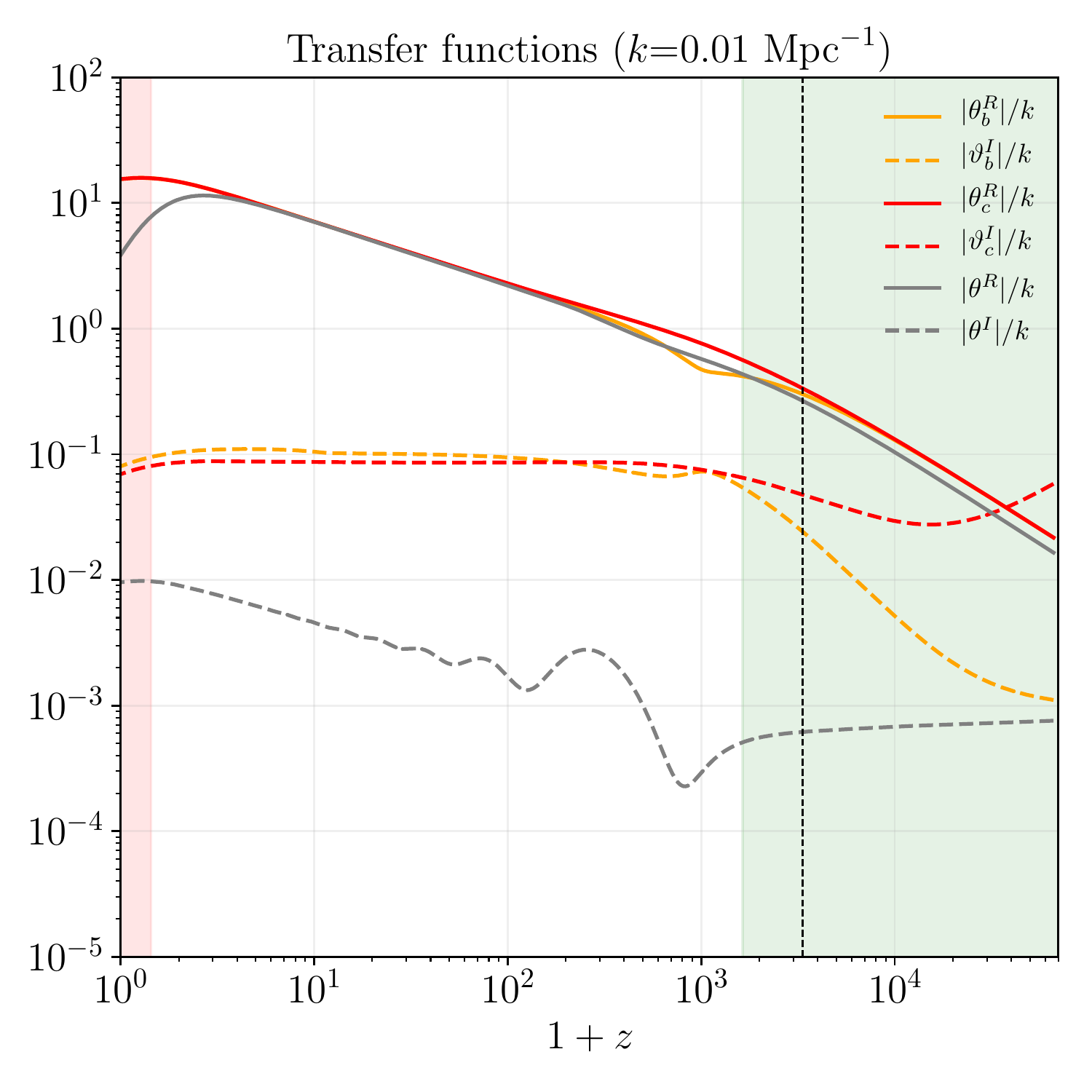}
			}
			\caption{Transfer functions, with initial curvature perturbation normalized to one, 
			    in the Newtonian gauge. Both panels represent the
			    evolution of the velocity divergence, where the imaginary parts have been redefined 
			    according to \eqref{eq:def_phys_velocities}. The imaginary parts are proportional to $\beta_0$.
			    We show the results for $\beta_0=10^{-3}$. (Left) The vertical line indicates the super-Hubble and non-linearity scales, respectively.
			    (Right) The vertical line marks the horizon crossing. In the red and green shaded regions
			    the RSA and TCA, respectively, are switched on.}
		    \label{fig:velocity_transfer}
        \end{figure}

\begin{figure}[htb]
			\subfigure{
			    \includegraphics[scale=0.5]{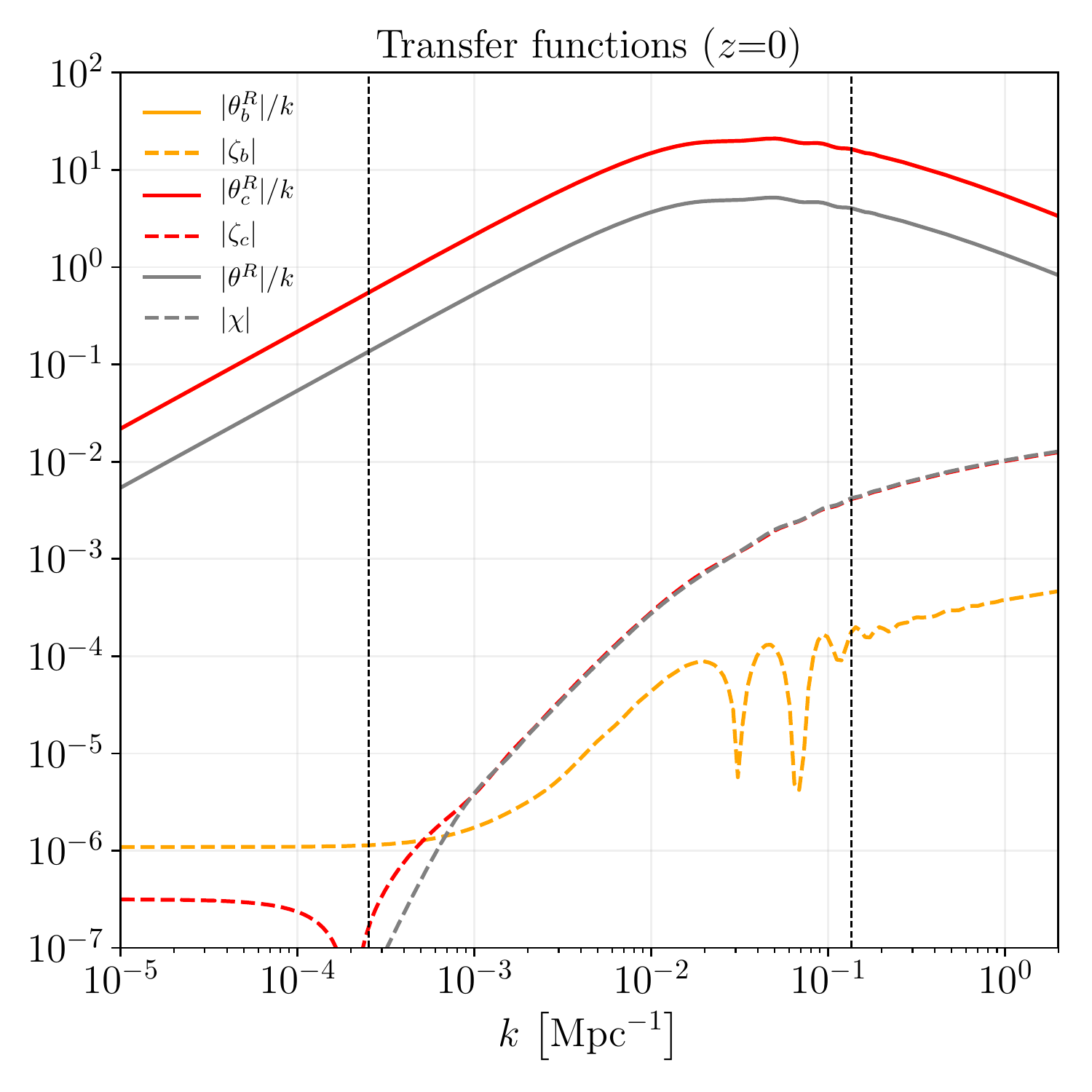}
			}
			\subfigure{
			    \includegraphics[scale=0.5]{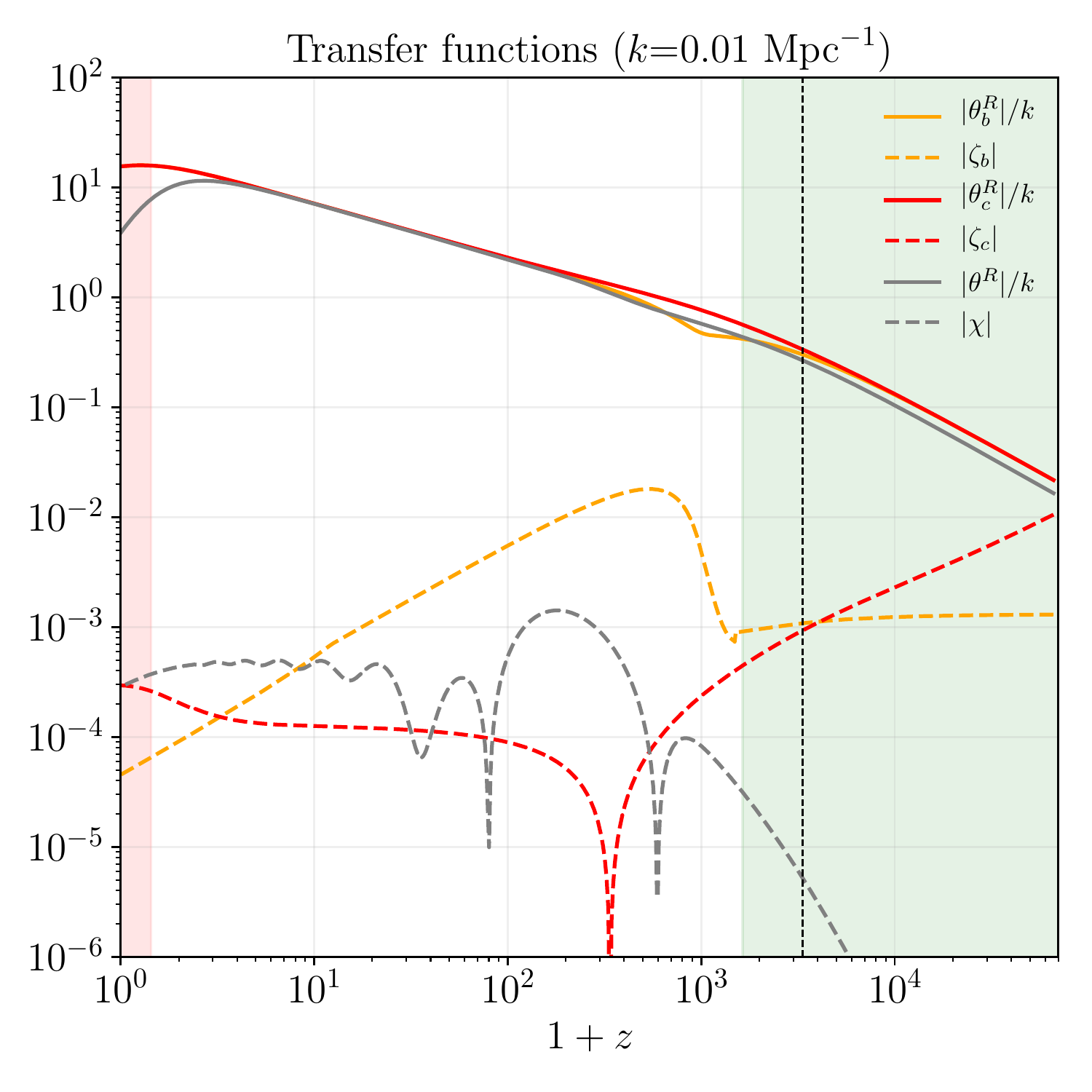}
			}
			\caption{Transfer functions, with initial curvature perturbation normalized to one, 
			    in the Newtonian gauge. Both panels represent the
			    evolution of the vorticity, compared with the velocity divergence. Again, the vorticity has
			    been redefined according to \eqref{eq:def_phys_velocities}. The vorticity is proportional to $\beta_0$.
			    We show the results for $\beta_0=10^{-3}$. (Left) The vertical line 
			    indicates the super-Hubble and non-linearity scales, respectively. (Right) The vertical 
			    line marks the horizon crossing. In the red and green shaded regions the RSA and TCA, 
			    respectively, are switched on.}
		    \label{fig:vorticity_transfer}
        \end{figure}

\begin{figure}[htb]
			\subfigure{
			    \includegraphics[scale=0.5]{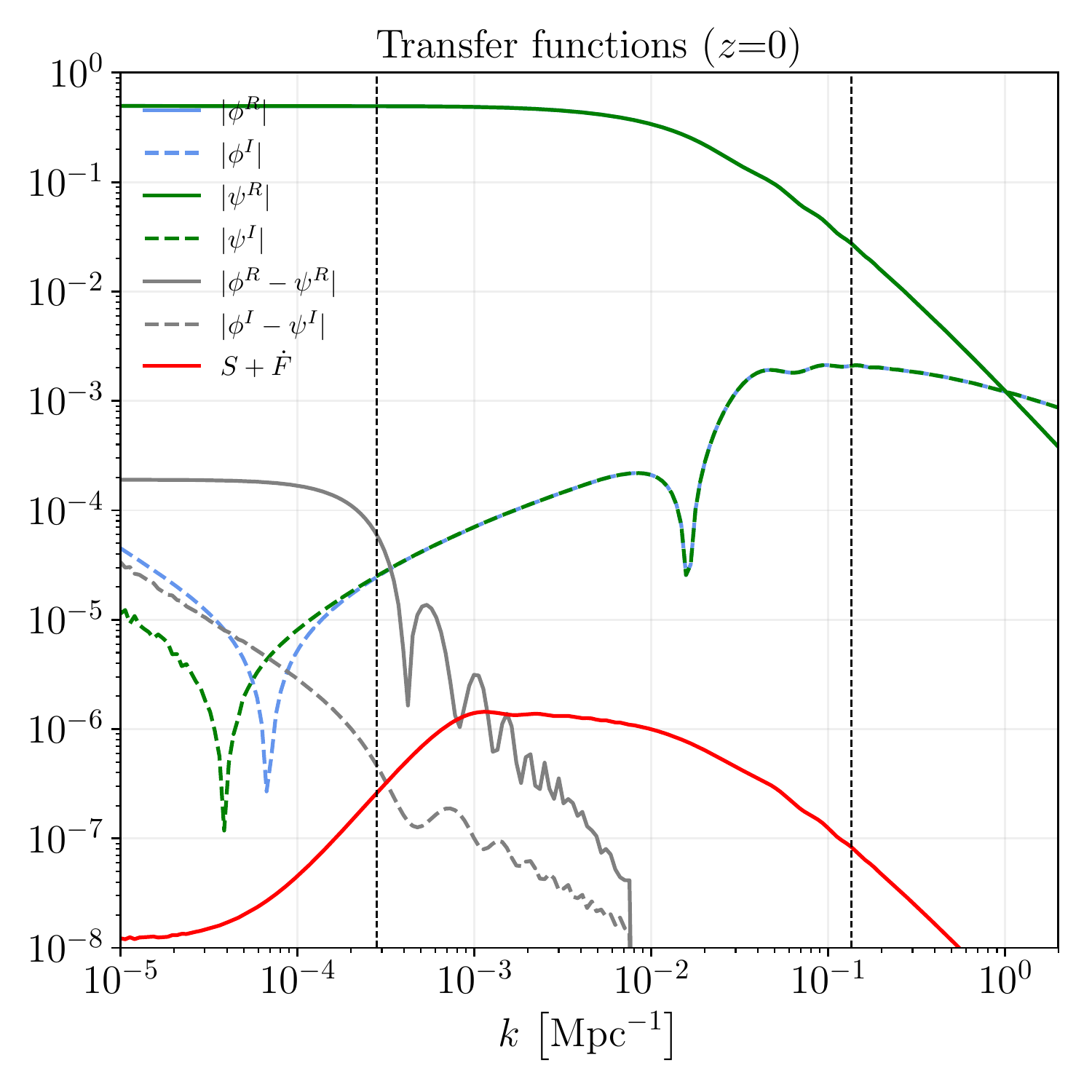}
			}
			\subfigure{
			    \includegraphics[scale=0.5]{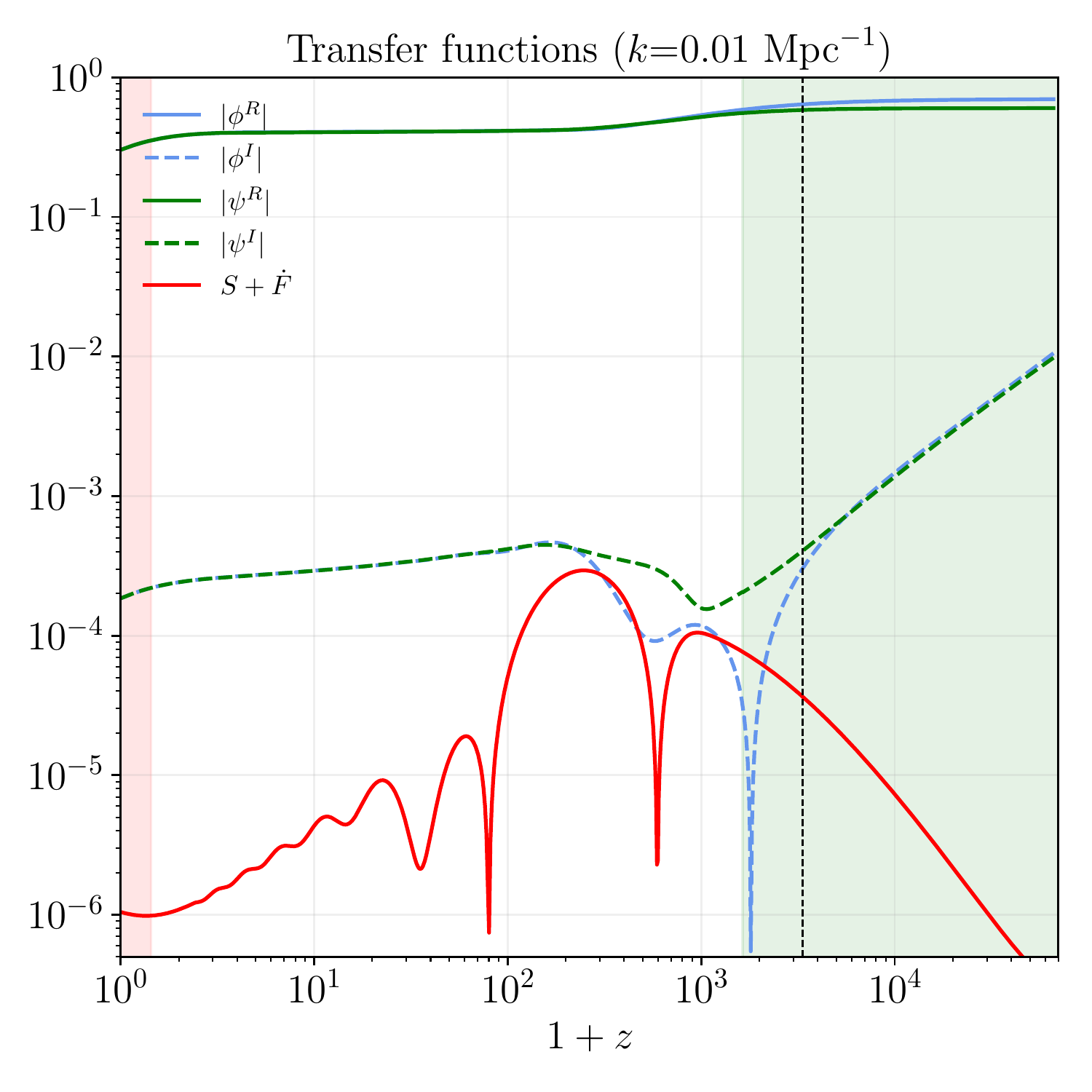}
			}
			\caption{Transfer functions, with initial curvature perturbation normalized to one, 
			    in the Newtonian gauge. Both panels represent the
			    evolution of the Newtonian potentials, their difference and the vector metric 
			    perturbation. The imaginary parts and the vector modes are proportional to $\beta_0$.
			    We show the results for $\beta_0=10^{-3}$. (Left) The vertical line indicates the super-Hubble and non-linearity scales, respectively.
			    The sharp drop in the difference between the Newtonian potentials at small scales is a consequence of the
			    RSA. On those scales we are setting the shear of ultrarelativistic species to zero, since it
			    has a negligible impact in our observables.			    
			    (Right) The vertical line marks the horizon crossing. In the red and green shaded regions
			    the RSA and TCA, respectively, are switched on.}
		    \label{fig:metric_transfer}
        \end{figure}

        Finally, we present the transfer functions evaluated today for a range of $k$ and their time 
        evolution for a fixed value $k=10^{-2}\ \text{Mpc}^{-1}$, as a sample from the full results for 
        the evolution of the perturbations. All the results 
        concerning cosmological perturbations are computed in the synchronous gauge and then transformed
        back to the Newtonian gauge, that can be more easily interpreted in the Newtonian limit
        \cite{Maartens:1998qw, Ellis:2001ms}. Figures \ref{fig:density_transfer}, \ref{fig:velocity_transfer},
        and \ref{fig:vorticity_transfer} contain the density and velocity of CDM, baryons and the full fluid. 
        Figure \ref{fig:metric_transfer} contains the metric variables, i.e. Newtonian potentials and
        vector metric perturbations. The quantities with an $R$ superscript follow the standard evolution 
        and are computed using {\sc class} with the \emph{Planck} 2018 \cite{Aghanim:2018eyx} input 
        values. The modified contributions, with an $I$ superscript, remain smaller than the standard 
        ones for most values of $k$, but not as small as could be expected. The difference at scales of 
        $0.1\ \text{Mpc}^{-1}$ is just one order of magnitude, instead of three as could be naively 
        anticipated from $\beta_0=10^{-3}$, and the modifications could grow even larger above the 
        non-linearity scale.        
        
    \subsection{Spectra}         
        \begin{figure}[htb]
			\subfigure{
			    \includegraphics[scale=0.5]{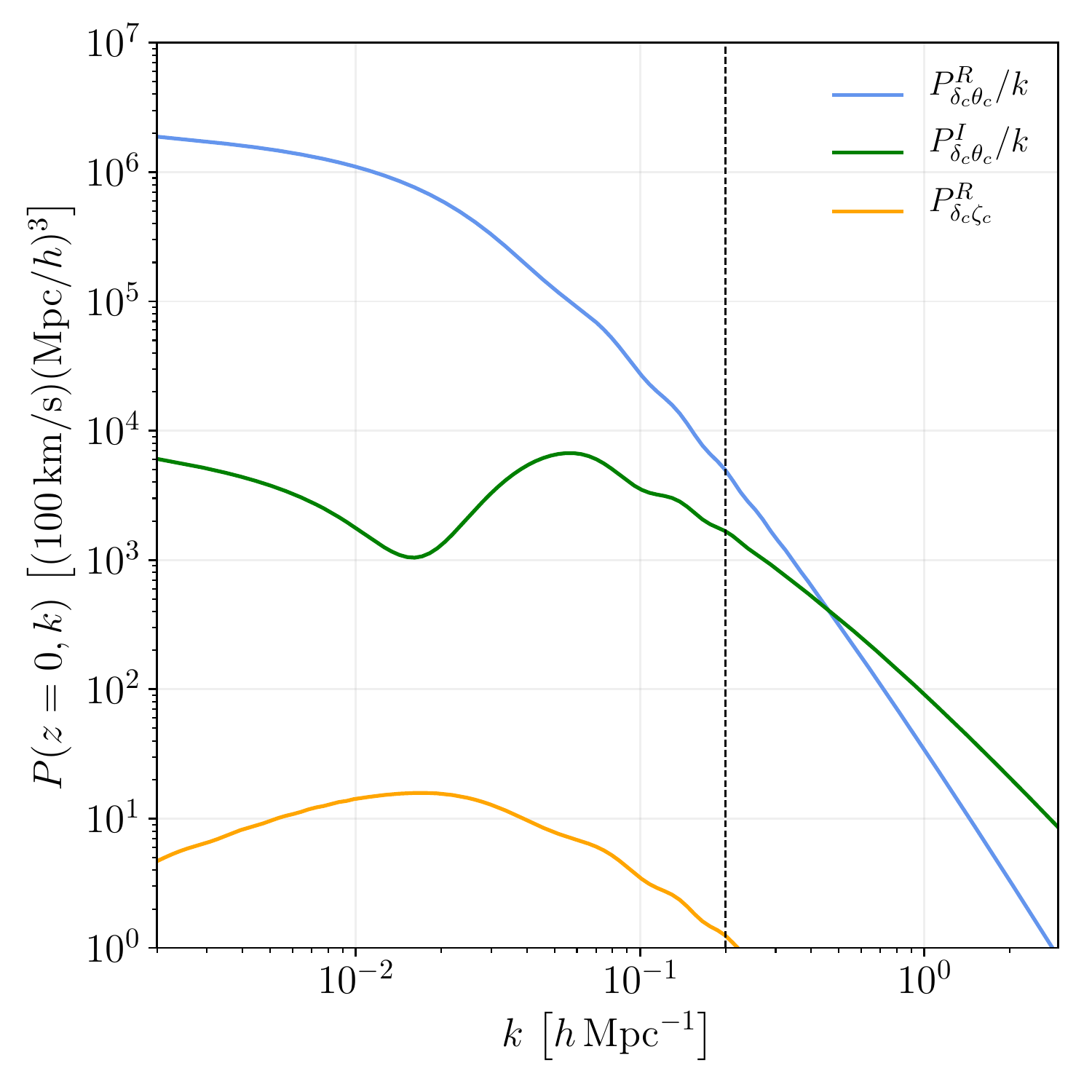}
			}
			\subfigure{
			    \includegraphics[scale=0.5]{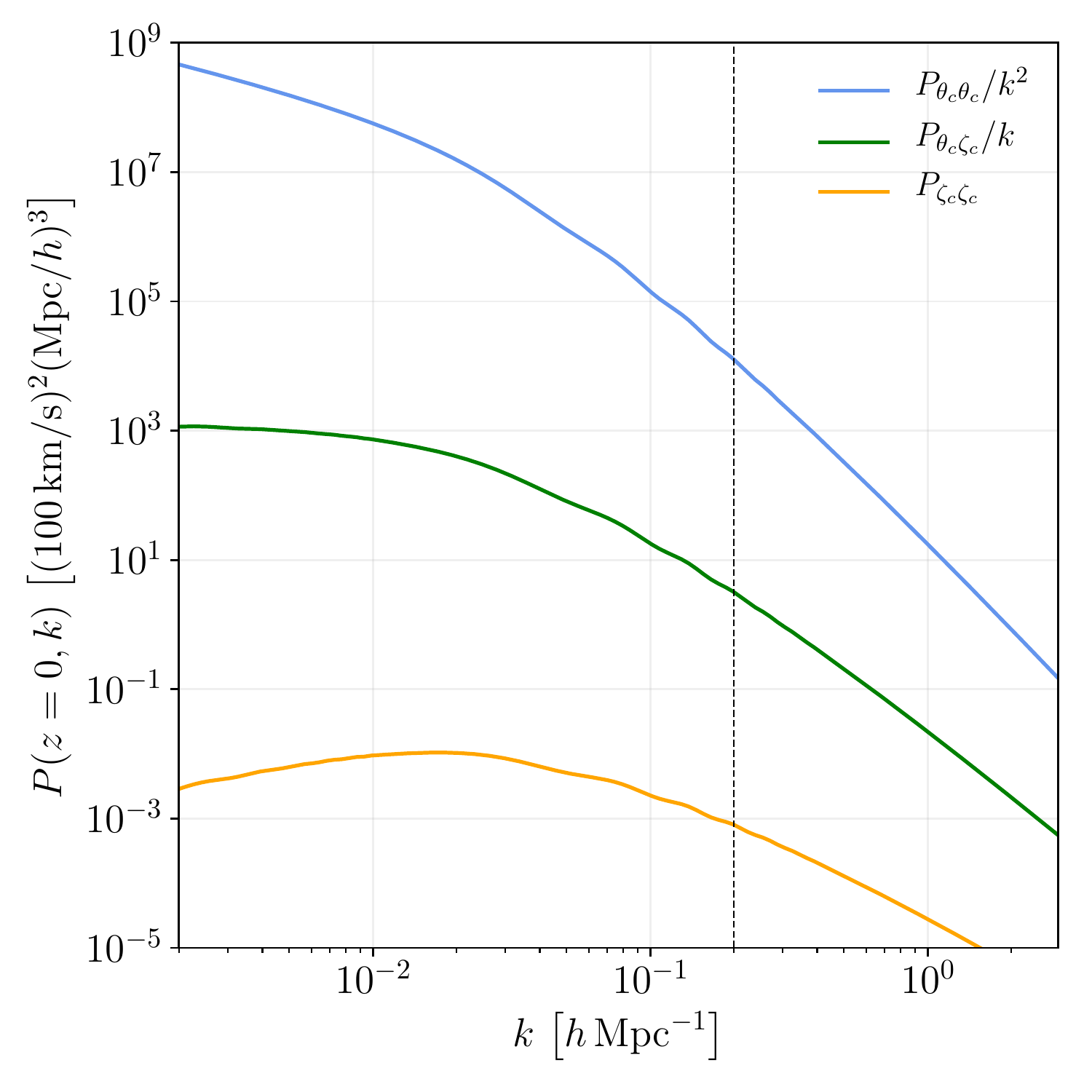}
			}
			\caption{The vertical line indicates	the scale of non-linearity. (Left) Cross-correlation spectra of dark matter densities
			    and velocities, including the standard as well as the dipolar contribution,
			    both for the divergence and for the vorticity. $P^I_{\delta_c\theta_c}$ and $P^R_{\delta_c\zeta_c}$ are given 
			    in units of $(\beta_0/10^{-3})(100\ \text{km}/\text{s})(\text{Mpc}/h)^3$,
			    i.e. the curves are plotted for $\beta_0=10^{-3}$ but the spectra are proportional to this value.
			    (Right) Autocorrelation spectra of dark 
			    matter velocity, as well as the cross-correlation spectrum between vorticity and the 
			    divergence of the velocity. $P_{\theta_c\zeta_c}$ and $P_{\zeta_c\zeta_c}$ are given
			    in units $(\beta_0/10^{-3})(100\ \text{km}/\text{s})^2(\text{Mpc}/h)^3$ 
			    and $(\beta_0/10^{-3})^2(100\ \text{km}/\text{s})^2(\text{Mpc}/h)^3$, respectively.}
		    \label{fig:spectra}
        \end{figure}
        \begin{figure}[htb]
			\includegraphics[scale=0.5]{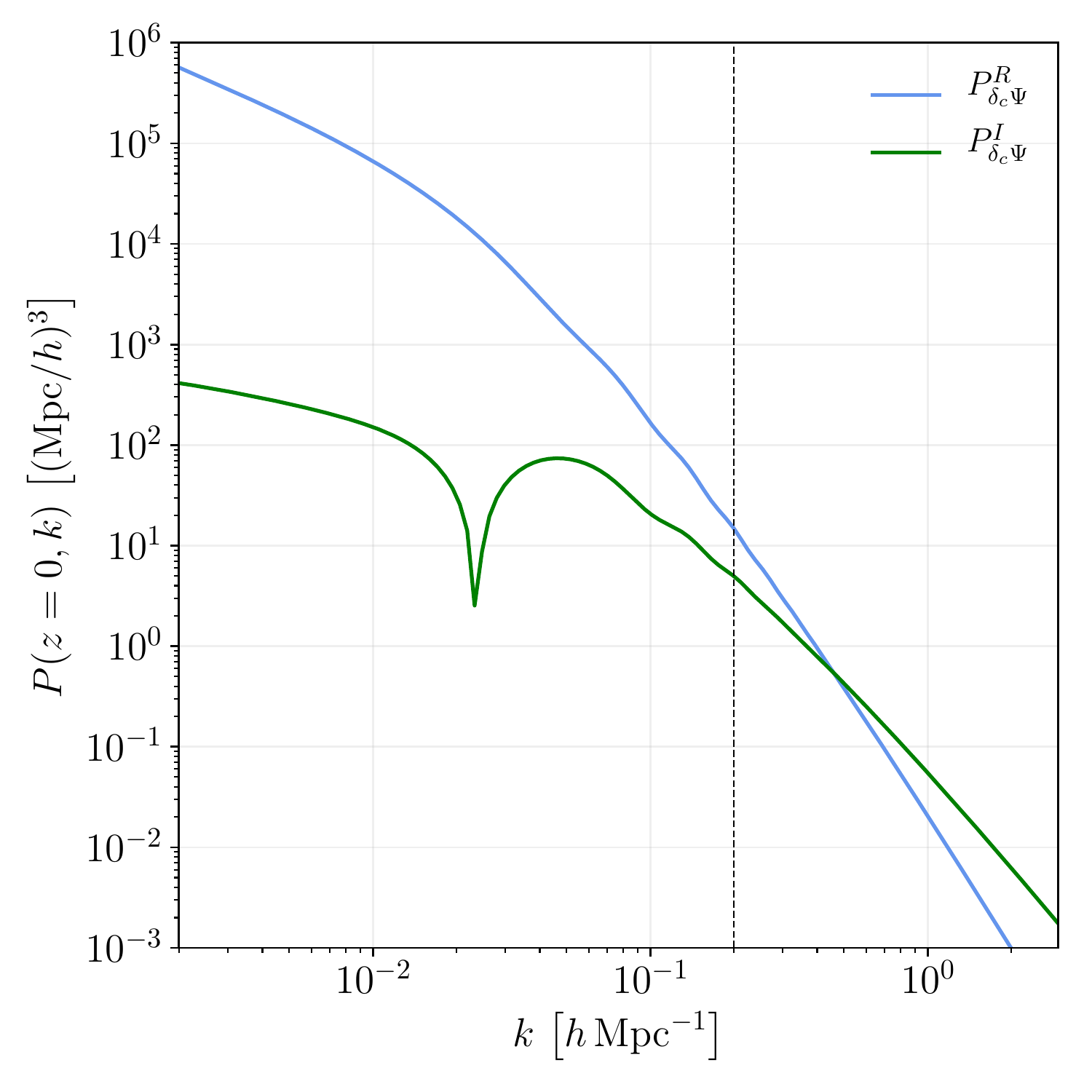}
			\caption{The vertical line indicates the scale of non-linearity.
			    Cross-correlation spectra of dark matter density and lensing potential 
			    $\Psi\equiv\phi+\psi$, including both the standard and the dipolar contribution.
			    $P_{\delta_c\Psi}^I$ is given in units $(\beta_0/10^{-3})(\text{Mpc}/h)^3$.}
		    \label{fig:spectra_lensing}
        \end{figure}        
        We stick to the following conventions for the definition of the spectra. In the
        first place, we define the spectrum of a single variable as
        \begin{align}
            \av{\delta(z, \v{k})\delta^*(z,\v{k}')} &\equiv \delta (\v{k}-\v{k}')P_{\delta\delta}(z,k)\\
                &= \delta(\v{k}-\v{k}')|\delta^R(z,k)|^2\frac{2\pi^2}{k^3}\mathcal{P}_\mathcal{R}(k)
                    +\Od{\beta^2}\ ,
        \end{align}
        where $\delta^R(z,k)$ is the real part of the transfer function and $\mathcal{P}_\mathcal{R}(k)$ is the usual
        nearly scale-invariant curvature spectrum. For the cross-correlations we define
        \begin{align}
            \text{Re}\av{\delta(z,\v{k})\theta^*(z,\v{k}')} &\equiv \delta(\v{k}-\v{k}')P^R_{\delta\theta}(z, k)\\
                &= \delta(\v{k}-\v{k}')\delta^R(z,k)\theta^R(z,k)\frac{2\pi^2}{k^3}\mathcal{P}_\mathcal{R}(k)\ + \Od{\beta^2},\\
            \text{Im}\av{\delta(z,\v{k})\theta^*(z,\v{k}')} &\equiv \delta(\v{k}-\v{k}')(\hat{\beta}\cdot\hat{k})P_{\delta\theta}^I(z, k)\\
                &= \delta(\v{k}-\v{k}')(\hat{\beta}\cdot\hat{k})\left(\delta^I(z,k)\theta^R(z,k)-\delta^R(z,k)\vartheta^I(z,k)\right)\frac{2\pi^2}{k^3}\mathcal{P}_\mathcal{R}(k)+\Od{\beta^2}\ .
        \end{align}
        The vorticity spectrum is defined in a similar way, according to definition 
        \eqref{eq:single_vrt_redef},
        \begin{align}
            \av{\v{\zeta}(z, \v{k})\cdot\v{\zeta}^*(z,\v{k}')} &\equiv \delta (\v{k}-\v{k}')\left(1-(\hat{\beta}\cdot\hat{k})^2\right)P_{\zeta\zeta}(z,k)\\
                &= \delta(\v{k}-\v{k}')\left(1-(\hat{\beta}\cdot\hat{k})^2\right)|\zeta(z,k)|^2\frac{2\pi^2}{k^3}\mathcal{P}_\mathcal{R}(k)+\Od{\beta^3}\ ,
        \end{align}
        The results for the velocity spectrum and the density cross-correlation for CDM are 
        represented in Figure \ref{fig:spectra}. In addition to this information, the velocity-density
        cross-correlation induced by our modification shows a distinctive dipolar pattern. In the
        same way, the vorticity autocorrelation, even though its amplitude is very small, deviates from
        statistical isotropy, with a quadrupole term in addition to the monopole. Figure 
        \ref{fig:spectra_lensing} shows the cross-correlation between the matter density and 
        the lensing potential, defined as $\Psi\equiv\phi+\psi$. This combination is observable 
        using weak-lensing information \cite{Durrer:2008eom, Schimd:2004nq}. Again, our additional 
        contribution becomes important at small scales, being just one order of magnitude below
        the standard result at scales $k=0.1\ \text{Mpc}^{-1}$ instead of three as might be inferred
        from $\beta_0 = 10^{-3}$.
        Both vorticity and deviations from statistical isotropy are absent in standard $\Lambda$CDM. 
        Their presence, with the structure proposed, is a testable effect that could be used to confirm, 
        or disprove, the non-comoving scenario.\\
        
        As we have seen, in our modified setting, velocity spectra are the most easily accessible 
        LSS observables that show significant deviations. Peculiar velocity surveys provide useful 
        complementary information but currently are not competitive with other cosmological observables 
        to constrain standard cosmology \cite{Tegmark:2003uf, Johnson:2014kaa}. Nonetheless, as the 
        precision of velocity surveys increase, such an effect might be seen or at least prove more 
        competitive than the dipole measurements in the CMB and galaxy distribution to 
        constrain $\beta_0$ \cite{Koda:2013eya}.

\section{Conclusions}\label{sec:Conclusions}
    In this work we have developed the theoretical framework needed to analyze the cosmology of
    non-comoving fluids. We have shown that it is possible to relax one of the underlying assumptions
    of $\Lambda$CDM, comoving CMB and matter frames, while retaining an homogeneous and isotropic
    universe. Even if the background behaviour is preserved, the evolution of the perturbations is
    modified, leading to new phenomenological signatures. To first order in the relative velocities,
    i.e. $\Od{\beta}$, we reproduce $\Lambda$CDM behaviour in the main cosmological observables, like
    the matter power spectrum and the CMB temperature power spectrum. We have postponed the full CMB
    analysis \cite{CMB_paper}, focusing instead on LSS observables in this work. As mentioned before, 
    we find that the effects on the autocorrelation spectra are $\Od{\beta^2}$ but there are 
    $\Od{\beta}$ effects on every cross-correlation between cosmological perturbations. Additionally, 
    these corrections present a dipolar pattern, producing deviations from statistical 
    isotropy. We have observed that the additional contributions to the cross-correlation spectra
    become important for small scales.\\
    
    Another distinctive feature of this model is the production of vorticity, which is absent in 
    $\Lambda$CDM. The relative motion of the fluids induces $\Od{\beta}$ couplings between scalar and 
    vector modes. This in its turn leads to the production of vorticity and vector metric perturbations, 
    sourced by the scalar modes. This vector contribution also leaves a characteristic imprint in the 
    velocity spectrum, with a statistically anisotropic quadrupolar modulation. No tensor modes are 
    excited to $\Od{\beta}$.\\
    
    Our only additional free parameter, $\beta_0$, is the initial velocity between the visible and the 
    dark sector, in the frame that observes a homogeneous and isotropic background. Measurements of
    large-scale bulk flows and the CMB dipole allow us to set a conservative limit
    \begin{equation}
        \beta_0 < 1.6\times 10^{-3}\ (95\%\,\text{CL})\ .
    \end{equation}
    Satisfying this constraint, our model is compatible with current observations, and yet it can
    have striking phenomenological consequences. There is a work under way to study the impact on
    the CMB and to give concrete predictions about violations of statistical isotropy
    \cite{CMB_paper}. In particular, it may alleviate the tension that seems to arise when interpreting 
    the anisotropic signal in the CMB as a pure kinematical effect \cite{Ade:2015hxq}. This tension corresponds
    to the dipolar modulation anomaly \cite{Buchert:2015wwr, Schwarz:2015cma} that arises when analyzing 
    low multipoles. The modulation points in a direction different from the kinematic dipole and seems to 
    have a larger amplitude, even if it is compatible with zero within $2\sigma$ \cite{Ade:2015hxq}.\\
    
    More importantly, the production of vorticity
    for the photon-baryon plasma opens an avenue for the creation of magnetic fields \cite{magnetic_paper}.
    The origin of the galactic magnetic fields is a long-standing open problem \cite{Durrer:2013pga}.
    The Harrison mechanism \cite{harrison1970generation} is a cosmological production mechanism that needs vorticity in the 
    photon-baryon plasma to operate, but unfortunately it is absent in $\Lambda$CDM to first order
    in cosmological perturbation theory. Several studies of second-order cosmological perturbation
    theory have proven that in this case vorticity, and thus magnetic fields, is created but with an
    amplitude far too small to act as seed fields for the galactic dynamo amplification mechanism 
    \cite{Takahashi:2005nd, Fenu:2010kh, Saga:2015bna}. Our setup is similar in some regards to a second 
    order computation, but our relevant scale $\beta_0$ is larger than a typical cosmological 
    perturbation and so it is our vorticity production. The associated magnetic fields in our case are 
    expected to be larger and their spectrum would have a different tensor structure as well, since we are 
    singling out a privileged direction \cite{magnetic_paper}.\\
    
    After setting up the formalism for a non-comoving
    cosmology and proving its viability, this work paves the way for the search of these exciting
    new signatures.
    
\begin{acknowledgments}
    This work has been supported by the MINECO (Spain) projects FIS2014-52837-P, 
    FIS2016-78859-P(AEI/FEDER, UE), and Consolider-Ingenio MULTIDARK CSD2009-00064.
    Some symbolic computations have been performed using {\sc SymPy} \cite{sympy}
    and {\sc cadabra} \cite{Peeters:2007wn, peeters2018cadabra2}.
\end{acknowledgments}

\clearpage
\appendix
\section{Geodesics}\label{sec:geodesics}
    In this appendix we will compute in detail the geodesics for a general perturbed RW metric
    \begin{equation}\label{eq:ds2_general}
        \di s^2 = a^2(\tau)\Big(-(1-A)\di\tau^2 + 2B_i\di\tau\di x^i + (\delta_{ij}+H_{ij})\di x^i\di x^j\Big)\ ,
    \end{equation}
    that is
    \begin{equation}
        g_{\mu\nu} = a^2\begin{pmatrix} -1+A & B_i\\B_i &\delta_{ij}+H_{ij} \end{pmatrix},\qquad
        g^{\mu\nu} = \frac{1}{a^2}\begin{pmatrix} -1-A & B^i\\B^i &\delta^{ij}-H^{ij} \end{pmatrix}\ ,
    \end{equation}
    and with the constraint
    \begin{equation}\label{eq:mass_shell_app}
        g^{\mu\nu}P_\mu P_\nu = -m^2\ .
    \end{equation}
    Defining the proper-time parameter as $\di\lambda\equiv \sqrt{-\di s^2}$, the standard definitions
    for the 4-velocity and 4-momentum are
    \begin{align}
        U^\mu &\equiv \d{x^\mu}{\lambda}\ ,\\
        P_\mu &\equiv m U_\mu\ ,\\
        P_0 &\equiv -\epsilon + \delta P_0\ .
    \end{align}
    The geodesics are given by
    \begin{equation}
        \d{U^\mu}{\lambda} +\Gamma\indices{^\mu_\nu_\rho}U^\nu U^\rho = 0\ ,
    \end{equation}
    that can be conveniently rewritten as
    \begin{equation}
        \d{U_\mu}{\lambda} = \frac{1}{2}\pd{g_{\nu\rho}}{x^\mu}U^\nu U^\rho\ .
    \end{equation}
    This last form is especially useful. Since the background metric is homogeneous, we can keep only
    the zero order in $U_\mu$ to compute the spatial part.
    Writing the evolution in terms of the conformal time we have
    \begin{equation}
        \d{P_i}{\tau} = \frac{m}{U^0}\d{U_i}{\lambda} = \frac{1}{2\epsilon}\left(\epsilon^2\partial_iA
            +2\epsilon P^j\partial_iB_j + P^jP^k\partial_iH_{jk}\right)\ .
    \end{equation}
    The spatial momentum is redefined as
    \begin{align}\label{eq:momentum_param}
        P_i &\equiv \left(\delta\indices{^j_i}+\frac{1}{2}H\indices{^j_i}\right)q_j\ ,\\
        a^2P^i &= -\epsilon B^i + \left(\delta^{ki}-\frac{1}{2}H^{ki}\right)q_k\ ,
    \end{align}
    where, from now on, every spatial index on a perturbed quantity is assumed to be raised or lowered 
    using $\delta_{ij}$. From the mass-shell condition \eqref{eq:mass_shell_app} we obtain
    \begin{align}
        \epsilon^2 &= m^2a^2 + q^2\ ,\\
        P_0 &= -\epsilon\left(1-\frac{1}{2}A\right) + q_iB^i \ ,\\
        P^0 &= \frac{\epsilon}{a^2}\left(1+\frac{1}{2}A\right)\ .
    \end{align}
    The geodesic equation can be written with this parameterization as
    \begin{equation}
        \d{q_i}{\tau} = \frac{1}{2}\epsilon\partial_iA + q^j\partial_iB_j + \frac{1}{2\epsilon}q^jq^k
            \left(\partial_iH_{jk}-\partial_kH_{ij}\right) - \frac{1}{2}q^j\dot{H}_{ij}\ .
    \end{equation}
    Further decomposing $q_i$ into direction $\hat{n}$ and magnitude $q$
    \begin{equation}
        q_i\equiv q\,n_i\qquad\to \qquad n_i\delta^{ij}n_j= 1\ ,\quad q^2 = \delta^{ij}q_iq_j\ ,
    \end{equation}
    we have
    \begin{subequations}
    \begin{align}
        \d{q}{\tau} &= n^i\d{q_i}{\tau}\ ,\\
        \d{n_i}{\tau} &= \frac{1}{q}\left(\delta\indices{^j_i}-n^jn_i\right)\d{q_j}{\tau}\ .
    \end{align}
    \end{subequations}
    Finally, using the following succint redefinition of metric variables
    \begin{align}
        C_{ij} &\equiv \partial_iB_j - \frac{1}{2}\dot{H}_{ij}\ ,\\
        D_{ijk} &\equiv \frac{1}{2}\left(\partial_iH_{jk}-\partial_kH_{ij}\right)\ ,
    \end{align}
    the final formulae needed to compute the geodesics, with the parameterization 
    \eqref{eq:momentum_param}, are
    \begin{subequations}
    \begin{align}
        \d{x^i}{\tau} &= \frac{q^i}{\epsilon}\left(1-\frac{1}{2}A\right)-B^i
            -\frac{1}{2\epsilon}H^i_kq^k\ ,\\
        \d{q^i}{\tau} &= \frac{1}{2}\epsilon\partial_iA + q^jC_{ij} + \frac{q^jq^k}{\epsilon}D_{ijk}\ ,\\
        \d{q}{\tau} &= \frac{1}{2}\epsilon\, n^i\partial_iA + q\,n^in^jC_{ij}\ ,\\
        \d{n_i}{\tau} &= \left(\delta\indices{^j_i}-n^jn_i\right)\left(
            \frac{\epsilon}{2q}\partial_jA + n^kC_{jk} + \frac{q}{\epsilon}n^kn^lD_{jkl}\right)\ .
    \end{align}
    \end{subequations}
    
    Some relevant metric quantities, written in terms of the scalar-vector-tensor decomposition and in
    Fourier space, are
    \begin{align}
        A &= -2\psi\ ,\\
        B_i &= \ii k_i B -S_i\ ,\\
	    H_{ij} &= -2\phi\delta_{ij} - 2k_ik_jE+ \ii\left(k_iF_j+k_jF_i\right) + h_{ij}\ ,\\
	    C_{ij} &= -k_ik_j(B-\dot{E})-\ii k_iS_j + \dot{\phi}\delta_{ij} - \frac{\ii}{2}\left(k_i\dot{F}_j
	        +k_j\dot{F}_i\right) - \frac{1}{2}\dot{h}_{ij}\ ,\\
	    D_{ijk} &= -\ii(k_i\delta_{jk}-k_k\delta_{ij})\phi - \frac{1}{2}\left(k_ik_jF_k-k_jk_kF_i\right)
	        +\frac{\ii}{2}\left(k_ih_{jk}-k_kh_{ij}\right)\ .
    \end{align}
    
\section{Full computation of the collision term}\label{sec:full_collision}
    Starting from the collision term \eqref{eq:collision_starting_point},
    \begin{equation}
        \mathcal{C}[f]=\frac{\sigma_T}{4\pi p}\int\tilde{p}'\di\tilde{p}'\di\tilde{\Omega}'
            \left[\tilde{n}_e^\text{full}\delta (\tilde{p}-\tilde{p}')+\tilde{n}_e\v{\tilde{u}}^\text{full}_e\cdot
            (\v{\tilde{p}}-\v{\tilde{p}}')\pd{\delta (\tilde{p}-\tilde{p}')}{\tilde{p}'}\right]
            \left(\bar{f}(\Lambda_\beta\Lambda_{\beta_e}^{-1}\v{\tilde{p}}')
            -\bar{f}(\Lambda_\beta\Lambda_{\beta_e}^{-1}\v{\tilde{p}})\right)\ ,
    \end{equation}
    we need to solve two integrals
    \begin{align}
        I_1 &= \int\tilde{p}'\di\tilde{p}'\di\tilde{\Omega}'\delta (\tilde{p}-\tilde{p}')
            \left(\bar{f}(\Lambda_\beta\Lambda_{\beta_e}^{-1}\v{\tilde{p}}')
            -\bar{f}(\Lambda_\beta\Lambda_{\beta_e}^{-1}\v{\tilde{p}})\right)\nonumber\\
            &= -4\pi\tilde{p}\bar{f}(\Lambda_\beta\Lambda_{\beta_e}^{-1}\v{\tilde{p}})
                +\tilde{p}\int\di\tilde{\Omega}'\,\bar{f}(\Lambda_\beta\Lambda_{\beta_e}^{-1}
                \v{\tilde{p}}')\rvert_{\tilde{p}'=\tilde{p}}\ ,\\
        \v{I}_2 &= \int\tilde{p}'\di\tilde{p}'\di\tilde{\Omega}' (\v{\tilde{p}}-\v{\tilde{p}}')
            \pd{\delta (\tilde{p}-\tilde{p}')}{\tilde{p}'}\left(\bar{f}(\Lambda_\beta
            \Lambda_{\beta_e}^{-1}\v{\tilde{p}}')-\bar{f}(\Lambda_\beta\Lambda_{\beta_e}^{-1}
            \v{\tilde{p}})\right)\nonumber\\
            &= -\tilde{p}\int\di\tilde{\Omega}'(\hat{\tilde{n}}-2\hat{\tilde{n}}')
                \left(\bar{f}(\Lambda_\beta\Lambda_{\beta_e}^{-1}\v{\tilde{p}}')\rvert_{\tilde{p}'=\tilde{p}}
                -\bar{f}(\Lambda_\beta\Lambda_{\beta_e}^{-1}\v{\tilde{p}})\right)
                +\tilde{p}^2\int\di\tilde{\Omega}'(\hat{\tilde{n}}-\hat{\tilde{n}}')
                \pd{\bar{f}(\Lambda_\beta\Lambda_{\beta_e}^{-1}\v{\tilde{p}})}{\tilde{p}'}
                \bigg\rvert_{\tilde{p}'=\tilde{p}}\ .
    \end{align}
    Now, we will integrate out the dependence on the momentum $p$ in the $\mathcal{O}$ frame, like we
    did with the left-hand side of the Boltzmann equation. First, for simplicity, we will assume that
    $\v{\beta}$ and $\v{\beta}_e$ point in the same direction, so we can obtain
    \begin{equation}
        \Lambda_\beta\Lambda_{\beta_e}^{-1} = \Lambda_{\Delta\beta}\ ,\qquad 
        \Delta\beta \equiv \frac{\beta-\beta_e}{1-\beta\beta_e}\ .
    \end{equation}
    This assumption simplifies the derivation for arbitrary values of $\beta$, but it is not needed and
    in fact the first order results are independent of it. Since the expressions are already
    quite cumbersome, and this will be the only physical configuration of interest, we will adopt this 
    assumption throughout this appendix. Some preliminary results and definitions are
    \begin{subequations}
    \begin{align}
        \int q^3\di q\,\bar{f}_0(\Lambda_\beta q) &= \frac{\tilde{\mathcal{N}}}{\gamma^4(1-\hat{n}\cdot\v{\beta})^4}\ ,\\
        \int\frac{\di^3\tilde{q}}{4\pi}\,\tilde{q}\,\bar{f}_0(\Lambda_{\Delta\beta}\tilde{q}) 
            &= \tilde{\mathcal{N}}\gamma^2_{\Delta\beta}\left(1+\frac{\Delta\beta^2}{3}\right)\ ,\\
        \int\frac{\di^3\tilde{q}}{4\pi}\,\tilde{q}^i\,\bar{f}_0(\Lambda_{\Delta\beta}\tilde{q}) 
            &= \frac{4}{3}\tilde{\mathcal{N}}\gamma^2_{\Delta\beta}\Delta\beta^i\ ,
    \end{align}
    \end{subequations}
    where $\tilde{\mathcal{N}}$ is defined in \eqref{eq:normalization_photon} and $\mathcal{P}^i_{e\,j}$, 
    $\gamma_e$ correspond to \eqref{eq:Lorentz_defs}, evaluated with $\beta_e$. We can proceed now to 
    compute the integrals term by term. First for $I_1$
    \begin{subequations}
    \begin{align}
        \int\di q\, q^2\tilde{q}\,\bar{f}(\Lambda_\beta\Lambda_{\beta_e}^{-1}\v{\tilde{q}})
            &= \tilde{\mathcal{N}}\gamma_e(1-\hat{n}\cdot\v{\beta}_e)\left(\mathcal{F}_\gamma 
                + \frac{1}{\gamma^4(1-\hat{n}\cdot\v{\beta})^4}\right)\ ,\\
        \int\di q\,q^2\tilde{q}\int\di\tilde{\Omega}
                \,\bar{f}(\Lambda_\beta\Lambda_{\beta_e}^{-1}\v{\tilde{q}})
            &=\frac{4\pi \tilde{\mathcal{N}}}{\gamma^3_e(1-\hat{n}\cdot\v{\beta}_e)^3}\left[\left(1
                +\frac{\Delta\beta^2}{3}\right)\gamma_{\Delta\beta}^2 
                +\gamma_e^2\int\frac{\di\Omega}{4\pi}\left(1-2\hat{n}\cdot\v{\beta}_e + (\hat{n}\cdot
                \v{\beta}_e)^2\right)\mathcal{F}_\gamma\right]\ .
    \end{align}
    \end{subequations}
    And for $\v{I}_2$
    \begin{align}
        \int q^2\di q\,\tilde{q}^i \bar{f}_0(\Lambda_{\Delta\beta}\tilde{q})
            &= \tilde{\mathcal{N}}\frac{\mathcal{P}^i_{e\,j}n^j-\gamma_e\beta^i_e}{\gamma^4(1-\hat{n}\cdot\v{\beta})^4}\ ,\\
        \int q^2\di q\,\tilde{q}^i \int\di\tilde{\Omega}\,\bar{f}_0(\Lambda_{\Delta\beta}\tilde{q})
            &= 4\pi \tilde{\mathcal{N}}\frac{\mathcal{P}^{i}_{e\,j}n^j-\gamma_e\beta^i_e}{\gamma^4_e(1-\hat{n}\cdot\v{\beta}_e)^4}
                \gamma^2_{\Delta\beta}\left(1+\frac{\Delta\beta^2}{3}\right)\ ,\\
        \int q^2\di q \int\di\tilde{\Omega}\,\tilde{q}^i\,\bar{f}_0(\Lambda_{\Delta\beta}\tilde{q})
            &= \frac{16\pi}{3} \tilde{\mathcal{N}}\Delta\beta^i\frac{\gamma_{\Delta\beta}^2}{\gamma_e^3(1-\hat{n}
                \cdot\v{\beta}_e)^3}\ ,\\
        \int q^2\di q\,\tilde{q}\,\tilde{q}^i\int\di\tilde{\Omega}\,
                \pd{\bar{f}_0(\Lambda_{\Delta\beta}\tilde{q})}{\tilde{q}}    
            &= -16\pi \tilde{\mathcal{N}}\frac{\mathcal{P}^i_{e\,j}n^j-\gamma_e\beta_e^i}{\gamma_e^4(1-\hat{n}\cdot
                \v{\beta}_e)^4}\gamma_{\Delta\beta}^2\left(1+\frac{\Delta\beta^2}{3}\right)\ ,\\
        \int q^2\di q\,\tilde{q}\int\di\tilde{\Omega}\,\tilde{q}^i\,
                \pd{\bar{f}_0(\Lambda_{\Delta\beta}\tilde{q})}{\tilde{q}}
            &= -\frac{64\pi}{3} \tilde{\mathcal{N}}\Delta\beta^i\frac{\gamma_{\Delta\beta}^2}{\gamma_e^3(1-\hat{n}
                \cdot\v{\beta}_e)^3}\ .
    \end{align}
    Finally, for the electron quantities,
    \begin{subequations}
    \begin{align}
        \tilde{n}_e^\text{full} &\equiv 2\int\frac{\di^3\tilde{p}_e}{(2\pi)^3}\tilde{f}_e(\v{\tilde{p}}_e) = \tilde{n}_e + \delta \tilde{n}_e\ ,\\
        \tilde{n}_e\v{\tilde{u}}^\text{full}_e &= 2\int\frac{\di^3\tilde{p}_e}{(2\pi)^3}\frac{\v{\tilde{p}}_e}{\tilde{E}_{p_e}}\tilde{f}_e(\v{\tilde{p}}_e)
            = 2\int\frac{\di^3\tilde{p}_e}{(2\pi)^3}\frac{\v{\tilde{p}}_e}{\tilde{E}_{p_e}}\delta \tilde{f}_e(\v{\tilde{p}}_e)
            = \tilde{n}_e\delta \v{\tilde{v}}_e\ ,
    \end{align}
    \end{subequations}
    where $\tilde{n}_e$ is the physical background electron number density, computed in its comoving frame as every
    other background quantity. Using the Lorentz transformation properties
    \begin{subequations}
    \begin{align}
        \delta \tilde{n}_e &= \gamma_e\left(\delta n_e - \delta v_e^j\beta_{e\, j}\right)\ ,\\
        \delta \tilde{v}_e^i &= \mathcal{P}^i_{e\,j}\delta v^j_e - \gamma_e\beta_e^i\delta n_e\ .
    \end{align}
    \end{subequations}
    The final results, to first order in $\beta$, are
    \begin{subequations}
    \begin{align}
        \int q^2 \di q\,I_1 &= -4\pi \tilde{\mathcal{N}}\left((1-\hat{n}\cdot\v{\beta}_e)\mathcal{F}_\gamma +4\hat{n}\cdot
            \Delta\v{\beta}-(1+3\hat{n}\cdot\v{\beta}_e)\int\frac{\di\Omega}{4\pi}\mathcal{F}_\gamma
                +2\v{\beta}_e\cdot\int\frac{\di\Omega}{4\pi}\hat{n}\,\mathcal{F}_\gamma\right)\ ,\\
            \int q^2 \di q\,\v{I}_2 &= 4\pi \tilde{\mathcal{N}}\left(-\frac{8}{3}\Delta\v{\beta}+4\hat{n}\,(\hat{n}\cdot\Delta\v{\beta})
                +4\left(\hat{n}-\v{\beta}_e+4\hat{n}\,(\hat{n}\cdot\v{\beta}_e)\right)\right)\ ,\\
            \delta\tilde{n}_e &= \delta n_e-\beta_{e\,j}\delta v^j_e\ ,\\
            \delta\tilde{v}^i_e &= \delta v^i_e-\beta^i_e\delta n_e\ .
    \end{align}
    \end{subequations}
    
\section{Gauge transformations with non-comoving fluids}\label{sec:gauge_transformations}
    After an infinitesimal gauge transformation
    \begin{equation}
        \Delta x^\mu = \epsilon^\mu\ , \qquad \epsilon^{\mu} = \Big(T(\tau,\v{x}), \v{L}(\tau,\v{x})\Big)\ ,
    \end{equation}
    a tensor changes as
    \begin{equation}
        \Delta T_{\mu\nu} = \epsilon^\rho\pd{T_{\mu\nu}}{x^\rho}+\epsilon^\rho_{,\mu}T_{\rho\nu}+\epsilon^\rho_{,\nu}T_{\mu\rho}=\mathcal{L}_\epsilon T_{\mu\nu}\ ,
    \end{equation}
    where $\mathcal{L}_\epsilon$ is the Lie derivative. Applying the result to the metric tensor, we have
    for the metric variables
    \begin{subequations}
    \begin{align}
        \Delta A &= -2(\dot{T}+\mathcal{H}T)\ ,\\
        \Delta B_i &= \dot{L}_i-\partial_iT\ ,\\
        \Delta H_{ij} &= 2\mathcal{H}T\delta_{ij}+\partial_iL_j+\partial_jL_i\ .
    \end{align}
    \end{subequations}
    On the other hand, for the perturbed fluid variables we get
    \begin{subequations}
    \begin{align}
        \Delta\delta\rho &= T\dot{\rho}-2(\rho+P)\beta^i\partial_iT\ ,\\
        \Delta\delta Q_i &= -(\rho+P)\partial_iT+T\partial_\tau\left(\beta_i(\rho+P)\right)
            +\frac{1}{2}(\rho+P)\beta^j\left(\partial_iL_j-\partial_jL_i\right)\ ,\\
        \Delta\delta P &= T\dot{P}-\frac{2}{3}\beta^i\partial_iT(\rho+P)\ ,\\
        \Delta\delta\Pi_{ij} &= -(\rho+P)\left(\beta_i\partial_jT+\beta_j\partial_iT-\frac{2}{3}\delta_{ij}\beta^k\partial_kT\right)\ ,
    \end{align}
    \end{subequations}
    where we are adopting the definitions \eqref{eq:def_perturbed_fluid}. With our previous definition
    for the scalar-vector-tensor decomposition of the metric variables \eqref{eq:metric_main_defs}, 
    we get, in Fourier space,
    \begin{alignat}{2}\label{eq:gauge_full_metric}
        \Delta\psi &= \dot{T}+\mathcal{H}T\ , &\qquad\qquad \Delta S_+ &= -\dot{L}_+\ ,\nonumber\\
        \Delta B &= -\frac{\ii}{k}\hat{k}\cdot\dot{\v{L}}-T\ , &\qquad\qquad \Delta F_+ &= L_+\ ,\nonumber\\
        \Delta\phi &= -\mathcal{H}T\ , &\qquad\qquad \Delta h_{++} &= 0\ ,\nonumber\\
        \Delta E &= -\frac{\ii}{k}\hat{k}\cdot\v{L}\ .
    \end{alignat}
    The results for the $-$ helicity can be obtained substituting $-\leftrightarrow +$
    in every sub and superscript.
    If we want to change from the synchronous to the Newtonian gauge, for scalar perturbations, the 
    following conditions must be satisfied
    \begin{subequations}
    \begin{align}
        \psi &= \dot{T}+\mathcal{H}T\ ,\\
        0 &= -\frac{\ii}{k}\hat{k}\cdot\dot{\v{L}}-T\ ,\\
        \phi-\eta &= -\mathcal{H}T\ ,\\
        \frac{1}{2k^2}(h+6\eta) &= -\frac{\ii}{k}\hat{k}\cdot\v{L}\ ,
    \end{align}
    \end{subequations}
    that can be solved to yield
    \begin{align}
        T &= \frac{1}{2k^2}\left(\dot{h}+6\dot{\eta}\right)\ ,\\
        \psi &= \dot{T}+\mathcal{H}T\ ,\\
        \phi &= \eta+\mathcal{H}T\ .
    \end{align}
    Finally, for the fluid variables that we use in the main part of the computations, we have the rules
    \begin{align}
        \delta (\text{Newt})-\delta (\text{Syn}) &= T\frac{\dot{\rho}}{\rho}-2\ii(1+w)(\v{\beta}\cdot\v{k})T\ ,\\
        \theta (\text{Newt})-\theta (\text{Syn}) &= k^2T+\ii(\dot{\v{\beta}}\cdot\v{k})T +\ii(\v{\beta}\cdot\v{k})\frac{\dot{\rho}+\dot{P}}{\rho+P}T\ ,\\
        \chi_+(\text{Newt})-\chi_+ (\text{Syn}) &= T\dot{\beta}_++T\beta_+\frac{\dot{\rho}+\dot{P}}{\rho+P}\ ,\\
        \delta P (\text{Newt})-\delta P(\text{Syn}) &= T\dot{P}-\frac{2\ii}{3}(\v{\beta}\cdot\v{k})T(\rho+P)\ ,\\
        \sigma (\text{Newt})-\sigma (\text{Syn}) &= \ii(\v{\beta}\cdot\v{k})(1+w)T\ ,\\
        \pi^\text{V}_+ (\text{Newt})-\pi^\text{V}_+ (\text{Syn}) &= -\beta_+(1+w)T\ ,\\
        \pi^\text{T}_{++} (\text{Newt})-\pi^\text{T}_{++} (\text{Syn}) &= 0\ ,
    \end{align}    
    where we have neglected terms $\beta L_+$, making use of the fact that, according to 
    \eqref{eq:gauge_full_metric}, the transverse part of $\v{L}$ under our assumptions can be at most 
    order $\beta$. Again, we are omitting the results for the $-$ helicity, that can be obtained 
    substituting $-\leftrightarrow +$ in every sub and superscript.
    
\section{Initial conditions}\label{sec:initial_conditions}
    In this appendix we will find the appropiate initial conditions for the system of scalar and
    vector modes in sections \ref{sec:scalar_modes} and \ref{sec:vector_modes}. We will consider the
    most general initial condition and then study the physical restrictions that we must impose.
    For $\beta=0$, our system reproduce the standard cosmology. This case has been extensively studied
    over the years and the relevant modes, i.e. one adiabatic and four isocurvature modes, have been
    identified \cite{Bucher:1999re}. In our setup, the presence of an external source gives rise to
    the existence of a new ``mode'' of the system, in the sense that we have a non-trivial evolution
    even if the usual adiabatic and isocurvature modes are absent. First, we will identify this particular
    solution, 
    setting to zero the other modes of the system. Note that the external sources only contain variables
    that evolve according to standard $\Lambda$CDM, so for these variables only adiabatic initial 
    conditions are considered. After identifying the effect of the sources, the most general perturbation
    can be constructed adding to the sourced mode the adiabatic and isocurvature modes. Finally, we must
    analyze what physical requirements constrain our choice of initial conditions. In particular, we
    impose that neutrinos and photons, being tightly coupled in the very early Universe, share a common
    initial velocity. Every other initial condition that is not fixed by this condition is set to zero.
    This programme is carried out in detail in the next sections.
    
    \subsection{Scalar modes}    
	    To obtain our results, we have analyzed the most general type of perturbation, reproducing the
	    results of \cite{Bucher:1999re} but with a slight change of notation. In the first place, we use
	    an alternative approach where we integrate \eqref{eq:ScaSys_full_final} instead of the dark matter
	    equation. In this setup, matter isocurvature modes appear when taking non-zero initial conditions 
	    for $\delta_\nu$, $\delta_b$ or $\dot{h}$. In the second place, even though it is perfectly 
	    equivalent, we parameterize the neutrino isocurvature velocity mode with the initial value of 
	    $\theta_\gamma$ instead of $\theta_\nu$.\\
	    
	    Before presenting the results, some shorthand definitions that will be used later are
	    \begin{align}
	        S_{\gamma\nu} &\equiv \Omega_\gamma + \Omega_\nu\ ,\\
	        \mathcal{R}_s &\equiv \Omega_s/S_{\gamma\nu}\ ,\qquad s=\gamma,\nu,b,c\ ,\\
	        \mathcal{R}_{bc} &\equiv \mathcal{R}_b+\mathcal{R}_c\ ,
	    \end{align}
	    Assuming a universe composed of radiation and matter, where $\tau$ stands for conformal time, 
	    \begin{align}
	        \mathcal{H} &= \frac{\frac{H_{0} \mathcal{R}_{b c} \sqrt{S_{\gamma \nu}} \tau}{2} + 1}{\tau \left(\frac{H_{0} \mathcal{R}_{b c} \sqrt{S_{\gamma \nu}} \tau}{4} + 1\right)}\ ,\\
	        a &= H_{0} \sqrt{S_{\gamma \nu}} \tau \left(\frac{H_{0} \mathcal{R}_{b c} \sqrt{S_{\gamma \nu}} \tau}{4} + 1\right)\ .
	    \end{align}
	    During the radiation-dominated phase, they can be expanded as
	    \begin{align}
	        a &= H_{0} \sqrt{S_{\gamma \nu}} \tau + O\left(H_0^2\mathcal{R}^2_{bc}S_{\gamma\nu}\tau^{2}\right)\ ,\\
	        \mathcal{H} &= \frac{1}{\tau} + \frac{H_{0} \mathcal{R}_{b c} \sqrt{S_{\gamma \nu}}}{4} - \frac{H_{0}^{2} \mathcal{R}_{b c}^{2} S_{\gamma \nu} \tau}{16} + O\left(H_0^2\mathcal{R}^2_{bc}S_{\gamma\nu}\tau^{2}\right)\ ,\\
	        \mathcal{H}^{2} &= \frac{1}{\tau^{2}} + \frac{H_{0} \mathcal{R}_{b c} \sqrt{S_{\gamma \nu}}}{2 \tau} - \frac{H_{0}^{2} \mathcal{R}_{b c}^{2} S_{\gamma \nu}}{16} + O\left(H_0^2\mathcal{R}^2_{bc}S_{\gamma\nu}\tau^{2}\right)\ ,\\
	        \dot{\mathcal{H}} &= - \frac{1}{\tau^{2}} - \frac{H_{0}^{2} \mathcal{R}_{b c}^{2} S_{\gamma \nu}}{16} + \frac{H_{0}^{3} \mathcal{R}_{b c}^{3} S_{\gamma \nu}^{\frac{3}{2}} \tau}{32} + O\left(H_0^2\mathcal{R}^2_{bc}S_{\gamma\nu}\tau^{2}\right)\ .
	    \end{align}
	    Now, if we look for regular super-Hubble solutions and expand every cosmological variable as 
	    \begin{equation}
	        \delta^I_{\gamma} = D^{(0)}_{\delta_\gamma} + D^{(1)}_{\delta_\gamma} \tau + D^{(2)}_{\delta_\gamma} \tau^{2} + D^{(3)}_{\delta_\gamma} \tau^{3} + \dots
	    \end{equation}
	    The results for the sourced mode only (setting the initial conditions for $\eta^I$, $\dot{h}^I$, 
	    $\delta_\nu^I$, $\delta_b^I$ and $\theta_\gamma^I$ to zero), can be written in terms of the initial
	    value of $\psi^R(\tau=0)=\Psi$ or
	    \begin{equation}
	        \eta^R(\tau=0) = \frac{4\mathcal{R}_\nu+15}{10}\Psi\ ,
	    \end{equation}
	    as
	    \begin{alignat}{2}
	        D^{(0)}_{\delta_\gamma} &= 0\ ,  &\qquad\qquad  D^{(1)}_{\delta_\gamma} &= \frac{4 \Psi \beta_{0} k \left(4 \mathcal{R}_\nu + 15\right)}{15}\ ,\\
	        D^{(0)}_{\delta_\nu} &= 0\ ,     &\qquad\qquad  D^{(1)}_{\delta_\nu} &= \frac{4 \Psi \beta_{0} k \left(\mathcal{R}_\nu - 1\right) \left(4 \mathcal{R}_\nu + 15\right)}{15 \mathcal{R}_\nu}\ ,\\
	        D^{(0)}_{\delta_b} &= 0\ ,       &\qquad\qquad  D^{(1)}_{\delta_b} &= \frac{\Psi \beta_{0} k \left(4 \mathcal{R}_\nu + 15\right)}{5}\ ,\\
	        D^{(0)}_{\delta} &= 0\ ,       &\qquad\qquad  D^{(1)}_{\delta} &= 0\ .
	    \end{alignat}
	
	    \begin{align}
	        D^{(2)}_{\delta_\gamma} &= - \frac{H_{0} \Psi \sqrt{S_{\gamma \nu}} \mathcal{R}_b \beta_{0} k \left(\mathcal{R}_\nu - 3\right) \left(4 \mathcal{R}_\nu + 15\right)}{20 \left(\mathcal{R}_\nu - 1\right)}\ ,\\
	        D^{(2)}_{\delta_\nu} &= - \frac{H_{0} \Psi \sqrt{S_{\gamma \nu}} \mathcal{R}_b \beta_{0} k \left(4 \mathcal{R}_\nu + 15\right)}{20}\ ,\\
	        D^{(2)}_{\delta_b} &= - \frac{3 H_{0} \Psi \sqrt{S_{\gamma \nu}} \mathcal{R}_b \beta_{0} k \left(\mathcal{R}_\nu - 3\right) \left(4 \mathcal{R}_\nu + 15\right)}{80 \left(\mathcal{R}_\nu - 1\right)}\ ,\\
	        D^{(2)}_{\delta} &= \frac{H_{0} \Psi \sqrt{S_{\gamma \nu}} \mathcal{R}_b \beta_{0} k \left(4 \mathcal{R}_\nu + 15\right)}{20}\ .
	    \end{align}
	
	    \begin{align}
	        D^{(3)}_{\delta_\gamma} &= \frac{H_{0}^{2} \Psi S_{\gamma \nu} \mathcal{R}_b \beta_{0} k \left(4 \mathcal{R}_\nu + 15\right) \left(5 \mathcal{R}_\nu \mathcal{R}_{bc} \left(\mathcal{R}_\nu - 1\right) - 3 \mathcal{R}_b \left(\mathcal{R}_\nu - 6\right)\right)}{300 \left(\mathcal{R}_\nu - 1\right)^{2}} + \frac{4 \Psi \beta_{0} k^{3} \left(44 \mathcal{R}_\nu + 65\right)}{225 \left(4 \mathcal{R}_\nu + 5\right)}\ ,\\
	        D^{(3)}_{\delta_\nu} &= - \frac{H_{0}^{2} \Psi S_{\gamma \nu} \mathcal{R}_b \beta_{0} k \left(4 \mathcal{R}_\nu + 15\right) \left(3 \mathcal{R}_b - 5 \mathcal{R}_{bc} \left(\mathcal{R}_\nu - 1\right)\right)}{300 \left(\mathcal{R}_\nu - 1\right)} + \frac{2 \Psi \beta_{0} k^{3} \left(32 \mathcal{R}_\nu + 45\right)}{225 \mathcal{R}_\nu}\ ,\\
	        D^{(3)}_{\delta} &= \frac{3 H_{0}^{2} \Psi S_{\gamma \nu} \mathcal{R}_b \beta_{0} k \left(4 \mathcal{R}_\nu + 15\right) \left(2 \mathcal{R}_b - 5 \mathcal{R}_{bc} \left(\mathcal{R}_\nu - 1\right)\right)}{400 \left(\mathcal{R}_\nu - 1\right)} + \frac{\Psi \beta_{0} k^{3} \left(80 \mathcal{R}_\nu^{2} + 392 \mathcal{R}_\nu + 545\right)}{450 \left(4 \mathcal{R}_\nu + 5\right)}\ .
	    \end{align}
	
	    \begin{alignat}{2}
	        D^{(0)}_{\theta_\gamma} &= 0\ ,  &\qquad\qquad  D^{(1)}_{\theta_\gamma} &= 0\ ,\\
	        D^{(0)}_{\theta_\nu} &= \frac{\Psi \beta_{0} k \left(4 \mathcal{R}_\nu + 15\right)}{5 \mathcal{R}_\nu} \ ,&\qquad\qquad  D^{(1)}_{\theta_\nu} &= 0\ ,\\
	        D^{(0)}_{\theta} &= 0\ ,   &\qquad\qquad  D^{(1)}_{\theta} &= 0\ .
	    \end{alignat}
	
	    \begin{align}
	        D^{(2)}_{\theta_\gamma} &= - \frac{\Psi \beta_{0} k^{3}}{6}\ ,\\
	        D^{(2)}_{\theta_\nu} &= - \frac{\Psi \beta_{0} k^{3} \left(44 \mathcal{R}_\nu^{2} + 151 \mathcal{R}_\nu + 135\right)}{30 \mathcal{R}_\nu \left(4 \mathcal{R}_\nu + 5\right)}\ ,\\
	        D^{(2)}_{\theta} &= - \frac{\Psi \beta_{0} k^{3} \left(\mathcal{R}_\nu + 2\right) \left(4 \mathcal{R}_\nu + 15\right)}{15 \left(4 \mathcal{R}_\nu + 5\right)}\ .
	    \end{align}
	
	    \begin{align}
	        D^{(0)}_{\sigma_\nu} &= 0\ ,\\
	        D^{(1)}_{\sigma_\nu} &= \frac{2 \Psi \beta_{0} k \left(\mathcal{R}_\nu + 2\right) \left(4 \mathcal{R}_\nu + 15\right)}{15 \mathcal{R}_\nu \left(4 \mathcal{R}_\nu + 5\right)}\ ,\\
	        D^{(2)}_{\sigma_\nu} &= \frac{2 H_{0} \Psi \sqrt{S_{\gamma \nu}} \mathcal{R}_{bc} \beta_{0} k \left(\mathcal{R}_\nu + 2\right)}{5 \left(4 \mathcal{R}_\nu + 5\right)}\ ,\\
	        D^{(3)}_{\sigma_\nu} &= \frac{H_{0}^{2} \Psi S_{\gamma \nu} \mathcal{R}_{bc}^{2} \beta_{0} k \left(\mathcal{R}_\nu + 2\right) \left(4 \mathcal{R}_\nu - 45\right)}{30 \left(2 \mathcal{R}_\nu + 15\right) \left(4 \mathcal{R}_\nu + 5\right)} - \frac{\Psi \beta_{0} k^{3} \left(32 \mathcal{R}_\nu^{4} + 224 \mathcal{R}_\nu^{3} + 914 \mathcal{R}_\nu^{2} + 2097 \mathcal{R}_\nu + 1620\right)}{270 \mathcal{R}_\nu \left(2 \mathcal{R}_\nu + 15\right) \left(4 \mathcal{R}_\nu + 5\right)}\ .
	    \end{align}
	
	    \begin{alignat}{2}
	        D^{(0)}_{\eta} &= 0\ ,          &\qquad\qquad  D^{(1)}_{\eta} &= - \frac{2 \Psi \beta_{0} k \left(\mathcal{R}_\nu + 2\right) \left(4 \mathcal{R}_\nu + 15\right)}{15 \left(4 \mathcal{R}_\nu + 5\right)}\ ,\\
	        D^{(0)}_{h} &= D^{(0)}_{h}\ ,   &\qquad\qquad  D^{(1)}_{h} &= 0\ .
	    \end{alignat}
	
	    \begin{align}
	        D^{(2)}_{\eta} &= \frac{H_{0} \Psi \sqrt{S_{\gamma \nu}} \beta_{0} k \left(- \mathcal{R}_b \left(4 \mathcal{R}_\nu + 5\right) \left(4 \mathcal{R}_\nu + 15\right) + 40 \mathcal{R}_{bc} \left(\mathcal{R}_\nu + 2\right)\right)}{80 \left(4 \mathcal{R}_\nu + 5\right)}\ ,\\
	        D^{(2)}_{h} &= \frac{3 H_{0} \Psi \sqrt{S_{\gamma \nu}} \mathcal{R}_b \beta_{0} k \left(4 \mathcal{R}_\nu + 15\right)}{40}\ .
	    \end{align}
	    
	    \begin{align}
		    D^{(3)}_{\eta} &= - \frac{H_{0}^{2} \Psi S_{\gamma \nu} \mathcal{R}_b^{2} \beta_{0} k \left(4 \mathcal{R}_\nu + 15\right)}{400 \left(\mathcal{R}_\nu - 1\right)} + \frac{H_{0}^{2} \Psi S_{\gamma \nu} \mathcal{R}_b \mathcal{R}_{bc} \beta_{0} k \left(4 \mathcal{R}_\nu + 15\right)}{240}\nonumber\\
		        &\qquad + \frac{H_{0}^{2} \Psi S_{\gamma \nu} \mathcal{R}_{bc}^{2} \beta_{0} k \left(\mathcal{R}_\nu + 2\right) \left(4 \mathcal{R}_\nu - 45\right)}{24 \left(2 \mathcal{R}_\nu + 15\right) \left(4 \mathcal{R}_\nu + 5\right)} + \frac{\Psi \beta_{0} k^{3} \left(- 80 \mathcal{R}_\nu^{3} + 568 \mathcal{R}_\nu^{2} + 4525 \mathcal{R}_\nu + 4950\right)}{1350 \left(2 \mathcal{R}_\nu + 15\right) \left(4 \mathcal{R}_\nu + 5\right)}\ ,\\
		    D^{(3)}_{h} &= \frac{H_{0}^{2} \Psi S_{\gamma \nu} \mathcal{R}_b \beta_{0} k \left(4 \mathcal{R}_\nu + 15\right) \left(3 \mathcal{R}_b - 5 \mathcal{R}_{bc} \left(\mathcal{R}_\nu - 1\right)\right)}{200 \left(\mathcal{R}_\nu - 1\right)} - \frac{\Psi \beta_{0} k^{3} \left(80 \mathcal{R}_\nu^{2} + 528 \mathcal{R}_\nu + 655\right)}{450 \left(4 \mathcal{R}_\nu + 5\right)}\ .
		\end{align}
	    
	    Once we have the new behaviour of the system, we need to evaluate the
	    assignment of initial conditions. It seems reasonable to give zero initial values to our modification
	    but there is one further physical requirement that we must impose. As mentioned in the main text, 
	    if neutrinos and photons were in thermal contact in the primeval Universe it is physically sensible
	    to impose that they shared the same velocity
	    \begin{equation}
	        \theta_\nu(\tau=0) = \theta_\gamma(\tau=0) = \theta_\gamma^{(0)}\ ,
	    \end{equation}
	    In the standard scenario this leads to $\theta_\gamma^{(0)}=0$ and to the absence of neutrino
	    velocity isocurvature modes. However, in our case, if we consider a neutrino isocurvature velocity
	    mode on top of the sourced mode, upon imposing this restriction we get
	    \begin{equation}
	        \theta_\gamma^{(0)} = \frac{4\mathcal{R}_\nu+15}{5}\Psi\beta_0 k\ .
	    \end{equation}
	    In order to obtain the correct initial conditions, we must consider the combination of the sourced
	    mode with a neutrino isocurvature velocity mode with the previous initial condition. The final 
	    results are
	    \begin{align}
	        \delta_\gamma^I &= \delta_\nu^I = \delta_b^I = \delta^I = 0 + \Od{\tau^3}\ ,\\
	        \theta_\gamma^I &= \frac{\Psi \beta_{0} k \left(4 \mathcal{R}_\nu + 15\right)}{5} 
	                + \frac{3 H_{0} \Psi \sqrt{S_{\gamma \nu}} \mathcal{R}_b \beta_{0} k \left(4 \mathcal{R}_\nu + 15\right)}{20 \left(\mathcal{R}_\nu - 1\right)}\tau\nonumber\\
	            &\quad+ \frac{3 H_{0}^{2} \Psi S_{\gamma \nu} \mathcal{R}_b \beta_{0} k \left(4 \mathcal{R}_\nu + 15\right) \left((\mathcal{R}_\nu-1) \mathcal{R}_{bc} + 3 \mathcal{R}_b\right)}{80 \left(\mathcal{R}_\nu - 1\right)^{2}} \tau^2
	                - \frac{2 \Psi \beta_{0} k^{3} \left(\mathcal{R}_\nu + 5\right)}{15}\tau^2 + \Od{\tau^3}\ ,\\
	        \theta_\nu^I &= \frac{\Psi \beta_{0} k \left(4 \mathcal{R}_\nu + 15\right)}{5}
	                - \frac{\Psi \beta_{0} k^{3} \left(8 \mathcal{R}_\nu^{2} + 62 \mathcal{R}_\nu + 95\right)}{15 \left(4 \mathcal{R}_\nu + 5\right)}\tau^2 + \Od{\tau^3}\ ,\\
	        \theta^I &= - \frac{\Psi \mathcal{R}_\nu \beta_{0} k^{3} \left(4 \mathcal{R}_\nu + 15\right)}{5 \left(4 \mathcal{R}_\nu + 5\right)}\tau^2 + \Od{\tau^3}\ ,\\
	        \sigma_\nu^I &= \frac{2 \Psi \beta_{0} k \left(4 \mathcal{R}_\nu + 15\right)}{5 \left(4 \mathcal{R}_\nu + 5\right)}\tau
	                + \frac{6 H_{0} \Psi \sqrt{S_{\gamma \nu}} \mathcal{R}_\nu \mathcal{R}_{bc} \beta_{0} k}{5 \left(4 \mathcal{R}_\nu + 5\right)}\tau^2 + \Od{\tau^3}\ ,\\
	        h^I &= 0 + \Od{\tau^3}\ ,\\
	        \eta^I &= - \frac{2 \Psi \mathcal{R}_\nu \beta_{0} k \left(4 \mathcal{R}_\nu + 15\right)}{5 \left(4 \mathcal{R}_\nu + 5\right)}\tau
	                + \frac{3 H_{0} \Psi \sqrt{S_{\gamma \nu}} \mathcal{R}_\nu \mathcal{R}_{bc} \beta_{0} k}{2 \left(4 \mathcal{R}_\nu + 5\right)}\tau^2 + \Od{\tau^3}\ .
	    \end{align}
    
    \subsection{Vector modes}
        Considering adiabatic perturbations in the scalar contributions, during TC and deep in the
        radiation era, the super-Hubble evolution is
        \begin{alignat}{2}
            D^{(0)}_{\chi_\gamma} &= D^{(0)}_{\chi \gamma}\ ,                                          &\qquad D^{(1)}_{\chi_\gamma} &= \frac{3 D^{(0)}_{\chi \gamma} H_{0} \sqrt{S_{\gamma \nu}} \mathcal{R}_b}{4 \left(\mathcal{R}_\nu - 1\right)}\ , \\
            D^{(0)}_{\chi_\nu} &= - \frac{\Psi \beta_{0} \left(4 \mathcal{R}_\nu + 15\right)}{10}\ ,   &\qquad D^{(1)}_{\chi_\nu} &= 0\ , \\
            D^{(0)}_{\chi} &= 0\ ,                                                                     &\qquad D^{(1)}_{\chi} &= 0\ ,
        \end{alignat}
        \begin{align}
            D^{(2)}_{\chi_\gamma} &= \frac{3 D^{(0)}_{\chi \gamma} H_{0}^{2} S_{\gamma \nu} \mathcal{R}_b \left(3 \mathcal{R}_b + \mathcal{R}_{bc} \left(\mathcal{R}_\nu - 1\right)\right)}{16 \left(\mathcal{R}_\nu - 1\right)^{2}} - \frac{\Psi \beta_{0} k^{2} \left(8 \mathcal{R}_\nu + 25\right)}{60}\ ,\\
            D^{(2)}_{\chi_\nu} &= - \frac{\Psi \beta_{0} k^{2} \left(8 \mathcal{R}_\nu + 25\right)}{60}\ ,\\
            D^{(2)}_{\chi} &= 0\ ,
        \end{align}
        \begin{align}
            D^{(3)}_{\chi} &= 0\ ,\\
            D^{(4)}_{\chi} &= \frac{5 \Psi \beta_{0} k^{4}}{8 \left(8 \mathcal{R}_\nu + 45\right)}\ ,
        \end{align}
        \begin{align}
            D^{(0)}_{\pi_\nu} &= 0\ ,\\
            D^{(1)}_{\pi_\nu} &= 0\ ,\\
            D^{(2)}_{\pi_\nu} &= 0\ ,\\
            D^{(3)}_{\pi_\nu} &= \frac{\Psi \beta_{0} k^{2} \left(32 \mathcal{R}_\nu^{2} + 268 \mathcal{R}_\nu + 375\right)}{270 \left(8 \mathcal{R}_\nu + 45\right)}\ ,
        \end{align}
        \begin{align}
            D^{(0)}_{S+\dot{F}} &= 0\ ,\\
            D^{(1)}_{S+\dot{F}} &= 0\ ,\\
            D^{(2)}_{S+\dot{F}} &= \frac{5 \Psi \beta_{0} k^{2}}{8 \mathcal{R}_\nu + 45}\ ,
        \end{align}
        Again, imposing the physical requirement that photons and neutrinos had the same velocity 
        in the very early Universe, we are led to
        \begin{equation}
            D^{(0)}_{\chi_\gamma} = - \frac{\Psi \beta_{0} \left(4 \mathcal{R}_\nu + 15\right)}{10}\ .
        \end{equation}
        The initial conditions provided for the numerical integration are
        \begin{align}
            \chi_\gamma &= - \frac{\Psi \beta_{0} \left(4 \mathcal{R}_\nu + 15\right)}{10}
                 - \frac{3 H_{0} \Psi \sqrt{S_{\gamma \nu}} \mathcal{R}_b \beta_{0} \left(4 \mathcal{R}_\nu + 15\right)}{40 \left(\mathcal{R}_\nu - 1\right)}\tau\ ,\nonumber\\
                 &\qquad- \frac{3 H_{0}^{2} \Psi S_{\gamma \nu} \mathcal{R}_b \beta_{0} \left(4 \mathcal{R}_\nu + 15\right) \left(\mathcal{R}_\nu \mathcal{R}_{bc} + 3 \mathcal{R}_b - \mathcal{R}_{bc}\right)}{160 \left(\mathcal{R}_\nu - 1\right)^{2}}\tau^2 - \frac{\Psi \beta_{0} k^{2} \left(8 \mathcal{R}_\nu + 25\right)}{60}\tau^2 +\Od{\tau^3}\ ,\\
            \chi_\nu &= - \frac{\Psi \beta_{0} \left(4 \mathcal{R}_\nu + 15\right)}{10} - \frac{\Psi \beta_{0} k^{2} \left(8 \mathcal{R}_\nu + 25\right)}{60}\tau^2 + \Od{\tau^3}\ ,\\
            \chi &= \frac{5 \Psi \beta_{0} k^{4}}{8 \left(8 \mathcal{R}_\nu + 45\right)}\tau^4 + \Od{\tau^5}\ ,\\
            \pi^\text{V}_\nu &= \frac{\Psi \beta_{0} k^{2} \left(4 \mathcal{R}_\nu \left(8 \mathcal{R}_\nu + 67\right) + 375\right)}{270 \left(8 \mathcal{R}_\nu + 45\right)}\tau^3 + \Od{\tau^4}\ ,\\
            S+\dot{F} &= \frac{5 \Psi \beta_{0} k^{2}}{8 \mathcal{R}_\nu + 45}\tau^2 +  \Od{\tau^3}\ .\\
        \end{align}        
            
\bibliography{MovingBiblio}

\begin{thebibliography}{66}
\expandafter\ifx\csname natexlab\endcsname\relax\def\natexlab#1{#1}\fi
\expandafter\ifx\csname bibnamefont\endcsname\relax
  \def\bibnamefont#1{#1}\fi
\expandafter\ifx\csname bibfnamefont\endcsname\relax
  \def\bibfnamefont#1{#1}\fi
\expandafter\ifx\csname citenamefont\endcsname\relax
  \def\citenamefont#1{#1}\fi
\expandafter\ifx\csname url\endcsname\relax
  \def\url#1{\texttt{#1}}\fi
\expandafter\ifx\csname urlprefix\endcsname\relax\def\urlprefix{URL }\fi
\providecommand{\bibinfo}[2]{#2}
\providecommand{\eprint}[2][]{\url{#2}}

\bibitem[{\citenamefont{Ade et~al.}(2016)}]{Ade:2015hxq}
\bibinfo{author}{\bibfnamefont{P.~A.~R.} \bibnamefont{Ade}}
  \bibnamefont{et~al.} (\bibinfo{collaboration}{Planck}),
  \bibinfo{journal}{Astron. Astrophys.} \textbf{\bibinfo{volume}{594}},
  \bibinfo{pages}{A16} (\bibinfo{year}{2016}), \eprint{1506.07135}.

\bibitem[{\citenamefont{Akrami et~al.}(2018)}]{Akrami:2018vks}
\bibinfo{author}{\bibfnamefont{Y.}~\bibnamefont{Akrami}} \bibnamefont{et~al.}
  (\bibinfo{collaboration}{Planck}) (\bibinfo{year}{2018}),
  \eprint{1807.06205}.

\bibitem[{\citenamefont{Kolatt and Lahav}(2001)}]{Kolatt:2000yg}
\bibinfo{author}{\bibfnamefont{T.~S.} \bibnamefont{Kolatt}} \bibnamefont{and}
  \bibinfo{author}{\bibfnamefont{O.}~\bibnamefont{Lahav}},
  \bibinfo{journal}{Mon. Not. Roy. Astron. Soc.}
  \textbf{\bibinfo{volume}{323}}, \bibinfo{pages}{859} (\bibinfo{year}{2001}),
  \eprint{astro-ph/0008041}.

\bibitem[{\citenamefont{Antoniou and Perivolaropoulos}(2010)}]{Antoniou:2010gw}
\bibinfo{author}{\bibfnamefont{I.}~\bibnamefont{Antoniou}} \bibnamefont{and}
  \bibinfo{author}{\bibfnamefont{L.}~\bibnamefont{Perivolaropoulos}},
  \bibinfo{journal}{JCAP} \textbf{\bibinfo{volume}{1012}}, \bibinfo{pages}{012}
  (\bibinfo{year}{2010}), \eprint{1007.4347}.

\bibitem[{\citenamefont{Beltran~Jimenez
  et~al.}(2015)\citenamefont{Beltran~Jimenez, Salzano, and
  Lazkoz}}]{Jimenez:2014jma}
\bibinfo{author}{\bibfnamefont{J.}~\bibnamefont{Beltran~Jimenez}},
  \bibinfo{author}{\bibfnamefont{V.}~\bibnamefont{Salzano}}, \bibnamefont{and}
  \bibinfo{author}{\bibfnamefont{R.}~\bibnamefont{Lazkoz}},
  \bibinfo{journal}{Phys. Lett.} \textbf{\bibinfo{volume}{B741}},
  \bibinfo{pages}{168} (\bibinfo{year}{2015}), \eprint{1402.1760}.

\bibitem[{\citenamefont{Kogut et~al.}(1993)}]{Kogut:1993ag}
\bibinfo{author}{\bibfnamefont{A.}~\bibnamefont{Kogut}} \bibnamefont{et~al.},
  \bibinfo{journal}{Astrophys. J.} \textbf{\bibinfo{volume}{419}},
  \bibinfo{pages}{1} (\bibinfo{year}{1993}), \eprint{astro-ph/9312056}.

\bibitem[{\citenamefont{Lineweaver et~al.}(1996)\citenamefont{Lineweaver,
  Tenorio, Smoot, Keegstra, Banday, and Lubin}}]{Lineweaver:1996xa}
\bibinfo{author}{\bibfnamefont{C.~H.} \bibnamefont{Lineweaver}},
  \bibinfo{author}{\bibfnamefont{L.}~\bibnamefont{Tenorio}},
  \bibinfo{author}{\bibfnamefont{G.~F.} \bibnamefont{Smoot}},
  \bibinfo{author}{\bibfnamefont{P.}~\bibnamefont{Keegstra}},
  \bibinfo{author}{\bibfnamefont{A.~J.} \bibnamefont{Banday}},
  \bibnamefont{and} \bibinfo{author}{\bibfnamefont{P.}~\bibnamefont{Lubin}},
  \bibinfo{journal}{Astrophys. J.} \textbf{\bibinfo{volume}{470}},
  \bibinfo{pages}{38} (\bibinfo{year}{1996}), \eprint{astro-ph/9601151}.

\bibitem[{\citenamefont{Aghanim et~al.}(2014)}]{Aghanim:2013suk}
\bibinfo{author}{\bibfnamefont{N.}~\bibnamefont{Aghanim}} \bibnamefont{et~al.}
  (\bibinfo{collaboration}{Planck}), \bibinfo{journal}{Astron. Astrophys.}
  \textbf{\bibinfo{volume}{571}}, \bibinfo{pages}{A27} (\bibinfo{year}{2014}),
  \eprint{1303.5087}.

\bibitem[{\citenamefont{Yasini and Pierpaoli}(2017)}]{Yasini:2016dnd}
\bibinfo{author}{\bibfnamefont{S.}~\bibnamefont{Yasini}} \bibnamefont{and}
  \bibinfo{author}{\bibfnamefont{E.}~\bibnamefont{Pierpaoli}},
  \bibinfo{journal}{Phys. Rev. Lett.} \textbf{\bibinfo{volume}{119}},
  \bibinfo{pages}{221102} (\bibinfo{year}{2017}), \eprint{1610.00015}.

\bibitem[{\citenamefont{Ellis and Baldwin}(1984)}]{ellis1984expected}
\bibinfo{author}{\bibfnamefont{G.}~\bibnamefont{Ellis}} \bibnamefont{and}
  \bibinfo{author}{\bibfnamefont{J.}~\bibnamefont{Baldwin}},
  \bibinfo{journal}{Monthly Notices of the Royal Astronomical Society}
  \textbf{\bibinfo{volume}{206}}, \bibinfo{pages}{377} (\bibinfo{year}{1984}).

\bibitem[{\citenamefont{Gibelyou and Huterer}(2012)}]{Gibelyou:2012ri}
\bibinfo{author}{\bibfnamefont{C.}~\bibnamefont{Gibelyou}} \bibnamefont{and}
  \bibinfo{author}{\bibfnamefont{D.}~\bibnamefont{Huterer}},
  \bibinfo{journal}{Mon. Not. Roy. Astron. Soc.}
  \textbf{\bibinfo{volume}{427}}, \bibinfo{pages}{1994} (\bibinfo{year}{2012}),
  \eprint{1205.6476}.

\bibitem[{\citenamefont{Condon et~al.}(1998)\citenamefont{Condon, Cotton,
  Greisen, Yin, Perley, Taylor, and Broderick}}]{Condon:1998iy}
\bibinfo{author}{\bibfnamefont{J.~J.} \bibnamefont{Condon}},
  \bibinfo{author}{\bibfnamefont{W.~D.} \bibnamefont{Cotton}},
  \bibinfo{author}{\bibfnamefont{E.~W.} \bibnamefont{Greisen}},
  \bibinfo{author}{\bibfnamefont{Q.~F.} \bibnamefont{Yin}},
  \bibinfo{author}{\bibfnamefont{R.~A.} \bibnamefont{Perley}},
  \bibinfo{author}{\bibfnamefont{G.~B.} \bibnamefont{Taylor}},
  \bibnamefont{and} \bibinfo{author}{\bibfnamefont{J.~J.}
  \bibnamefont{Broderick}}, \bibinfo{journal}{Astron. J.}
  \textbf{\bibinfo{volume}{115}}, \bibinfo{pages}{1693} (\bibinfo{year}{1998}).

\bibitem[{\citenamefont{Itoh et~al.}(2010)\citenamefont{Itoh, Yahata, and
  Takada}}]{Itoh:2009vc}
\bibinfo{author}{\bibfnamefont{Y.}~\bibnamefont{Itoh}},
  \bibinfo{author}{\bibfnamefont{K.}~\bibnamefont{Yahata}}, \bibnamefont{and}
  \bibinfo{author}{\bibfnamefont{M.}~\bibnamefont{Takada}},
  \bibinfo{journal}{Phys. Rev.} \textbf{\bibinfo{volume}{D82}},
  \bibinfo{pages}{043530} (\bibinfo{year}{2010}), \eprint{0912.1460}.

\bibitem[{\citenamefont{Amendola et~al.}(2018)}]{Amendola:2016saw}
\bibinfo{author}{\bibfnamefont{L.}~\bibnamefont{Amendola}}
  \bibnamefont{et~al.}, \bibinfo{journal}{Living Rev. Rel.}
  \textbf{\bibinfo{volume}{21}}, \bibinfo{pages}{2} (\bibinfo{year}{2018}),
  \eprint{1606.00180}.

\bibitem[{\citenamefont{Maartens et~al.}(2015)\citenamefont{Maartens, Abdalla,
  Jarvis, and Santos}}]{Maartens:2015mra}
\bibinfo{author}{\bibfnamefont{R.}~\bibnamefont{Maartens}},
  \bibinfo{author}{\bibfnamefont{F.~B.} \bibnamefont{Abdalla}},
  \bibinfo{author}{\bibfnamefont{M.}~\bibnamefont{Jarvis}}, \bibnamefont{and}
  \bibinfo{author}{\bibfnamefont{M.~G.} \bibnamefont{Santos}}
  (\bibinfo{collaboration}{SKA Cosmology SWG}), \bibinfo{journal}{PoS}
  \textbf{\bibinfo{volume}{AASKA14}}, \bibinfo{pages}{016}
  (\bibinfo{year}{2015}), \eprint{1501.04076}.

\bibitem[{\citenamefont{King and Ellis}(1973)}]{King:1972td}
\bibinfo{author}{\bibfnamefont{A.~R.} \bibnamefont{King}} \bibnamefont{and}
  \bibinfo{author}{\bibfnamefont{G.~F.~R.} \bibnamefont{Ellis}},
  \bibinfo{journal}{Commun. Math. Phys.} \textbf{\bibinfo{volume}{31}},
  \bibinfo{pages}{209} (\bibinfo{year}{1973}).

\bibitem[{\citenamefont{Coley and Tupper}(1986)}]{coley1986two}
\bibinfo{author}{\bibfnamefont{A.}~\bibnamefont{Coley}} \bibnamefont{and}
  \bibinfo{author}{\bibfnamefont{B.}~\bibnamefont{Tupper}},
  \bibinfo{journal}{Journal of Mathematical Physics}
  \textbf{\bibinfo{volume}{27}}, \bibinfo{pages}{406} (\bibinfo{year}{1986}).

\bibitem[{\citenamefont{Turner}(1991)}]{Turner:1991dn}
\bibinfo{author}{\bibfnamefont{M.~S.} \bibnamefont{Turner}},
  \bibinfo{journal}{Phys. Rev.} \textbf{\bibinfo{volume}{D44}},
  \bibinfo{pages}{3737} (\bibinfo{year}{1991}).

\bibitem[{\citenamefont{Maroto}(2006)}]{Maroto:2005kc}
\bibinfo{author}{\bibfnamefont{A.~L.} \bibnamefont{Maroto}},
  \bibinfo{journal}{JCAP} \textbf{\bibinfo{volume}{0605}}, \bibinfo{pages}{015}
  (\bibinfo{year}{2006}), \eprint{astro-ph/0512464}.

\bibitem[{\citenamefont{Beltran~Jimenez and
  Maroto}(2007)}]{BeltranJimenez:2007rsj}
\bibinfo{author}{\bibfnamefont{J.}~\bibnamefont{Beltran~Jimenez}}
  \bibnamefont{and} \bibinfo{author}{\bibfnamefont{A.~L.}
  \bibnamefont{Maroto}}, \bibinfo{journal}{Phys. Rev.}
  \textbf{\bibinfo{volume}{D76}}, \bibinfo{pages}{023003}
  (\bibinfo{year}{2007}), \eprint{astro-ph/0703483}.

\bibitem[{\citenamefont{Harko and Lobo}(2013)}]{Harko:2013wsa}
\bibinfo{author}{\bibfnamefont{T.}~\bibnamefont{Harko}} \bibnamefont{and}
  \bibinfo{author}{\bibfnamefont{F.~S.~N.} \bibnamefont{Lobo}},
  \bibinfo{journal}{JCAP} \textbf{\bibinfo{volume}{1307}}, \bibinfo{pages}{036}
  (\bibinfo{year}{2013}), \eprint{1304.0757}.

\bibitem[{\citenamefont{Kashlinsky et~al.}(2009)\citenamefont{Kashlinsky,
  Atrio-Barandela, Kocevski, and Ebeling}}]{Kashlinsky:2008ut}
\bibinfo{author}{\bibfnamefont{A.}~\bibnamefont{Kashlinsky}},
  \bibinfo{author}{\bibfnamefont{F.}~\bibnamefont{Atrio-Barandela}},
  \bibinfo{author}{\bibfnamefont{D.}~\bibnamefont{Kocevski}}, \bibnamefont{and}
  \bibinfo{author}{\bibfnamefont{H.}~\bibnamefont{Ebeling}},
  \bibinfo{journal}{Astrophys. J.} \textbf{\bibinfo{volume}{686}},
  \bibinfo{pages}{L49} (\bibinfo{year}{2009}), \eprint{0809.3734}.

\bibitem[{\citenamefont{Ade et~al.}(2014)}]{Ade:2013opi}
\bibinfo{author}{\bibfnamefont{P.~A.~R.} \bibnamefont{Ade}}
  \bibnamefont{et~al.} (\bibinfo{collaboration}{Planck}),
  \bibinfo{journal}{Astron. Astrophys.} \textbf{\bibinfo{volume}{561}},
  \bibinfo{pages}{A97} (\bibinfo{year}{2014}), \eprint{1303.5090}.

\bibitem[{\citenamefont{Scrimgeour et~al.}(2016)}]{Scrimgeour:2015khj}
\bibinfo{author}{\bibfnamefont{M.~I.} \bibnamefont{Scrimgeour}}
  \bibnamefont{et~al.}, \bibinfo{journal}{Mon. Not. Roy. Astron. Soc.}
  \textbf{\bibinfo{volume}{455}}, \bibinfo{pages}{386} (\bibinfo{year}{2016}),
  \eprint{1511.06930}.

\bibitem[{\citenamefont{Schwarz et~al.}(2016)\citenamefont{Schwarz, Copi,
  Huterer, and Starkman}}]{Schwarz:2015cma}
\bibinfo{author}{\bibfnamefont{D.~J.} \bibnamefont{Schwarz}},
  \bibinfo{author}{\bibfnamefont{C.~J.} \bibnamefont{Copi}},
  \bibinfo{author}{\bibfnamefont{D.}~\bibnamefont{Huterer}}, \bibnamefont{and}
  \bibinfo{author}{\bibfnamefont{G.~D.} \bibnamefont{Starkman}},
  \bibinfo{journal}{Class. Quant. Grav.} \textbf{\bibinfo{volume}{33}},
  \bibinfo{pages}{184001} (\bibinfo{year}{2016}), \eprint{1510.07929}.

\bibitem[{\citenamefont{Cembranos
  et~al.}(2019{\natexlab{a}})\citenamefont{Cembranos, Maroto, and
  Villarrubia-Rojo}}]{CMB_paper}
\bibinfo{author}{\bibfnamefont{J.~A.~R.} \bibnamefont{Cembranos}},
  \bibinfo{author}{\bibfnamefont{A.~L.} \bibnamefont{Maroto}},
  \bibnamefont{and}
  \bibinfo{author}{\bibfnamefont{H.}~\bibnamefont{Villarrubia-Rojo}}
  (\bibinfo{year}{2019}{\natexlab{a}}), \eprint{work in preparation}.

\bibitem[{\citenamefont{Ma and Bertschinger}(1995)}]{Ma:1995ey}
\bibinfo{author}{\bibfnamefont{C.-P.} \bibnamefont{Ma}} \bibnamefont{and}
  \bibinfo{author}{\bibfnamefont{E.}~\bibnamefont{Bertschinger}},
  \bibinfo{journal}{Astrophys. J.} \textbf{\bibinfo{volume}{455}},
  \bibinfo{pages}{7} (\bibinfo{year}{1995}), \eprint{astro-ph/9506072}.

\bibitem[{\citenamefont{Bernstein}(1988)}]{Bernstein:1988bw}
\bibinfo{author}{\bibfnamefont{J.}~\bibnamefont{Bernstein}},
  \emph{\bibinfo{title}{{Kinetic Theory in the Expanding Universe}}}
  (\bibinfo{publisher}{Cambridge University Press},
  \bibinfo{address}{Cambridge, U.K.}, \bibinfo{year}{1988}).

\bibitem[{\citenamefont{Dodelson}(2003)}]{Dodelson:2003ft}
\bibinfo{author}{\bibfnamefont{S.}~\bibnamefont{Dodelson}},
  \emph{\bibinfo{title}{{Modern Cosmology}}} (\bibinfo{publisher}{Academic
  Press}, \bibinfo{address}{Amsterdam}, \bibinfo{year}{2003}).

\bibitem[{\citenamefont{Kodama and Sasaki}(1984)}]{Kodama:1985bj}
\bibinfo{author}{\bibfnamefont{H.}~\bibnamefont{Kodama}} \bibnamefont{and}
  \bibinfo{author}{\bibfnamefont{M.}~\bibnamefont{Sasaki}},
  \bibinfo{journal}{Prog. Theor. Phys. Suppl.} \textbf{\bibinfo{volume}{78}},
  \bibinfo{pages}{1} (\bibinfo{year}{1984}).

\bibitem[{\citenamefont{Mukhanov et~al.}(1992)\citenamefont{Mukhanov, Feldman,
  and Brandenberger}}]{Mukhanov:1990me}
\bibinfo{author}{\bibfnamefont{V.~F.} \bibnamefont{Mukhanov}},
  \bibinfo{author}{\bibfnamefont{H.~A.} \bibnamefont{Feldman}},
  \bibnamefont{and} \bibinfo{author}{\bibfnamefont{R.~H.}
  \bibnamefont{Brandenberger}}, \bibinfo{journal}{Phys. Rept.}
  \textbf{\bibinfo{volume}{215}}, \bibinfo{pages}{203} (\bibinfo{year}{1992}).

\bibitem[{\citenamefont{Durrer}(2008)}]{Durrer:2008eom}
\bibinfo{author}{\bibfnamefont{R.}~\bibnamefont{Durrer}},
  \emph{\bibinfo{title}{{The Cosmic Microwave Background}}}
  (\bibinfo{publisher}{Cambridge University Press},
  \bibinfo{address}{Cambridge}, \bibinfo{year}{2008}).

\bibitem[{\citenamefont{Arfken and Weber}(1999)}]{arfken1999mathematical}
\bibinfo{author}{\bibfnamefont{G.~B.} \bibnamefont{Arfken}} \bibnamefont{and}
  \bibinfo{author}{\bibfnamefont{H.~J.} \bibnamefont{Weber}},
  \emph{\bibinfo{title}{Mathematical methods for physicists}}
  (\bibinfo{year}{1999}).

\bibitem[{\citenamefont{Seljak and Zaldarriaga}(1996)}]{Seljak:1996is}
\bibinfo{author}{\bibfnamefont{U.}~\bibnamefont{Seljak}} \bibnamefont{and}
  \bibinfo{author}{\bibfnamefont{M.}~\bibnamefont{Zaldarriaga}},
  \bibinfo{journal}{Astrophys. J.} \textbf{\bibinfo{volume}{469}},
  \bibinfo{pages}{437} (\bibinfo{year}{1996}), \eprint{astro-ph/9603033}.

\bibitem[{\citenamefont{Lewis and Bridle}(2002)}]{Lewis:2002ah}
\bibinfo{author}{\bibfnamefont{A.}~\bibnamefont{Lewis}} \bibnamefont{and}
  \bibinfo{author}{\bibfnamefont{S.}~\bibnamefont{Bridle}},
  \bibinfo{journal}{Phys. Rev.} \textbf{\bibinfo{volume}{D66}},
  \bibinfo{pages}{103511} (\bibinfo{year}{2002}), \eprint{astro-ph/0205436}.

\bibitem[{\citenamefont{Blas et~al.}(2011)\citenamefont{Blas, Lesgourgues, and
  Tram}}]{Blas:2011rf}
\bibinfo{author}{\bibfnamefont{D.}~\bibnamefont{Blas}},
  \bibinfo{author}{\bibfnamefont{J.}~\bibnamefont{Lesgourgues}},
  \bibnamefont{and} \bibinfo{author}{\bibfnamefont{T.}~\bibnamefont{Tram}},
  \bibinfo{journal}{JCAP} \textbf{\bibinfo{volume}{1107}}, \bibinfo{pages}{034}
  (\bibinfo{year}{2011}), \eprint{1104.2933}.

\bibitem[{\citenamefont{Hu et~al.}(1995)\citenamefont{Hu, Scott, Sugiyama, and
  White}}]{Hu:1995fqa}
\bibinfo{author}{\bibfnamefont{W.}~\bibnamefont{Hu}},
  \bibinfo{author}{\bibfnamefont{D.}~\bibnamefont{Scott}},
  \bibinfo{author}{\bibfnamefont{N.}~\bibnamefont{Sugiyama}}, \bibnamefont{and}
  \bibinfo{author}{\bibfnamefont{M.~J.} \bibnamefont{White}},
  \bibinfo{journal}{Phys. Rev.} \textbf{\bibinfo{volume}{D52}},
  \bibinfo{pages}{5498} (\bibinfo{year}{1995}), \eprint{astro-ph/9505043}.

\bibitem[{\citenamefont{Hu}(1998)}]{Hu:1998kj}
\bibinfo{author}{\bibfnamefont{W.}~\bibnamefont{Hu}},
  \bibinfo{journal}{Astrophys. J.} \textbf{\bibinfo{volume}{506}},
  \bibinfo{pages}{485} (\bibinfo{year}{1998}), \eprint{astro-ph/9801234}.

\bibitem[{\citenamefont{Liddle and Lyth}(2000)}]{Liddle:2000cg}
\bibinfo{author}{\bibfnamefont{A.~R.} \bibnamefont{Liddle}} \bibnamefont{and}
  \bibinfo{author}{\bibfnamefont{D.~H.} \bibnamefont{Lyth}},
  \emph{\bibinfo{title}{{Cosmological inflation and large scale structure}}}
  (\bibinfo{publisher}{Cambridge, UK: Univ. Pr. (2000) 400 p},
  \bibinfo{year}{2000}).

\bibitem[{\citenamefont{Bucher et~al.}(2000)\citenamefont{Bucher, Moodley, and
  Turok}}]{Bucher:1999re}
\bibinfo{author}{\bibfnamefont{M.}~\bibnamefont{Bucher}},
  \bibinfo{author}{\bibfnamefont{K.}~\bibnamefont{Moodley}}, \bibnamefont{and}
  \bibinfo{author}{\bibfnamefont{N.}~\bibnamefont{Turok}},
  \bibinfo{journal}{Phys. Rev.} \textbf{\bibinfo{volume}{D62}},
  \bibinfo{pages}{083508} (\bibinfo{year}{2000}), \eprint{astro-ph/9904231}.

\bibitem[{\citenamefont{Naoz and Barkana}(2005)}]{Naoz:2005pd}
\bibinfo{author}{\bibfnamefont{S.}~\bibnamefont{Naoz}} \bibnamefont{and}
  \bibinfo{author}{\bibfnamefont{R.}~\bibnamefont{Barkana}},
  \bibinfo{journal}{Mon. Not. Roy. Astron. Soc.}
  \textbf{\bibinfo{volume}{362}}, \bibinfo{pages}{1047} (\bibinfo{year}{2005}),
  \eprint{astro-ph/0503196}.

\bibitem[{\citenamefont{Lewis}(2007)}]{Lewis:2007zh}
\bibinfo{author}{\bibfnamefont{A.}~\bibnamefont{Lewis}},
  \bibinfo{journal}{Phys. Rev.} \textbf{\bibinfo{volume}{D76}},
  \bibinfo{pages}{063001} (\bibinfo{year}{2007}), \eprint{0707.2727}.

\bibitem[{\citenamefont{Seager et~al.}(2000)\citenamefont{Seager, Sasselov, and
  Scott}}]{Seager:1999km}
\bibinfo{author}{\bibfnamefont{S.}~\bibnamefont{Seager}},
  \bibinfo{author}{\bibfnamefont{D.~D.} \bibnamefont{Sasselov}},
  \bibnamefont{and} \bibinfo{author}{\bibfnamefont{D.}~\bibnamefont{Scott}},
  \bibinfo{journal}{Astrophys. J. Suppl.} \textbf{\bibinfo{volume}{128}},
  \bibinfo{pages}{407} (\bibinfo{year}{2000}), \eprint{astro-ph/9912182}.

\bibitem[{\citenamefont{Kolb and Turner}(1990)}]{Kolb:1990vq}
\bibinfo{author}{\bibfnamefont{E.~W.} \bibnamefont{Kolb}} \bibnamefont{and}
  \bibinfo{author}{\bibfnamefont{M.~S.} \bibnamefont{Turner}},
  \bibinfo{journal}{Front. Phys.} \textbf{\bibinfo{volume}{69}},
  \bibinfo{pages}{1} (\bibinfo{year}{1990}).

\bibitem[{\citenamefont{Hu and Sugiyama}(1996)}]{Hu:1995en}
\bibinfo{author}{\bibfnamefont{W.}~\bibnamefont{Hu}} \bibnamefont{and}
  \bibinfo{author}{\bibfnamefont{N.}~\bibnamefont{Sugiyama}},
  \bibinfo{journal}{Astrophys. J.} \textbf{\bibinfo{volume}{471}},
  \bibinfo{pages}{542} (\bibinfo{year}{1996}), \eprint{astro-ph/9510117}.

\bibitem[{\citenamefont{Pant et~al.}(2018)\citenamefont{Pant, Rotti, Bengaly,
  and Maartens}}]{Pant:2018smd}
\bibinfo{author}{\bibfnamefont{N.}~\bibnamefont{Pant}},
  \bibinfo{author}{\bibfnamefont{A.}~\bibnamefont{Rotti}},
  \bibinfo{author}{\bibfnamefont{C.~A.~P.} \bibnamefont{Bengaly}},
  \bibnamefont{and} \bibinfo{author}{\bibfnamefont{R.}~\bibnamefont{Maartens}}
  (\bibinfo{year}{2018}), \eprint{1808.09743}.

\bibitem[{\citenamefont{Tully et~al.}(2008)\citenamefont{Tully, Shaya,
  Karachentsev, Courtois, Kocevski, Rizzi, and Peel}}]{Tully:2007ue}
\bibinfo{author}{\bibfnamefont{R.~B.} \bibnamefont{Tully}},
  \bibinfo{author}{\bibfnamefont{E.~J.} \bibnamefont{Shaya}},
  \bibinfo{author}{\bibfnamefont{I.~D.} \bibnamefont{Karachentsev}},
  \bibinfo{author}{\bibfnamefont{H.~M.} \bibnamefont{Courtois}},
  \bibinfo{author}{\bibfnamefont{D.~D.} \bibnamefont{Kocevski}},
  \bibinfo{author}{\bibfnamefont{L.}~\bibnamefont{Rizzi}}, \bibnamefont{and}
  \bibinfo{author}{\bibfnamefont{A.}~\bibnamefont{Peel}},
  \bibinfo{journal}{Astrophys. J.} \textbf{\bibinfo{volume}{676}},
  \bibinfo{pages}{184} (\bibinfo{year}{2008}), \eprint{0705.4139}.

\bibitem[{\citenamefont{Tegmark et~al.}(2004)}]{Tegmark:2003uf}
\bibinfo{author}{\bibfnamefont{M.}~\bibnamefont{Tegmark}} \bibnamefont{et~al.}
  (\bibinfo{collaboration}{SDSS}), \bibinfo{journal}{Astrophys. J.}
  \textbf{\bibinfo{volume}{606}}, \bibinfo{pages}{702} (\bibinfo{year}{2004}),
  \eprint{astro-ph/0310725}.

\bibitem[{\citenamefont{Landau and Lifshitz}(1978)}]{LandauFluidMechanics}
\bibinfo{author}{\bibfnamefont{L.~D.} \bibnamefont{Landau}} \bibnamefont{and}
  \bibinfo{author}{\bibfnamefont{E.~M.} \bibnamefont{Lifshitz}},
  \emph{\bibinfo{title}{{Fluid mechanics}}} (\bibinfo{publisher}{Pergamon
  Press}, \bibinfo{address}{Oxford}, \bibinfo{year}{1978}).

\bibitem[{\citenamefont{Weinberg}(1972)}]{Weinberg:1972kfs}
\bibinfo{author}{\bibfnamefont{S.}~\bibnamefont{Weinberg}},
  \emph{\bibinfo{title}{{Gravitation and Cosmology}}} (\bibinfo{publisher}{John
  Wiley and Sons}, \bibinfo{address}{New York}, \bibinfo{year}{1972}).

\bibitem[{\citenamefont{Maartens}(1998)}]{Maartens:1998qw}
\bibinfo{author}{\bibfnamefont{R.}~\bibnamefont{Maartens}},
  \bibinfo{journal}{Phys. Rev.} \textbf{\bibinfo{volume}{D58}},
  \bibinfo{pages}{124006} (\bibinfo{year}{1998}), \eprint{astro-ph/9808235}.

\bibitem[{\citenamefont{Ellis et~al.}(2001)\citenamefont{Ellis, van Elst, and
  Maartens}}]{Ellis:2001ms}
\bibinfo{author}{\bibfnamefont{G.~F.~R.} \bibnamefont{Ellis}},
  \bibinfo{author}{\bibfnamefont{H.}~\bibnamefont{van Elst}}, \bibnamefont{and}
  \bibinfo{author}{\bibfnamefont{R.}~\bibnamefont{Maartens}},
  \bibinfo{journal}{Class. Quant. Grav.} \textbf{\bibinfo{volume}{18}},
  \bibinfo{pages}{5115} (\bibinfo{year}{2001}), \eprint{gr-qc/0105083}.

\bibitem[{\citenamefont{Aghanim et~al.}(2018)}]{Aghanim:2018eyx}
\bibinfo{author}{\bibfnamefont{N.}~\bibnamefont{Aghanim}} \bibnamefont{et~al.}
  (\bibinfo{collaboration}{Planck}) (\bibinfo{year}{2018}),
  \eprint{1807.06209}.

\bibitem[{\citenamefont{Schimd et~al.}(2005)\citenamefont{Schimd, Uzan, and
  Riazuelo}}]{Schimd:2004nq}
\bibinfo{author}{\bibfnamefont{C.}~\bibnamefont{Schimd}},
  \bibinfo{author}{\bibfnamefont{J.-P.} \bibnamefont{Uzan}}, \bibnamefont{and}
  \bibinfo{author}{\bibfnamefont{A.}~\bibnamefont{Riazuelo}},
  \bibinfo{journal}{Phys. Rev.} \textbf{\bibinfo{volume}{D71}},
  \bibinfo{pages}{083512} (\bibinfo{year}{2005}), \eprint{astro-ph/0412120}.

\bibitem[{\citenamefont{Johnson et~al.}(2014)}]{Johnson:2014kaa}
\bibinfo{author}{\bibfnamefont{A.}~\bibnamefont{Johnson}} \bibnamefont{et~al.},
  \bibinfo{journal}{Mon. Not. Roy. Astron. Soc.}
  \textbf{\bibinfo{volume}{444}}, \bibinfo{pages}{3926} (\bibinfo{year}{2014}),
  \eprint{1404.3799}.

\bibitem[{\citenamefont{Koda et~al.}(2014)\citenamefont{Koda, Blake, Davis,
  Magoulas, Springob, Scrimgeour, Johnson, Poole, and
  Staveley-Smith}}]{Koda:2013eya}
\bibinfo{author}{\bibfnamefont{J.}~\bibnamefont{Koda}},
  \bibinfo{author}{\bibfnamefont{C.}~\bibnamefont{Blake}},
  \bibinfo{author}{\bibfnamefont{T.}~\bibnamefont{Davis}},
  \bibinfo{author}{\bibfnamefont{C.}~\bibnamefont{Magoulas}},
  \bibinfo{author}{\bibfnamefont{C.~M.} \bibnamefont{Springob}},
  \bibinfo{author}{\bibfnamefont{M.}~\bibnamefont{Scrimgeour}},
  \bibinfo{author}{\bibfnamefont{A.}~\bibnamefont{Johnson}},
  \bibinfo{author}{\bibfnamefont{G.~B.} \bibnamefont{Poole}}, \bibnamefont{and}
  \bibinfo{author}{\bibfnamefont{L.}~\bibnamefont{Staveley-Smith}},
  \bibinfo{journal}{Mon. Not. Roy. Astron. Soc.}
  \textbf{\bibinfo{volume}{445}}, \bibinfo{pages}{4267} (\bibinfo{year}{2014}),
  \eprint{1312.1022}.

\bibitem[{\citenamefont{Buchert et~al.}(2016)\citenamefont{Buchert, Coley,
  Kleinert, Roukema, and Wiltshire}}]{Buchert:2015wwr}
\bibinfo{author}{\bibfnamefont{T.}~\bibnamefont{Buchert}},
  \bibinfo{author}{\bibfnamefont{A.~A.} \bibnamefont{Coley}},
  \bibinfo{author}{\bibfnamefont{H.}~\bibnamefont{Kleinert}},
  \bibinfo{author}{\bibfnamefont{B.~F.} \bibnamefont{Roukema}},
  \bibnamefont{and} \bibinfo{author}{\bibfnamefont{D.~L.}
  \bibnamefont{Wiltshire}}, \bibinfo{journal}{Int. J. Mod. Phys.}
  \textbf{\bibinfo{volume}{D25}}, \bibinfo{pages}{1630007}
  (\bibinfo{year}{2016}), \eprint{1512.03313}.

\bibitem[{\citenamefont{Cembranos
  et~al.}(2019{\natexlab{b}})\citenamefont{Cembranos, Maroto, and
  Villarrubia-Rojo}}]{magnetic_paper}
\bibinfo{author}{\bibfnamefont{J.~A.~R.} \bibnamefont{Cembranos}},
  \bibinfo{author}{\bibfnamefont{A.~L.} \bibnamefont{Maroto}},
  \bibnamefont{and}
  \bibinfo{author}{\bibfnamefont{H.}~\bibnamefont{Villarrubia-Rojo}}
  (\bibinfo{year}{2019}{\natexlab{b}}), \eprint{work in preparation}.

\bibitem[{\citenamefont{Durrer and Neronov}(2013)}]{Durrer:2013pga}
\bibinfo{author}{\bibfnamefont{R.}~\bibnamefont{Durrer}} \bibnamefont{and}
  \bibinfo{author}{\bibfnamefont{A.}~\bibnamefont{Neronov}},
  \bibinfo{journal}{Astron. Astrophys. Rev.} \textbf{\bibinfo{volume}{21}},
  \bibinfo{pages}{62} (\bibinfo{year}{2013}), \eprint{1303.7121}.

\bibitem[{\citenamefont{Harrison}(1970)}]{harrison1970generation}
\bibinfo{author}{\bibfnamefont{E.}~\bibnamefont{Harrison}},
  \bibinfo{journal}{Monthly Notices of the Royal Astronomical Society}
  \textbf{\bibinfo{volume}{147}}, \bibinfo{pages}{279} (\bibinfo{year}{1970}).

\bibitem[{\citenamefont{Takahashi et~al.}(2005)\citenamefont{Takahashi, Ichiki,
  Ohno, and Hanayama}}]{Takahashi:2005nd}
\bibinfo{author}{\bibfnamefont{K.}~\bibnamefont{Takahashi}},
  \bibinfo{author}{\bibfnamefont{K.}~\bibnamefont{Ichiki}},
  \bibinfo{author}{\bibfnamefont{H.}~\bibnamefont{Ohno}}, \bibnamefont{and}
  \bibinfo{author}{\bibfnamefont{H.}~\bibnamefont{Hanayama}},
  \bibinfo{journal}{Phys. Rev. Lett.} \textbf{\bibinfo{volume}{95}},
  \bibinfo{pages}{121301} (\bibinfo{year}{2005}), \eprint{astro-ph/0502283}.

\bibitem[{\citenamefont{Fenu et~al.}(2011)\citenamefont{Fenu, Pitrou, and
  Maartens}}]{Fenu:2010kh}
\bibinfo{author}{\bibfnamefont{E.}~\bibnamefont{Fenu}},
  \bibinfo{author}{\bibfnamefont{C.}~\bibnamefont{Pitrou}}, \bibnamefont{and}
  \bibinfo{author}{\bibfnamefont{R.}~\bibnamefont{Maartens}},
  \bibinfo{journal}{Mon. Not. Roy. Astron. Soc.}
  \textbf{\bibinfo{volume}{414}}, \bibinfo{pages}{2354} (\bibinfo{year}{2011}),
  \eprint{1012.2958}.

\bibitem[{\citenamefont{Saga et~al.}(2015)\citenamefont{Saga, Ichiki,
  Takahashi, and Sugiyama}}]{Saga:2015bna}
\bibinfo{author}{\bibfnamefont{S.}~\bibnamefont{Saga}},
  \bibinfo{author}{\bibfnamefont{K.}~\bibnamefont{Ichiki}},
  \bibinfo{author}{\bibfnamefont{K.}~\bibnamefont{Takahashi}},
  \bibnamefont{and} \bibinfo{author}{\bibfnamefont{N.}~\bibnamefont{Sugiyama}},
  \bibinfo{journal}{Phys. Rev.} \textbf{\bibinfo{volume}{D91}},
  \bibinfo{pages}{123510} (\bibinfo{year}{2015}), \eprint{1504.03790}.

\bibitem[{\citenamefont{Meurer et~al.}(2017)}]{sympy}
\bibinfo{author}{\bibfnamefont{A.}~\bibnamefont{Meurer}} \bibnamefont{et~al.},
  \bibinfo{journal}{PeerJ Computer Science} \textbf{\bibinfo{volume}{3}},
  \bibinfo{pages}{e103} (\bibinfo{year}{2017}).

\bibitem[{\citenamefont{Peeters}(2007)}]{Peeters:2007wn}
\bibinfo{author}{\bibfnamefont{K.}~\bibnamefont{Peeters}}
  (\bibinfo{year}{2007}), \eprint{hep-th/0701238}.

\bibitem[{\citenamefont{Peeters}(2018)}]{peeters2018cadabra2}
\bibinfo{author}{\bibfnamefont{K.}~\bibnamefont{Peeters}},
  \bibinfo{journal}{Journal of Open Source Software.}
  \textbf{\bibinfo{volume}{3}}, \bibinfo{pages}{1118} (\bibinfo{year}{2018}).

\end{thebibliography}

\end{document}